\documentclass[preprint,showpacs,preprintnumbers,amsmath,amssymb,nofootinbib]{revtex4}

\usepackage{amsthm}
\newtheorem{thm}{Theorem}[section]
\newtheorem{corollary}
{Corollary}[section]

\newtheorem{definition}{Definition}[section]
\usepackage{mathrsfs}
\usepackage{etex}
\usepackage{amssymb,amsthm,amscd,amsbsy,array}
\usepackage{bm}
\usepackage{amsmath,dsfont}
\usepackage{soul} 
\usepackage{graphics,graphicx,xcolor}

\usepackage{tikz}
\usetikzlibrary{shapes,calc,arrows,arrows.meta,decorations.markings,angles,math}

\usepackage[colorlinks=true, pdfstartview=FitV, linkcolor=blue, citecolor=blue, urlcolor=blue]{hyperref} 

\newcommand{\magenta}{\textcolor{magenta}}
\newcommand{\red}{\textcolor{red}}

\newcommand{\blue}{\textcolor{blue}}

\newcommand{\orange}{\textcolor{orange}}

\newcommand{\gb}{\quad\colorbox{green}}

\newcommand{\dgreen}{\textcolor[rgb]{0,0.5,0}}

\newenvironment{redtext}{\color{red}}
{\ignorespacesafterend}
\newenvironment{bluetext}{\color{blue}}{\ignorespacesafterend}
\newenvironment{greentext}{\color{green}}{\ignorespacesafterend}
\newenvironment{magentatext}{\color{magenta}}{\ignorespacesafterend}
\newenvironment{cyantext}{\color{cyan}}{\ignorespacesafterend}
\newenvironment{orangetext}{\color{orange}}
{\ignorespacesafterend}

\newcommand{\bmagenta}{\begin{magentatext}}
\newcommand{\emagenta}{\end{magentatext}}
\newcommand{\bcyan}{\begin{cyantext}}
\newcommand{\ecyan}{\end{cyantext}}
\newcommand{\bblue}{\begin{bluetext}}
\newcommand{\eblue}{\end{bluetext}}
\newcommand{\bred}{\begin{redtext}}
\newcommand{\ered}{\end{redtext}}

\newcommand{\bgreen}{\begin{greentext}}
\newcommand{\egreen}{\end{greentext}}
\newcommand{\borange}{\begin{orangetext}}
\newcommand{\eorange}{\end{orangetext}}

\numberwithin{equation}{section}
\renewcommand{\theequation}{\thesection.\arabic{equation}}
\let\ssection=\section
\renewcommand{\section}{\setcounter{equation}{0}\ssection}
\newcommand{\beq}{\begin{equation}}
\newcommand{\eeq}{\end{equation}}
\newcommand{\bec}{\begin{center}}
\newcommand{\ec}{\end{center}}

\newcommand{\ex}{{\mathrm{exo}}}
\newcommand{\ma}{{\mathrm{mag}}}

\newcommand{\GW}{{gravitational wave\,}}
\newcommand{\ba}{{\mathbf{a}}}
\newcommand{\bA}{{\mathbf{A}}}
\newcommand{\bB}{{\mathbf{B}}}
\newcommand{\cA}{{\mathcal{A}}}

\newcommand{\bsigma}{\boldsymbol{\sigma}}

\newcommand{\Ad}{\mathrm{Ad}}

\newcommand{\bb}{{\mathbf{b}}}

\newcommand{\bbeta}{\boldsymbol{\beta}}
\newcommand{\bpi}{\boldsymbol{\pi}}

\newcommand{\bx}{{\bm{x}}}
\newcommand{\bX}{{\bm{X}}}

\newcommand{\mB}{{\mathscr{B}}}
\newcommand{\mT}{{\mathscr{T}}}
\newcommand{\mH}{{\mathscr{H}}}
\newcommand{\mE}{{\mathscr{E}}}

\newcommand{\gE}{\mathfrak{E}}
\newcommand{\gA}{\mathfrak{A}}
\newcommand{\gG}{\mathfrak{G}}
\newcommand{\gB}{\mathfrak{B}}
\newcommand{\gC}{\mathfrak{C}}
\newcommand{\gP}{\mathfrak{P}}

\newcommand{\ICl}{{{$\cC_{t_0}$\;}}} 

\newcommand{\bc}{{\mathbf{c}}}

\newcommand{\cC}{{\mathcal{C}}}

\newcommand{\rC}{{\mathrm{C}}}

\newcommand{\Coad}{\mathrm{Coad}}

\newcommand{\carr}{{\mathfrak{carr}}}

\def\aand{{\quad\text{\small and}\quad}}
\def\where{{\quad\text{\small where}\quad}}
\def\with{{\quad\text{\small with}\quad}}
\def\ie{{\;\text{\small i.e.}\;}}

\newcommand{\bd}{{\mathbf{D}}}
\newcommand{\bD}{{\mathbf{D}}}

\renewcommand{\d}{\mathrm{d}}

\newcommand{\rB}{{\mathrm{B}}}

\newcommand{\diag}{\mathrm{diag}}

\newcommand{\cE}{{\mathcal{E}}}
\newcommand{\cB}{{\mathcal{B}}}

\newcommand{\bg}{{\mathbf{g}}}

\newcommand{\rg}{\mathrm{g}}

\newcommand{\fg}{\mathfrak{g}}

\newcommand{\bgamma}{\boldsymbol{\gamma}}

\newcommand{\cH}{{\mathcal{H}}}

\newcommand{\cL}{{\mathscr{L}}}
\newcommand{\cS}{{\mathscr{S}}}

\newcommand{\cM}{\mathcal{M}}

\newcommand{\cO}{{\mathcal{O}}}

\newcommand{\bomega}{{\boldsymbol{\omega}}}

\newcommand{\bp}{{\mathbf{p}}}

\newcommand{\bP}{{\bf P}}
\newcommand{\cP}{\mathcal{P}}

\newcommand{\bq}{{\bf q}}
\newcommand{\bQ}{{\bf Q}}
\newcommand{\cQ}{\mathcal{Q}}

\newcommand{\bu}{\mathbf{u}}

\newcommand{\br}{{\bm{r}}}

\newcommand{\SE}{\mathrm{SE}}
\newcommand{\SL}{\mathrm{SL}}

\newcommand{\SO}{\mathrm{SO}}

\newcommand{\rso}{\mathfrak{so}}

\newcommand{\Conf}{{\mathrm{Conf}}}

\newcommand{\cT}{\mathcal{T}}

\newcommand{\bv}{{\bf v}}

\def\bnabla{{\bm{\nabla}}}
\newcommand{\wG}{{\gB}}

\def\smallover\#1/\#2{\hbox{$\textstyle\frac{\#1}{\#2}$}} %

\def\bu{{\bm{u}}}
\def\bv{{\bm{v}}}
\def\bp{{\bm{p}}}
\def\bP{{\bm{P}}}

\def\bq{{\bm{q}}}

\def\parag{\hfil\break} 
\def\kikezd{\parag\underbar}

\def\bequ{\begin{enumerate}}
\def\eenu{\end{enumerate}}
\def\bitem{\begin{itemize}}
\def\eitem{\end{itemize}}

\def\beq{\begin{equation}}
\def\eeq{\end{equation}}
\def\beqa{\begin{eqnarray}}
\def\eeqa{\end{eqnarray}}
\def\nn{\nonumber}
\def\barray{\left(\begin{array}}
\def\earray{\end{array}\right)}
\def\barraynb{\begin{array}}
\def\earraynb{\end{array}}

\def\IR{{\mathbb{R}}} 
\def\II{{\mathbb{I}}} 
\def\IF{{\mathbb{F}}} 

\def\IC{{\mathbb{C}}} %
\def\IS{{\mathbb{S}}} 
\def\SL{{\rm SL}}
\def\GW{{gravitational wave\;}}
\def\GWs{{gravitational waves\;}}

\def\?{{\,\gb{\fbox{\texttt{??}}\;}}\,}

\def\p{{\partial}}

\def\Rarrow{{\quad\Rightarrow\quad}}

\def \p{{\partial}}

\newcommand{\bE}{{\mathbf{E}}}
\def\bP{\mathbb P}

\def\benu{\begin{enumerate}}
\def\eenu{\end{enumerate}}
\def\bitem{\begin{itemize}}
\def\eitem{\end{itemize}}

\def\bP{{\rm{\bf P}}}

\newcommand{\const}{\mathop{\rm const.}\nolimits}
\newcommand{\half }{\smallover{1}/{2}}

\newcommand{\la}{{\langle}}
\newcommand{\ra}{{\rangle}}
\def\smallover#1/#2{\hbox{$\textstyle\frac{#1}{#2}$}} %
\def\smallcirc{{\raise 0.5pt \hbox{$\scriptstyle\circ$}}}
\def\cabove(#1){\stackrel{\smallcirc}{#1}}
\def\ccabove(#1){\,\stackrel{\smallcirc\smallcirc}{#1}\,}
\def\cccabove(#1){\stackrel{\,\smallcirc\smallcirc\smallcirc}{#1}\,}
\def\2{{\smallover1/2}}

\def\cA{{\cal A}}
\def\boxit#1{
\vbox{\hrule\hbox{\vrule\kern4pt
\vbox{\kern5pt#1\kern5pt}\kern4pt\vrule}\hrule}
} 

\newcommand{\bigbox}[1]{\fbox{%
\rule[-20pt]{0pt}{45pt}$\;\;\displaystyle{#1}\;\;$}
}
\newcommand{\medbox}[1]{\fbox{%
\rule[-10pt]{0pt}{25pt}$\;\;\displaystyle{#1}\;\;$}%
}
\renewcommand{\theequation}{\thesection.\arabic{equation}}
\let\ssection=\section
\renewcommand{\section}
{\setcounter{equation}{0}\ssection}

\def\besub{\begin{subequations}}
\def\esub{\end{subequations}}

\begin{document}

\preprint{\texttt{arXiv:2212.02360v4}}

\title{Hall motions in Carroll dynamics\footnote{Research initiated by Christian Duval \cite{Duval95} and dedicated to his memory.}}
\author{L. Marsot$^{1,2,3}$\footnote{marsot3@mail.sysu.edu.cn},
P.-M. Zhang$^{4}$\footnote{corresponding author.  zhangpm5@mail.sysu.edu.cn},
M. Chernodub$^{5}$\footnote{mailto:maxim.chernodub@univ-tours.fr}
and
P.~A. Horvathy$^{5}$\footnote{horvathy@lmpt.univ-tours.fr}
}

\affiliation{
${}^1$School of Science, Shenzhen campus of Sun Yat-sen University, Shenzhen, Guangdong 518107, P.R. China
\\
${}^2$Aix Marseille Univ, Universit\'e de Toulon\\
 CNRS, CPT, Marseille, France
\\
${}^3$Department of Physics, Babe\c{s}-Bolyai University, Kogalniceanu Street, 400084 Cluj-Napoca, (Romania)
\\
${}^4$ School of Physics and Astronomy, Sun Yat-sen University, Zhuhai 519082, (China)
\\
${}^5$ Institut Denis Poisson CNRS/UMR 7013 - Universit\'e de Tours - Universit\'e d'Orl\'eans Parc de Grandmont, 37200, Tours, (France)
 \\
}
\date{\today}


\begin{abstract}
``Do Carroll particles move?'' The answer depends on the characteristics of the particle such as its mass, spin, electric charge, and magnetic moment. A massive Carroll particle (closely related to fractons) does not move; its immobility follows from Carroll boost symmetry which implies dipole conservation, but not conversely. A  massless Carroll particle may propagate by following the Hall law, consistently with the partial breaking of the Carroll boost symmetry. The framework is extended to Carroll field theory. In $d=2$ space dimensions, the Carroll group has a two-fold central extension which allows us to generalize the dynamics to massive and massless particles, including anyons. The anyonic spin and magnetic moment combine with the doubly-extended structure parameterized by two Casimir invariants interpreted as intrinsic magnetization and non-commutativity parameter. The extended Carroll particle subjected to an electromagnetic background field moves following a generalized Hall law which includes a Zeeman force. This theory is illustrated by massless, uncharged anyons with doubly-centrally extended structure we call exotic photons, which move on the horizon of a Black Hole, giving rise to an anyonic spin-Hall Effect.
\bigskip

\end{abstract}
\maketitle

\tableofcontents
\goodbreak

\section{Introduction}\label{Intro}

The Carroll algebra $\gC$, as first introduced by L\'evy-Leblond and Sen Gupta \cite{Leblond,SenGupta}, is a contraction of the Poincar\'e algebra $\gP$ in which the velocity of light is sent to zero instead of going to infinity as in the Galilean limit  \cite{Leblond,SenGupta,BacryLL,AnkerZiegler,SouriauSDSIV}. In the Carroll limit massive particles do not move, because they would otherwise violate the causality constraint imposed by special relativity \cite{Leblond,SenGupta,BacryLL,SouriauSDSIV,Ancille,Carrollvs,Bergshoeff14}.

Interest in Carrollian dynamics has long been limited, precisely, due to the lack of motion.

The situation started to change after physical applications were found \cite{Dautcourt,Henneaux,Thoughts}.
The celebrated  Bondi-Metzner-Sachs (BMS) group of General Relativity \cite{BMS} is, for example, a conformal extension of the Carroll group \cite{Bagchi,BMSCarr,ConfCarr,Duval2015,MBHhorizon}. Carrollian hydrodynamics has attracted considerable recent attention \cite{DonnayM,hydro,Freidel22}. Further applications include the memory effect in a plane gravitational wave \cite{Sou73,GenMemory,Carroll4GW,DGHMemory,Soft}.

The geometric approach is nicely complemented from the Lie algebra point of view \cite{Jose,Bergshoeff22,Figueroa-OFarrill:2018ilb}. Non elativistic (including Carrollian) gravitation has become a popular topic \cite{deBoer20,Andringa,Oling}.

A  sprain to the ``no-motion-for-Carroll'' tenet comes from  recent study of an ``exotic photon'' i.e., a massless Carroll particle with spin, magnetic moment but with no electric charge on the \emph{horizon of a Kerr-Newman black hole} (see sec. \ref{ToyMods}).  In \cite{Marsot21,MZHLett} we found that an ``exotic photon'' \emph{can move},  namely by following a \emph{Hall-type law}.

Further developments which overturn the negative assessment comes also from \emph{condensed matter physics}:  the \emph{limited mobility} of the quasiparticles called \emph{fractons} \cite{Pretkofractons,PretkoCY,Gromov:2018nbv,Seiberg,Bidussi,JainJensen,Ozkan,Grosvenor:2021hkn,Surowka,Zoo}
\footnote{The term ``fractons'', which is now widely  associated with immobile excitations in condensed matter systems, has been used earlier to denote particles that carry fractional charges~\cite{khlopov} and collective quantized phonon-like excitations on a substrate with a fractal structure~\cite{Alexander:1983}. These two objects should not be confused with the immobile excitations which are also called ``fractons''~\cite{Zoo,Pretkofractons,PretkoCY,Gromov:2018nbv,Seiberg,JainJensen,Grosvenor:2021hkn,Bidussi,Surowka}.}
follows from that they are indeed Carroll particles \cite{Bidussi,Surowka}: the {original defect} of Carrollian physics has become a virtue!

A link between these notions appear from the principal property of the fracton: it is a (quasi)particle which, if set in isolation, is unable to move in response to any applied force. However, fractons can move a collective excitations by forming combinations -- that distantly resemble bound states -- which that can propagate through the medium~\cite{Pretkofractons,PretkoCY}. More details on fractions can be found in Sec.~\ref{fractonSec}.

From the theoretical point of view, the \emph{move-or-not-to-move-dilemma} is related to the symmetry under shifting Carroll time  we denote by $s$,
\beq
\bx \to \bx,
\qquad
s \to s - \bb\cdot\bx\,,
\label{Cboost}
\eeq
where $\bx$ is the position.
what we shall call (see sec.\ref{genCarrollSec}) a Carroll or a \emph{C-boost} \cite{Carrollvs,Bidussi,MZHLett}.

This paper is devoted to a comprehensive study of the Carroll dynamics and its applications  to condensed matter physics and to general relativity.
\goodbreak

C-boosts are related to the \emph{dipole symmetry}  of  fractons~\cite{Seiberg,Bidussi,JainJensen,Grosvenor:2021hkn,Surowka,Ozkan}.
Close similarities can indeed be established between the Carroll algebra and the space-time symmetry group of fracton theories once the generator of  C-boosts is identified with the dipole generator,  whereas the Carroll-time translation operator (Hamiltonian) becomes the electric charge generator~\cite{Hartong:2015xda,Copetti:2019ooe,Pena-Benitez:2021ipo}.

\medskip
Historically, Hall-type motion arises when a charged particle is put into combined perpendicular electric and  magnetic fields in such a way that the forces compensate and the particle to moves forcelessly with the Hall velocity~\cite{EHall,QHE,Stone,Ezawa},
\beq
\left(\frac{dx^i}{dt\;}\right)^{\mathrm{Hall}} =\epsilon_{ij}\frac{E_j}{B}\,, \qquad i,j = 1,2, \qquad B \equiv B_3\,.
\label{Halllaw}
\eeq
(The motion is considered in a $d=2$ plane normal to the direction of magnetic field ${\boldsymbol B} = (0,0,B_3)$).
To fix our terminology, we shall agree that a Hall motion is one which satisfies relation~\eqref{Halllaw}.

Many years after Hall's original discovery, an anomalous Hall Effect (i.e. one with no magnetic field) was observed in certain ferromagnetic crystals and explained by an anomalous current \cite{KarpLutt}. A similar effect was proposed,  later, for massive particles with spin \cite{SPINHALL} and then further extended to light \cite{LightHall,DHFermat,SpinOptics} and to chiral fermions \cite{StoneAnomVel,DHchiral,ZHchiral,OanceaKumar}. Spin-Hall effects in curved space were considered in~\cite{Saturnini,GravSpin,DMS18,HarteOancea}.

Reviewing 150 years of research of Hall-type effects goes  beyond our scope here. The interested reader is advised to consult the literature, see, e.g. \cite{QHE,Stone,Ezawa}.
Here we just mention that a semi-classical explanation based on the two-parameter  ``exotic"  \emph{central extension} of the planar Galilei group  \cite{LLGal,Galexo,DH01,NCLandau,HPl} was proposed using a Berry phase--extended framework \cite{ChangNiu,AHEPLA,DHHMS}.

The r\^ole of \emph{central extensions} for physics has been first recognized by Bargmann in his seminal paper \cite{Barg54}: for a massive non-relativistic system it is not the Galilei group $\gG$ itself, but its \emph{$1$-parameter central extension} by the mass (called the Bargmann group \cite{SSD} and denoted by $\gB$), which is physically relevant.

Unexpectedly, the dynamics is indeed even richer in the plane: Galilean systems admit a second  (``exotic'') central extension \cite{LLGal,Duval95,BGGK,LSZ} which yields  \emph{non-commuting  position coordinates}
 \cite{Duval95,BGGK,LSZ,Galexo,DH01,Inzunza:2021vlb},
\beq
\big\{x,y\big\} =\frac{\kappa}{m^2} = \theta\,,
\label{NCtheta}
\eeq
where $\kappa$ is the extension parameter.

Returning to Carroll, a curious fact is that while in $d\geq3$ space dimensions the group has no nontrivial central extension,  in $d=2$  it does admit one  with \emph{two central parameters}, $\kappa_{\ex}$ and $\kappa_{\ma}$~\cite{Ancille,Azcarraga,Marsot21}
\footnote{We changed our notation
with respect to Ref.~\cite{Marsot21}:
$q_1 \to -\half \kappa_{\ma}$
and
$q_2 \to  +\half \kappa_{\ex}\,$. We provide also a ``dictionary'' from Souriau's notations to a more commun language \cite{Marsden,MarsdenRatiu}.} which escaped attention until recently.
 The doubly extended planar Carroll group will be denoted
by $\widehat{\gC}$. The associated dynamics has been worked out in \cite{Marsot21}.

In our investigations we are led by the remarkable analogy (called in \cite{Carrollvs} a ``duality") between extended planar Carrollian dynamics  outlined in Sec.~\ref{ExoCarrSec}  with its  ``exotic'' Galilean counterpart  \cite{Duval95,LLGal,Galexo,DH01,NCLandau}, highlighted by that the central charges combine, in both cases, with the external magnetic field, $B$, into an effective mass:
\beq
m^* = \left\{\barraynb{clcll} m\left(1-\displaystyle\theta \,eB\right)
& \quad \hbox{Galilei\,}\;\; \gG
\\[10pt]
m\left(1-\displaystyle\frac{\kappa_{\ex}}{m^2}\, B^* \right)\,
& \quad \hbox{Carroll}\;\; \gC\,
\earraynb\right.
\where
    B^* = eB + \kappa_{\ma}\,.
    \label{eq_B_star}
\end{equation}
The comparison of these equations allows us to clarify the physical sense of the double central extension: the parameter $\kappa_{\ex}$ in the Carrollian dynamics corresponds to the noncommutativity parameter $\theta$ of its Galilean counterpart (with the identification $\theta \leftrightarrow \kappa_{\ex} / m^2$) while the parameter $\kappa_{\ma}$ generates a magnetic-field background by shifting the physical background field\footnote{Notice that the field $B^*$ in Eq.~\eqref{eq_B_star} is defined without the standard electric charge factor $e$. This intentional omission keeps the field $B^*$ finite in the neutral-charge limit $e \to 0$.} as in \eqref{eq_B_star}.

The relationship between $B^*$ and $B$ is reminiscent the relationship between the magnetic fields $\boldsymbol{B}$ and $\boldsymbol{H}$, and the magnetization ${\boldsymbol M}$ inside a medium: ${\boldsymbol B} = {\boldsymbol H} + 4 \pi {\boldsymbol M}$, respectively, where we used Gaussian units. Within this interpretation, the parameter $\kappa_{\ma}$ corresponds to the $z$-component of the intrinsic magnetization of the material, $\kappa_{\ma} = 4 \pi M_z$.

The double-extension of the Carroll system has a peculiar relation with the anomalous Hall Effect (AHE) which emerges in the absence of the background magnetic field~$B$~\cite{KarpLutt}.

One could naively think that the AHE is caused by the intrinsic magnetization $\kappa_\ma$ which could play the role of a background magnetic $B$ inside the system even when the external magnetic field vanishes, $B=0$.
However, the Casimir invariant $\kappa_\ma$ is \emph{not} the origin of the anomalous Hall effect. It is the other invariant, $\kappa_\ex$, which is responsible for the AHE: it generates the non-commutativity in the phase space,~\eqref{NCtheta} with $\theta \neq 0$, which induces an anomalous velocity/momentum relation \eqref{Gvelrel} and leads in turn to Hall-type motion even in the absence of the background magnetic field \cite{Galexo,DH01,NCLandau,AHEPLA,DHchiral,ZHchiral,HPl,StoneAnomVel}. See also sec.\ref{exom0C}.

In solid-state language, the AHE is produced by the Berry curvature of occupied electronic bands which appears due to the topology of Fermi surfaces in particular ferromagnets when spin-orbit coupling is included~\cite{AHE}.

When the effective mass vanishes,  $m^*=0$, the system becomes singular and requires Hamiltonian (alias ``Faddeev-Jackiw'') reduction \cite{Galexo,DH01,NCLandau,FaJa}. In the Galilean case the only allowed motions then obey the Hall law \eqref{Halllaw}. The Carroll case is more subtle, as it is  explained in sections \ref{ExoCarrSec}
and \ref{BHh}.

Our fundamental tool to find a classical system with a given symmetry group is to apply ``(coadjoint) orbit method'' called, with a slight abuse, the  Kirillov, Kostant, and Souriau (KKS) method \cite{SSD,Kir,Kost}.

Our review is organized as follows. After a short review of the relevant Lie algebras and a pedestrian introduction to Souriau's presymplectic framework in section \ref{GeoZooSec}, we recall how  ``exotic'' (i.e., doubly extended) Galilean particles \cite{LLGal,BGGK,Duval95,LSZ,Galexo,DH01} are constructed by the KKS algorithm
 \cite{Eisenhart,DBKP,DGH91,BCG,Morand,CiambelliNull}.

Sec.\ref{genCarrollSec} discusses Carroll particles with no central extension  \cite{Leblond,SenGupta,BacryLL,Dautcourt,Henneaux,Carrollvs,Bergshoeff14,Ancille}. The unusual form of the C-boost momentum $\bd^{\mathrm{Gal}}$ in \eqref{mCboost}, whose conservation implies immobility, comes from
absence of the conventional kinetic term in \eqref{emHam} - \eqref{emCLag}.

Massless Carroll particles have a richer dynamics, though \ref{s:massless_carr}.
The relation of immobility and Carroll boost symmetry is highlighted in subsec.\ref{mCpart}.

Fractons, to which section \ref{fractonSec} is devoted, attract much current attention in condensed matter physics  \cite{Pretkofractons,PretkoCY,Gromov:2018nbv,Seiberg,Bidussi,JainJensen,Grosvenor:2021hkn,Surowka,Zoo}.
Their much-discussed immobility follows from the conservation of their \emph{dipole momentum} \cite{Bidussi,JainJensen,Surowka,Ozkan},
\beq
\bd = m\bx \,
\label{mdipolemom}
\eeq
which follows in turn from the relation of fractons with the Carrollian model, highlighted in sec.\ref{fieldCarrSec}. The wave equation of free fractons, eqn.~\eqref{freefrac2}, is indeed identical to the Carrollian one studied in \cite{Marsot21}. The intimate relation of dipole and Carroll symmetries  is explained in sec.\ref{fracsymmSec}.
The absence of spatial derivative terms from the Carroll wave equations  \eqref{DiracCarr} or \eqref{Carrfeq2} is the field-theory equivalent of absence of the usual kinetic term.

Doubly-extended Carroll particles \cite{Ancille,Azcarraga,Marsot21,MZHLett} are studied in sec. \ref{ExoCarrSec} and then extended to massless particles. Theorems \ref{ThmFreeNoExtendedMotion}-\ref{ThmIV.7} address the recurrent ``moving or not?'' question with answers listed in Tables \ref{CarrollTable1} and \ref{CarrollTable2}, and illustrated in sec.\ref{ToyMods}.

The Carrollian model is related to relativistic anyons in sec.~\ref{CarrPoinSec},
which are particles in two spatial dimensions with statistics described by a continuous parameter. The standard paradigm implies that, in three spatial dimensions, local particle fields can possess only two types of statistics: integer spins are associated with bosonic particles characterized by commuting fields, while half-integer spins describe fermionic particles described by anti-commuting fields, as follows from the spin-statistics theorem~\cite{Finkelstein:1968hy}. In two spatial dimensions, the spin-statistics theorem does not work anymore because the group of rotations becomes Abelian, and the value of particle spin can take any value~\cite{Leinaas:1977fm}. The latter property supports the existence of the third kind of particles, which was coined in Ref.~\cite{Wilczek:1982wy} ``the anyons''. Similarly to bosons and fermions, statistics plays a principal role in the physics of many-particle anyonic systems affecting their thermodynamical properties (for a recent account, see Ref.~\cite{Myers:2021}) as well as transport phenomena that include electric (super)conductivity~\cite{Chen:1989xs} and heat flow transport~\cite{Banerjee:2017} as well as Hall-type effects~\cite{Stern:2007}, not mentioning a vital role of anyons in topological quantum computations essential for the construction of a fault-tolerant quantum computer~\cite{Nayak:2008}.

In sec. \ref{GWmotion} the role of Carroll symmetry is highlighted by the geodesic motion in a plane gravitational wave \cite{Sou73,Carroll4GW}.
Sect.\ref{BHh} illustrates the general results by a comprehensive study of motion of an hypothetical massless particle with anyonic magnetic moment on the \emph{horizon of a black hole} \cite{MBHhorizon,Marsot21,MZHLett}.

Our notations are:  $\cC$ (for Carroll), $\cB$ (for Bargmann), etc are space-time manifolds whose isometry groups $\gC,\, \gB$, etc act through the matrices denoted by $\rC,\, \rB$, etc.
\goodbreak

\section{From Euclide to Carroll: a guided tour in the Symmetry Zoo}
\label{GeoZooSec}

In this section, we review the different geometries and symmetries which will appear through the paper. The presymplectic framework used to derive classical models from symmetries is also recalled.

\subsection{Galilean, Carrollian, and Aristotle geometries}\label{GCAgeo}

A relativistic structure is given by a couple $(\cM, g)$ : the manifold $\cM$ is endowed  with a Lorentzian metric $g$. Geometries that give rise to Galilean or Carrollian symmetries are less straightforward, though.
Before proceeding we recall, for future reference, a few definitions.

\begin{definition}
\textit{A \emph{Galilean structure} on a (d+1)-dimensional manifold $\cM$ is \cite{Cartan23,Trautman63,Havas64,Trautman67,Kunzle72}
given by a triple $(\cM,\gamma^{ab},\theta)$ where,}
\begin{enumerate}
\item $\gamma^{ab}$ ($a,b=0,1,\ldots,d$) \textit{is a twice-contravariant symmetric tensor field of signature} $(0,+,\ldots,+)$ ;
\item closed ``clock'' one-form $\tau=\tau_adx^a$  which spans
 the $1$-dimensional kernel of $\gamma^{ab}$.
\end{enumerate}
\label{GalSdef}
\end{definition}

The coordinates on $\cM$ are typically denoted by $(x^i,t)$  and are thought of as position and Galilean time. The flat structure is given by the clock $\tau = dt$ and the ``metric'' $\gamma=\delta^{ij}\partial_i \otimes \partial_j$.

\begin{definition}
\textit{A \emph{Newton-Cartan structure}} \cite{Cartan23,Trautman63,Havas64,Trautman67,Kunzle72,DHNewtonCartan} \textit{on a (d+1)-dimensional manifold $\cM$ with coordinates $x^i,t$ is
given by a quadruple $(\cM,\gamma^{ab},\theta,\Gamma)$, where}
\begin{enumerate}
 \item $(\cM,\gamma^{ab},\theta)$ is a Galilean structure as in definition  \ref{GalSdef} above;
 \item
 $\Gamma$ is a Galilean connection, i.e., a symmetric linear connection compatible with $\gamma$ and~$\tau$.
 \end{enumerate}
 \label{NCSdef}
\end{definition}

Note that the connection $\Gamma$ is \textit{not}
uniquely determined by the Galilei  structure $(M,\gamma,\theta)$, one really has an equivalence class of connections.

\begin{definition}
\textit{A \emph{weak Carroll structure} \cite{Henneaux} is a triple $(\cC, g, \xi)$ composed of a manifold $\cC$ of dimension $d+1$, endowed with:
\begin{enumerate}
\item a degenerate twice symmetric covariant tensor $(g_{\mu\nu})$ such that $\dim \ker (g_{\mu\nu})\!=\!1$\,;
\item a nowhere vanishing vector field $(\xi^\mu)$
in the kernel, $g_{\mu\nu}\xi^{\nu}=0$\,.
\end{enumerate}
\label{CarrWSdef}
}
\end{definition}

The coordinates are typically denoted by $(x^i,s)$ with $s$ referred to as ``Carrollian time". $\xi=\p_s$.

\begin{definition}
\textit{A \emph{strong Carroll structure} \cite{DGH91,Dautcourt,BMSCarr,ConfCarr,Carrollvs} is a quadruple $(\cC, g, \xi, \Gamma)$ composed of a manifold $\cC$ of dimension $d+1$, where
\begin{enumerate}
\item $(\cC, g, \xi)$ is a weak Carroll structure as in def. \ref{CarrWSdef} ;
\item $\Gamma$ is a Carrollian connection, i.e., a symmetric linear connection compatible with $g$ and $\xi$ (much like for Newton-Cartan structures in definition \eqref{NCSdef}).
\end{enumerate}
\label{CarrSSdef}
}
\end{definition}

\begin{definition}
\textit{An \emph{Aristotelian structure} \cite{Penrose67,JMSPG} is given by a $(d+1)$ dimensional manifold $\cM$ carrying
\begin{enumerate}
\item
a ``clock'' $1$-form $\tau=(\tau_{\mu})$
\item a symmetric tensor $g_{\mu\nu}$ of euclidean signature  such that
\item the $1$-dimensional kernel of $g_{\mu\nu}$ is spanned by
a vector $\xi^\mu$ dual to the clock, $\tau_{\mu}\xi^{\nu}=\delta_{\mu}^{\;\;\nu}$,
$
g_{\mu\nu}\xi^\mu=0\,.
$
\end{enumerate}
\label{ArisSdef}
}
\end{definition}
An Aristotelian structure thus combines the Galilean and the Carrollian definitions.

\begin{definition}
\textit{A \emph{Bargmann structure} \cite{Eisenhart,DBKP,DGH91,BCG,JMSPG} is a triple $(\cB, g, \xi)$ composed of a  $(d+2)$ dimensional manifold $\cB$ carrying
\begin{enumerate}
\item a Lorentzian metric $g$
\item a nowhere vanishing covariantly constant null vector field $\xi$
\end{enumerate}
\label{BargSdef}
}
\end{definition}
The key features of a Bargmann structure are that:
\begin{enumerate}
\item
A Newton-Cartan structure (definition \eqref{NCSdef}) is obtained by projection $\cM = \cB/\IR\xi$ \cite{DBKP,DHNewtonCartan}
\item
The $t = \const$ hypersurfaces $\cC_t$ of a Bargmann manifold $\cB$ are
Carroll manifolds \cite{Carrollvs}. More generally, any null-submanifold of a Lorentzian spacetime is a Carroll manifold \cite{Morand,CiambelliNull}.
\end{enumerate}

Most of our subsequent investigations focus at spacetimes in two spatial dimensions, therefore we collect the various ``non-Einsteinian'' symmetry groups in $d=2$. However these groups (except central extensions in general) can be defined at higher dimensions, so we keep the notation $d$. We only list the nonzero Lie brackets. Non-trivial central extensions are also considered as they play an important role to build models for elementary particles \cite{LevyLeblond69,Kir,Kost,SSD}.
With some mild confusion between Lie groups and their generating algebras, we follow \cite{Penrose67,SSD}.

\benu
\item
The  \emph{Euclidean} group $\gE = \SE(d) \simeq \SO(d) \ltimes \IR^d$, whose action on space is,
\begin{equation}
x^i \mapsto R^i{}_j x^j + c^i\,.
\end{equation}
For $d = 2$, $\gE$ is generated by  $(J, P_i)$ s.t.\footnote{In arbitrary spatial dimension $d$, one can write the generators as $(M_{ij}, P_i)$ where $M$ is a skewsymmetric $d \times d$ matrix for rotation and $P$ for translation such that
\begin{align*}
[M_{ij}, M_{kl}] & = \delta_{jk} M_{il} - \delta_{ik} M_{jl} - \delta_{jl} M_{ik} + \delta_{il} M_{jk}\,,\\
[M_{ij}, P_k] & = \delta_{ik} P_j - \delta_{jk} P_i
\end{align*}
However, since the present review is mainly about $d = 2$ physics, we will prefer the notation valid in only a specific dimension with $J=M_{12}$, here \eqref{Euccomm}.
}
\beq
[J, P_i] = \epsilon_{ij} P_j\,;
\label{Euccomm}
\eeq
\item
For $d = 2$, $\gE$ admits a 1-parameter ``magnetic'' central extension $\widehat{\gE} \simeq \gE \ltimes \IR$ \cite{MarsdenRatiu} with generators  $(J, P_i, A_{\ma})$ which satisfy \footnote{$A_{\ma}$, etc are elements of the extended Lie algebra while $\kappa_{\ma}$ etc  belong to the dual algebra.}:
\beq
[J, P_i] = \epsilon_{ij} P_j\,,
\quad
[P_i, P_j] = \epsilon_{ij}A_{\ma}\,;
\label{extEuccomm}
\eeq
\item  The  \emph{Aristotle} group \cite{Penrose67,SSD} $\gA \simeq \SE(d) \times \IR$ is obtained by trivially extending the Euclidean group $\gE$ by time translations, which commute with the $\gE$ subgroup. This yields an action on spacetime,
\begin{equation}
\left(\begin{matrix}
x^i \\
t
\end{matrix}\right)
\mapsto
\left(\begin{matrix}
R^i{}_j x^j + c^i \\
t + h
\end{matrix}\right)
\label{AristAct}
\end{equation}
$\gA$, generated by
$(J, P_i, P_0)$,
\beq
[J, P_i] = \epsilon_{ij} P_j\,,
\label{Aristcomm}
\eeq
 is the group of automorphism of flat Aristotle structures (definition \eqref{ArisSdef}).
\item
The \emph{magnetic Aristotle} group is the central extension $\widehat{\gA} \simeq \gA \ltimes \IR$, for $d = 2$. It has generators $(J, P_i, P_0, A_{\ma})$ s.t.
\beq
[J, P_i] = \epsilon_{ij} P_j\,,
\quad
[P_i, P_j] = \epsilon_{ij} A_{\ma}\,.
\label{extAristcomm}
\eeq
Particles with planar Aristotle symmetry were constructed in \cite{AncilleAristotle}.

\item
The \emph{Galilei group} $\gG \simeq (\SO(d) \times \IR) \ltimes (\IR^d \times \IR^d)$,  is the group of automorphisms of the flat Newton-Cartan structure (see the definition \eqref{NCSdef}) and acts as,
\begin{equation}
\label{galilei group}
\left(\begin{matrix}
x^i \\
t
\end{matrix}\right)
\mapsto
\left(\begin{matrix}
R^i{}_j x^j + b^i t + c^i \\
t + h
\end{matrix}\right)\,.
\end{equation}
The generators  $(J, P_i, P_0, K_i)$ satisfy,
\beq
[J, K_i] = \epsilon_{ij} K_j\,, \quad
[J, P_i] = \epsilon_{ij} P_j\,,\quad
[K_i, P_0] = P_i\,.
\label{Galcomm}
\eeq

\item
The Galilei group has, in arbitrary dimension, a $1$-parameter central extension by the mass, $M$, \cite{Barg54} called the \emph{Bargmann} group. $\gB \simeq \gG \ltimes \IR$ is generated by  $(J_i, P_i, P_0, K_i, M)$,

\beq\barraynb{lll}
[J, K_i] = \epsilon_{ij} K_j\,,
&[J, P_i] = \epsilon_{ij} P_j\,,
\\[4pt]
[K_i, P_j] = \delta_{ij} M\,,
&[K_i, P_0] = P_i\,,
\earraynb
\label{Bargcomm}
\eeq

The elements of the Bargmann group can be conveniently represented by the $(d+3)\times (d+3)$ matrices \cite{S92},
\begin{equation}
\label{bargGroup}
\rB=
\left(\begin{matrix}
1 & -b_k R^k{}_j & -\bb^2/2 & b_0 \\
0 & R^i{}_j & \bb & \bc \\
0 & 0 & 1 & h\\
0 & 0 & 0 & 1
\end{matrix}\right)
\in \wG\,
\end{equation}
with $(R_{ij}) \in \SO(d),\, \bb,\bc\in \IR^d, h, b_0\in \IR$ representing rotations, Galilean boosts, space translations, Galilean  time translations and ``vertical'' translations along the central extension.
A $\gB$-matrix  acts affinely on Bargmann spacetime
 represented by the column vectors
$
\tiny {\begin{pmatrix}
	  s\\ x^j\\t\\1
\end{pmatrix}}
$
 by matrix multiplication, yielding,
\begin{equation}
\left(\begin{matrix}
s \\
x^i \\
t
\end{matrix}\right)
\mapsto
\begin{pmatrix}
s - b_i R^i{}_j x^j + b_0\\
R^i{}_j x^j + b^i t + c^i \\
t + h
\end{pmatrix}\,.
\label{Bargaction}
\end{equation}

\item
The \emph{planar Galilei} group has a 2-parameter ``exotic'' central extension $\widehat{\gB} \simeq \gG \ltimes (\IR \times \IR)$ \cite{Leblond,Galexo}, with
 generators  $(J, P_i, P_0, K_i, M, A_{\ex})$ :
\beq\barraynb{lll}
[J, K_i] = \epsilon_{ij} K_j\,,
&[K_i, K_j] = \epsilon_{ij} A_{\ex}\,,
&[J, P_i] = \epsilon_{ij} P_j\,,
\\[4pt]
[K_i, P_j] = \delta_{ij} M\,,
&[K_i, P_0] = P_i\,,
\earraynb
\label{exoGalcomm}
\eeq
which includes both the mass-extension ($M$) above and also the ``exotic'' extension specific for $d=2$, discussed below in sec.\ref{GalSec}.

\item
The \emph{Carroll} group $\gC \simeq \SE(d) \ltimes (\IR^d \times \IR)$ is the group of isometries of a flat strong Carroll structure (definition \eqref{CarrSSdef}), and has generators  $(J, P_i, P_0, K_i)$ with non-zero Lie brackets
\beq
[J, K_i] = \epsilon_{ij}K_j\,, \quad
[J, P_i] = \epsilon_{ij} P_j\,,\quad
[K_i, P_j] = \delta_{ij} P_0\,;
\label{Carrcomm}
\eeq

A representation of $\gC$ is readily obtained from that of the Bargmann group, \eqref{bargGroup}. The Carroll group
  is a subgroup of the latter, obtained by reducing the dimension  of the representation  by erasing the line and column which correspond to $t$ and providing us with $(d+2) \times (d+2)$  matrices,
\beq
\rC = \barray{lcl}
1 & -b^k R_{kj} & b_0
\\
0 &  R^i{}_j & c^i
\\
0 & 0& 1
\earray\, \in \gC\,,
\label{BCmatrix}
\eeq
where $(R_{ij})\in \SO(d)$ is a rotation, $\bc$ is a space translation, $\bb$ is a (Carroll or) C-boost, and $b_0=\const\in\IR$ is a ``Carrollian time translation''
(In order to distinguish it from Galilean time,  $t$,
we denoted the Carroll time by $s$).
 The affine action of  $\rC$ on $\tiny {\begin{pmatrix}
	  s\\ \bx \\1
  \end{pmatrix}}$ induces one on Carroll space-time $\cC = \big\{(s,\bx)\big\}$,
\beq
\barray{rc}
s\\ \bx
\earray
\mapsto
\barray{c}
s + b_0 - \bb\cdot R\bx\\
R\bx+\bc
\earray\,.
\label{sCarrollaction}
\eeq
This is the implementation we used for plane \GWs \cite{Carroll4GW}, see sec.\ref{GWmotion}.

We mention for completeness that the  Carroll group as introduced by L\'evy-Leblond \cite{Leblond} acts instead on
\emph{usual spacetime},
\beq
\begin{pmatrix}
	  \bx\\t
  \end{pmatrix}\to
  \begin{pmatrix}
	  R\bx+\bc\\{}t+\bb\cdot\bx+h
  \end{pmatrix}\,,
  \label{tCarrollaction}
\eeq
which can be obtained from the affine action of the matrix
\beq
\widetilde{\rC} = \barray{cll}
R^i{}_j &0 & c^i
\\
b_kR^k{}_j &1 &h
\\
0 & 0& 1
\earray \in \widetilde{\gC}\,
\label{tCarrollmatrix}
\eeq
which should to be compared with \eqref{sCarrollaction}.
It is {this}, ``$t$-based'' implementation that is considered in the BMS context \cite{BMSCarr, DonnayM,hydro,Freidel22}, and it is also the one that the authors of
 \cite{Bidussi} call the Carroll action. Eqns.
  \eqref{BCmatrix}-\eqref{sCarrollaction} correspond in turn to the ``internal'' or ``dipole symmetry'' of condensed matter physicists, see sec.\ref{fractonSec}.

 The ``$t$-based'' Aristotle group $\widetilde{\gA}$ involves $t$-translations by $h$ and is a subgroup of both of the $t$-based Carroll ($\widetilde{\gC}$) and Galilei groups. It is obtained by switching boosts off in \eqref{tCarrollaction}. Likewise,  an ``$s$-based'' Aristotle subgroup of $\gC$,  obtained by trading $t$ for $s$, i.e., by eliminating C-boosts from \eqref{BCmatrix}, involves $s$-translations by $b_0$.

Both implementations \eqref{sCarrollaction} and \eqref{tCarrollaction}  highlight the characteristic feature of Carroll structures: C-boosts leave the position invariant and shift instead time --- either $s$ or $t$, cf. \eqref{Cboost} and \eqref{tCarrollaction}.

While \eqref{BCmatrix} is, properly speaking, ``the'' Carroll group, we will call  ``(generalized) Carroll group'' any (conformal) isometry group of a chosen Carroll structure.

\item The group of isometries of a flat weak Carroll structure defined in \eqref{CarrWSdef} is infinite dimensional, and is denoted by
$\gA^\infty \simeq \SE(d) \ltimes C^\infty(\IR^d, \IR)$\footnote{The Galilean counterpart of this group (not used in this paper) is the \emph{Coriolis group} $C^\infty(\IR, \SE(d)) \ltimes \IR$.}. It acts on spacetime as
\beq
\barray{rc}
\bx\\ s
\earray
\mapsto
\barray{c}
R\bx+\bc\\
s + T(\bx)
\earray\,.
\label{Supercarrollaction}
\eeq
Compared to the Carroll group, we see that the $s$-translation and Carroll boosts are promoted to shifts by an  arbitrary function $T(\bx)$
and yields an infinite dimensional version of the Aristotle group.
While we may call  this group ``super Aristotle'' or ``infinitely extended Aristotle'', do note that it spans the isometries of a flat weak Carroll structure, and not of an Aristotle structure.

The shift in \eqref{Supercarrollaction} is analogous to the celebrated Bondi - van der Burg - Metzner - Sachs (BMS) group \cite{BMS}, which is indeed the conformal isometry group of a Carroll structure with topology $\IS^2\times\IR$ \cite{BMSCarr}. Hence, following the BMS literature, $s\to s+ T(\bx)$ will be called a \emph{supertransl}. Its zeroth-order term can be thought of as an $s$-translation, and the first-order term as a C-boost \eqref{Cboost}.

From the algebraic point of view, the Carroll group lies, ``at halfway" between the Aristotle and the super-Aristotle'' groups,
\begin{equation}
\gA \subset \gC \subset \gA^\infty\,.
\label{ACAinf}
\end{equation}
The Bargmann group $\gB$  contains \emph{two different copies of the Aristotle group} $\gA$: one projecting on Galilean space, and one acting on ``vertical'' Carroll space.  Indeed, we readily see from the Bargmann action on spacetime \eqref{Bargaction} that the first copy is the subgroup of $\gB$ without boosts and Carroll time translations, $b_i = 0$ and $b_0 = 0$, and projects on  Galilei spacetime  as in \eqref{AristAct},
while the second one is the subgroup of $\gB$ without boosts and Galilean time translations, yielding the action \eqref{sCarrollaction} on Carroll spacetime with $b_i = 0$  and $h = 0$,

\item
In the plane, for $d = 2$, we have a \emph{doubly extended Carroll} group $\widehat{\gC} \simeq \gC \ltimes (\IR \times \IR)$, which  is a 2-parameter central extension
with generators  $(J, P_i, P_0, K_i, A_{\ma}, A_{\ex})$ :
\beq\barraynb{lll}
[J, K_i] = \epsilon_{ij} K_j\,, \qquad
&[K_i, K_j] = \epsilon_{ij} A_{\ex}\,,
&
\\[4pt]
[J, P_i] = \epsilon_{ij}P_j\,,\quad
&[K_i, P_j] = \delta_{ij} P_0\,, \quad
&[P_i, P_j] = \epsilon_{ij} A_{\ma}\,.
\earraynb
\label{extCarrcomm}
\eeq
\eenu

At this point we mention a curious fact.
The planar Euclidean ($\gE$), Aristotle ($\gA$), and Carroll ($\gC$), groups all admit a one-parameter ``magnetic'' central extension, defined by the commutator of space translations, cf.  \eqref{extEuccomm}-\eqref{extAristcomm}-\eqref{extCarrcomm},
which could be interpreted as an additional, ``internal'' magnetic field. But this central extension does \emph{not} appear in the Galilei algebra $\gG$ in \eqref{exoGalcomm}, obtained by further extension by Galilean boosts: the magnetic extension is broken when passing from $\gA$ to $\gG$ (or from $\gC$ to $\gB$).
 One may wonder where the difference comes from.
Let us recall that the bracket of a Lie algebra should respect the Jacobi identity, which imposes strong constraints.
For the Galilei algebra the Jacobi identity for a time translation, a boost in one, and a spatial translation in the other direction,
$$
0=[P_0, [K_1,P_2]]+[K_1,[P_2,P_0]]+[P_2,[P_0, K_1]]=[P_2,\underbrace{[P_0, K_1]}_{K_1}]=
[P_2,P_1]
$$
requires the magnetic extension to vanish.
For the Carroll algebra instead, the identity is trivially satisfied  because  Carroll time translations commute with boosts, $[P_0, K_1] = 0$.
\goodbreak

\subsection{The (pre)symplectic framework and notations}
\label{s:presymplectic}

\kikezd{From Lagrange mechanics to presymplectic geometry}%

Classical mechanics is usually teached through Lagrangian or Hamiltonian mechanics. While these tools are perfectly fine to study simple enough systems, the results shown in this review call for a more advanced framework, which is (pre)symplectic geometry \cite{SSD}. While one may still find it possible to work in terms of Lagrangians, and most of the time we will write the corresponding Lagrangian, deriving them, through \textit{e.g.} the Kirillov-Kostant-Souriau theorem, or computing their symmetries, and so on, is a lot more convenient in the (pre)symplectic framework.

The following is a quick outline of \cite[\S 7]{SSD} with more standard notations. For more details, especially around subtle mathematical points, see \cite{Iglesias03} (in French).

Lagrangians are functions $L \in C^\infty(R \times TM)$, where $M$ is the space manifold, with $\dim M = n$, which are used to compute an action between $t_a$ and $t_b$,
\begin{equation}
A = \int^{t_b}_{t_a} L(t, \bx(t), \bv(t)) dt,
\end{equation}
where bold symbols are for vectors on $M$ (\ie \emph{spatial} vectors). While $\bv$ is, at this point, set to be $\bv(t) = d\bx(t)/dt$, this condition will be relaxed later on, and $\bv(t)$ will be considered as $n$ additional degrees of freedom. The variables describing the evolution of a dynamical system can thus be denoted $(t, \bx, \bv) \in \cE$, where $\cE$ will be called the ``evolution space''.

By introducing a new parameter $s$ to describe the curve
\begin{equation}
s \mapsto \left(t(s), \bx(s), \bv(s)\right)\,,
\end{equation}
and generalized coordinates,
\begin{equation}
q = \left(\begin{array}{c}
\bx \\
t
\end{array}\right) \in Q\,,
\qquad
\dot{q} = \left(\begin{array}{c}
\bv \\
\dot{t}
\end{array}\right) \in T_qQ\,,
\end{equation}
one can reparametrize the Lagrangian as\footnote{Note that it na\"ively looks like the Lagrangian is now defined as $l \in C^\infty(TQ)$, where $\dim TQ = 2n+2 = \dim(R \times TM) + 1$. This seemingly additional degree of freedom is however not relevant. One can always choose to write $\dot{t} = 1$, the other gauges (other than the forbidden $\dot{t} = 0$) being equivalent, see \cite{Iglesias03} for details.},
\begin{equation}
\label{homogeneized lagrangian}
l(q, \dot{q}) = L(t, \bx, \frac{\bv}{\dot{t}}) \dot{t}\,
\end{equation}
where the dot is the derivative with respect to $s$.

As is well-known, the Euler-Lagrange equations of motion are obtained by asking that the action is stationary,
\begin{equation}
\label{var_action}
\delta A = 0 \Leftrightarrow \left[\frac{\partial l}{\partial \dot{q}^\mu} \delta q^\mu \right]^{s_b}_{s_a} + \int^{s_b}_{s_a} \left(\frac{\partial l}{\partial q^\mu} - \frac{d}{ds} \frac{\partial l}{\partial \dot{q}^\mu} \right) \delta q^\mu ds = 0,
\end{equation}
where $\mu, \nu$ are indices running over the space $Q$.

The (canonical) momentum is defined as,
\begin{equation}
\label{def_p_from_l2}
p_\mu = \frac{\partial l}{\partial \dot{q}^\mu}
\end{equation}

Since Lagrangians of the form \eqref{homogeneized lagrangian} are homogeneous of degree one in velocity per their definition, \ie they obey $l(q, \lambda \dot{q}) = \lambda l(q, \dot{q})$, they can always be written $l(q, \dot{q}) = p_\mu \dot{q}^\mu$. Differentiating this latter expression leads to
\begin{equation}
\label{d_lp}
\frac{\partial l}{\partial q^\mu} dq^\mu = dp_\mu\dot{q}^\mu
\end{equation}

Using \eqref{def_p_from_l2} and \eqref{d_lp}, the variation of the action \eqref{var_action} can be written,
\begin{equation}
\label{var_action2}
\delta A = 0 \Leftrightarrow \left[p_\mu \delta q^\mu\right]^{s_b}_{s_a} + \int^{s_b}_{s_a} \left(\delta p_\mu  d\dot{q}^\mu -  dp_\mu \delta q^\mu \right) = 0
\end{equation}

Now, if one defines, with some abuse of notation, $(y^a) = \left(\begin{array}{c}
q^\mu \\
p_\mu
\end{array}\right) \in Y \simeq \cE$ \footnote{Recall that, given the above notations, $\dim \cE = 2n+1$, and coordinates on $\cE$ are $(\bx, \bv, t)$.}, where $a, b$ are indices running over the evolution space $\cE$, and the 1-form on $\cE$ \footnote{The basis of the space $T^*_y\cE$ is $dy^a = (dq^\mu, dp_\mu)$.},
\begin{equation}
\varpi = p_\mu dq^\mu \in T^*_y \cE
\end{equation}
then, contracting $\varpi$ with a variation $\delta y = (\delta q^\mu, \delta p_\mu)$ yields $\varpi_a \delta y^a = p_\mu \delta q^\mu$. This 1-form will be called ``Cartan 1-form''. In fact, since $l = p_\mu \dot{q}^\mu$, the Cartan 1-form is merely the Lagrangian ``up to $ds$'' (or $dt$ once one sets $\dot{t} = 1$) \cite{HPA79},
\beq
\int\!\varpi=\int\! l\, ds\,.
\eeq

The variation of the action \eqref{var_action2} can finally be written (again with some slight abuse of notation),
\begin{equation}
\delta A = 0 \Leftrightarrow \left[\varpi_a \delta y^a\right]^{s_b}_{s_a} + \int^{s_b}_{s_a} \half (d \varpi)_{ab} \delta y^a  \frac{dy^b}{ds} ds = 0\,,
\end{equation}
where $d\varpi$ is the exterior derivative of the 1-form $\varpi$, \ie $d\varpi = dp_\mu \wedge dq^\mu = dp_\mu \otimes dq^\mu - dq^\mu \otimes dp_\mu$. This 2-form will be denoted $\sigma = d\varpi$, and is called the \emph{presymplectic} 2-form, or \emph{Souriau form}.\footnote{Souriau calls his 2-form $\sigma$ the ``Lagrange form''. He claims indeed to have found his framework hidden in sect. V. of the second (1811) edition of  {\sl M\'ecanique Analytique} of Lagrange \cite{SouriauLagrange}. Opticians \cite{B59,H67} call it the Lagrange integral invariant or Lagrange bracket. }

Now, for the above equation to be true for all variations $\delta y$,
\begin{equation}
\sigma_{ab} \frac{dy^b}{ds} = 0
\end{equation}
must hold. In other words, $\frac{dy^b}{ds}$ must be in the kernel of $\sigma$ \footnote{The dimension of the kernel must be at least 1. While in standard classical mechanics the kernel has usually dimension 1, a physical example with a $3$-dimensional kernel is given by \emph{massless relativistic particles with spin} \cite{SSD,Saturnini,DHchiral}.}. The $\frac{dy^b}{ds}$ then yield the (Hamilton) equations of motion of the variational problem\footnote{Note that since we can always work with $\dot{t} = 1$, having the equations for $\frac{dy^b}{ds}$ means having the equations for $\frac{d\bx}{dt}$ and $\frac{d\bv}{dt}$.}.

As the above shows, the description of a dynamical system by a Lagrangian implies that it is possible to describe the system in terms of a 2-form $\sigma$ (the converse is not necessarily true). While the former description yields equations of motion after writing down the Euler-Lagrange equations, the latter description yields equations of motion after studying the kernel of the 2-form. The main selling point of the presymplectic description is that classical mechanics is recast in geometrical terms, which allows for a precise and powerful framework to prove theorems, such as the Kirillov-Kostant-Souriau theorem, which we will recall later on.

While the above derivation of presymplectic mechanics was done with Galilean mechanics in mind, \ie on a spacetime with a clear separation between space and time, it is more general. Any space $\cE$ endowed with a closed 2-form (of constant rank) is called a presymplectic manifold \cite{SSD}.

The last notion of presymplectic geometry we need to introduce for the understanding of this review is its version of Noether's theorem, which says, \cite[\S 11]{SSD},

\begin{thm}
\label{thm:noether}
\textit{\small An (infinitesimal) symmetry is given by a vector field $X$ on the evolution space $\cE$ which leaves the Souriau form \eqref{sigmaform} invariant, $L_X\sigma=0$, where $L_X = d \circ i_X + i_X \circ d$ is the Lie derivative along the direction defined by $X$. Then we have the conserved quantity $f_X$ \footnote{\label{fn: moment map}Provided that $i_X \sigma$ is an exact form. Also, in fact, $f_X = \mu \cdot \widetilde{X}$ where $\mu : \cE \rightarrow \fg^*$ is called the \emph{moment map} and $\widetilde{X} \in \fg$ is the Lie algebra element corresponding to its action $X \in T_y\cE$ on evolution space.}
\beq
i_X\sigma = -df_X\,.
\label{sigmaNoether}
\eeq
In terms of the Cartan form $\varpi$, the symmetry condition and conserved quantity are}
\beq
L_X\varpi = dK_X\,,
\qquad
f_X = i_X\varpi-K_X\,.
\label{varpiNoether}
\eeq
\label{sympNoether}
\end{thm}\vskip-8mm
In conventional terms, this requires that the Lagrangian changes by a total derivative.
Applications of the presymplectic framework are presented in the Appendix.

\medskip
For later use we recall also the field theory version of Noether's theorem using Lagrangian mechanics:

\begin{thm}\textit{\small If the action
$\cS\!=\!\displaystyle\int\!\cL\left(\phi \right)\!\sqrt{g}\,d^{d+1}\bx$,
where $g$ is the determinant of the metric, changes under a transformation $\phi \to \phi + \delta \phi$ by a surface term, $\delta\cS = \displaystyle\int\!\partial_{\alpha}K^{\alpha }d^{d+1}\bx$, then we have the conserved current%
\beq
J^{\alpha}=-\frac{\delta\left(\sqrt{g}\cL\right)}{\delta \left(\partial_{\alpha}\phi\right)}\delta\phi+K^{\alpha},
\qquad
\partial_{\alpha}J^{\alpha }=0\,,
\label{Ncurrent}%
\eeq
and thus a conserved charge,}%
\beq
Q=\!\int\!\left(-\frac{\delta\left(\sqrt{g}\cL\right) }{\delta \left( \partial
_{0}\phi \right) }\delta\phi +K^{0}\right) d^{3}\vec{x}\,.
\label{concharge}
\eeq
\label{fieldNoether}
\end{thm}

\kikezd{A Galilean example}%

As an example of the above presymplectic construction, consider standard Galilean mechanics with its well-known Lagrangian, $L(t, \bx, \bv) = \half m \bv^2 - U(\bx)$, which may also be written $l(q, \dot{q}) = \half m \bv^2/\dot{t} - U \dot{t}$.

One finds for the canonical momentum,
\begin{equation}
p = \frac{\partial l}{\partial \dot{q}}
 = \left(\begin{array}{c}
m \bv/\dot{t} \\
- \half m (\bv/\dot{t})^2 - U
\end{array}
\right)
=:
\left(\begin{array}{c}
\bp \\
- E
\end{array}\right)\,,
\end{equation}
and we easily verify that $p_i \dot{q}^i = m \dot{\bx}^2 / \dot{t} - \half m (\dot{\bx}/\dot{t})^2 \dot{t} - U \dot{t} = l(q, \dot{q})$.

The 1-form $\varpi = p_\mu dq^\mu$ is then,
\begin{equation}
\label{varpi_galilean_mechanics}
\varpi = \bp \cdot d\bx - \frac{\bp^2}{2m} dt - U dt\,,
\end{equation}
where the coefficient in front of the $dt$ is usually denoted $\cH = \frac{\bp^2}{2m} dt + U$, it is the Hamiltonian of the system. The 2-form $\sigma = d\varpi$ is then,
\begin{equation}
\label{sigma_galilean_mechanics}
\sigma = d\bp \wedge d\bx - \frac{\bp}{m} \cdot d\bp \wedge dt - \frac{\partial U}{\partial x^i} dx^i \wedge dt\,.
\end{equation}

We are interested in the equations of motion $\frac{dy^a}{ds} = \xi^a$ which are in the kernel of $\sigma$, where we will denote $(\xi^a) = (\xi_\bx^i, \xi^\bv_i, \xi_t)$. We want,
\begin{equation}
i_\xi \sigma = \sigma(\xi, \cdot) = \left(\xi^\bv_i + \xi_t \frac{\partial U}{\partial x^i}\right) dx^i + \left(\xi_t \frac{p^i}{m} - \xi_\bx^i\right) dp_i - \left(\frac{p^i \xi^\bv_i}{m} + \xi_\bx^i \frac{\partial U}{\partial x^i}\right) dt
\end{equation}
to vanish, which yields,
\besub
\begin{align}
& \xi^\bv_i + \xi_t \frac{\partial U}{\partial x^i} = 0\,, \\
& \xi_t \frac{p^i}{m} - \xi_\bx^i = 0\,, \\
& \frac{p^i \xi^\bv_i}{m} + \xi_\bx^i \frac{\partial U}{\partial x^i} = 0\,.
\end{align}
\esub

The first equation gives $\frac{dp_i}{ds} = - \frac{dt}{ds} \frac{\partial U}{\partial x^i}$, while the second is $\frac{dt}{ds} \frac{p^i}{m} = \frac{dx^i}{ds}$. The third equation is automatically satisfied. These two equations yield the usual
\besub
\begin{align}
\frac{dp_i}{dt} & = - \frac{\partial U}{\partial x^i} = F_i\,, \\
\frac{dx^i}{dt} & = \frac{p^i}{m} \,,
\end{align}
\esub
when accounting for the fact that we can work with $\dot{t} = 1$.

\kikezd{Phase space, Poissons brackets, symplectic geometry}%

Now, there is one more step in this framework which is to define the ``phase space'' $\cM$ of the system, or \emph{space of motions} according to the terminology in \cite{SSD}. It is obtained as the quotient $\cM = \cE / \ker \sigma$, and it is endowed with a \emph{symplectic form} $\Omega$ such that $\sigma = \pi^* \Omega$ where $\pi$ is the projection from $\cE$ to $\cM$. A symplectic form is a regular closed 2-form, in the sense that its kernel has dimension 0. A symplectic manifold is then a manifold $\cM$ endowed with a symplectic 2-form.

The evolution space $\cE$ contains the variables that describe the system along its worldline. For instance with the previous Galilean mechanics example, coordinates on $\cE$ are $(\bx, \bv, t)$, and following the trajectory of a system means following a line (assuming $\dim \sigma = 1$) $(\bx(s), \bv(s), t(s))$ in $\cE$, just like in Lagrangian mechanics. Hence, the evolution space is foliated by trajectories. Computing the quotient of $\cE$ by $\ker \sigma$ (this kernel representing the equations of motion) means ``reducing'' each trajectories down to a point. The space of motions $\cM$ thus holds its name, in that each point in $\cM$ represents a complete trajectory.

In a less rigorous, but more explicit way, consider a Galilean system, where $\dim \ker \sigma = 1$, parametrized by non-relativistic time $t$. The restriction to $t=t_0=\const$ of the evolution space, $\cE_{0}\equiv \cE_{t_0} = \{\bx, \bp, t_0\}$, is (a chart of) the ``phase space $\cM_0 = \{\bx, \bp\}$ at $t_0$''  and the restriction  of $\sigma$ to $\cM_{0}$ is then a  2-form  $\Omega_{0}=\sigma_{|_{\cM_0}}$ which we call the ``symplectic form  at $t_0$''. In a somewhat sloppy way, the choice of $t_0$ is neglected and one speaks of ``the'' phase space $\cM$ endowed ``the'' symplectic form $\Omega$.
 After the tacit choice of $t_0$ (say $t_0=0$), the Souriau 2-form can be split as
\beq
\sigma= \Omega - d\mathscr{H} \wedge dt\,,
\label{sigmaform}
\eeq
where $\mathscr{H}$ is the Hamiltonian. Compare with the example given in \eqref{sigma_galilean_mechanics}.
Note that while this decomposition does depend on the choice of $t_0$, different choices generate equivalent Hamiltonian structures.

Poisson brackets can now be introduced. Since the symplectic 2-form $\Omega=\half \Omega_{\alpha\beta} dz^\alpha \wedge dz^\beta$ (where the $z_\alpha$ are coordinates on $\cM$) is regular, $\Omega_{\alpha\beta}$ is invertible, $\Omega_{\alpha\beta}\Omega^{\beta\gamma} = \delta_{\alpha}^{\gamma}$. The Poisson bracket between two functions $f, g \in C^\infty(\cM)$ is,
\beq
\{f,g\} = \Omega^{\alpha\beta}\frac{\p f}{\p z^\alpha}
\frac{\p g}{\p z^\beta}\,.
\label{PBdef}
\eeq

\kikezd{The orbit method, central extensions, minimal coupling}%

A powerful result of (pre)symplectic geometry is the Kirillov-Kostan-Souriau (KKS) theorem \cite{SSD,Kir,Kost}, which is the cornerstone of the present review. This theorem states that
\begin{thm}
\label{thm:kks}
\textit{
Any coadjoint orbit of a Lie group is a symplectic manifold. Conversely, any $G$-homogeneous symplectic manifold is (locally) symplectomorph to a coadoint orbit of $G$ or of a central extension of $G$.
}
\end{thm}

As we have seen, the space of motions $\cM$ of a dynamical system is a symplectic manifold. A symplectic manifold is said to be $G$-homogeneous when a Lie group $G$ acts on it such that $\forall x, y \in \cM, \exists g \in G, y = g x$. In other words, it is $G$-homogeneous when the transformations in the group $G$ allow us to visit the whole manifold $\cM$ from any starting point. The physical intuition for homogeneity of the space of motions is that we can always relate two trajectories by a transformation, \textit{e.g.} we can always send one (free) Galilean trajectory to another Galilean trajectory by a suitable Galilean transformation, provided the Galilei group is a symmetry of the system.

A coadjoint orbit is an orbit of the coadjoint action of a group $G$ on the dual of its Lie algebra $\fg^*$. Let us first emphasize the physical relevance of $\fg^*$. There is a natural pairing between each symmetry, \ie between each generator of the Lie algebra $\fg$, and an element $\mu$ in the dual, which is called a  \emph{moment} \cite{SSD}. This can particularly be seen in Noether's theorem \ref{thm:noether}. As such, the dual of the Lie algebra is parametrized by physical quantities which describe the system. The coadjoint action of a group $G$ will then transform a moment, \ie a set of quantities describing a physical state, into another.

For instance for the Galilei group, elements $Z$ in the algebra are parametrized as $Z = (\bomega, \bbeta, \bgamma, \epsilon)$, for $\bomega$ infinitesimal rotations, $\bbeta$ boosts, $\bgamma$ spatial translations and $\epsilon$ time translation. Consequently, elements in the dual of the Galilei algebra are parametrized as $\mu = (\boldsymbol{\ell}, \bg, \bp, E)$, with $\boldsymbol{\ell}$ the angular momentum, $\bg$ the center of mass (or boost momentum), $\bp$ the linear momentum, and $E$ the energy, with the pairing $\mu \cdot Z = \boldsymbol{\ell} \cdot \bomega - \bg \cdot \bbeta + \bp \cdot \bgamma - E \epsilon \in \IR$ \footnote{The signs are choosen so that the moments have the expected sign later on. For instance, we want the energy $E$ to be positive.}. The coadjoint action of the Galilei group then tells us how a moment $\mu$, meaning a set of physical quantities $(\boldsymbol{\ell}, \bg, \bp, E)$, transforms to another, \ie how the momentum, the energy, etc transform when we change reference frame.

In order to define the coadjoint representation, one first needs to define the adjoint representation. The \emph{adjoint representation} $\Ad(g) : \fg \rightarrow \fg$ is such that
\begin{equation*}
\Ad(g) Z = g Z g^{-1}, \quad \forall g \in G, Z \in \fg\,.
\end{equation*}

Then, the \emph{Coadjoint representation} (or coadjoint action) $\Coad(g) : \fg^* \rightarrow \fg^*$, is such that,
\begin{equation*}
\left(\Coad(g) \mu \right) \cdot Z = \mu \cdot \left(\Ad(g^{-1}) Z\right), \quad \forall g \in G, \mu \in \fg^*, \forall Z \in \fg\,,
\end{equation*}
where the dot $\cdot : \fg^* \times \fg \rightarrow \IR$ is the pairing between the algebra and its dual.

Consider, again, the Galilei group. Knowing the group law, which can be immediately found from the action of the group on spacetime \eqref{galilei group}, we can compute the coadjoint representation of the Galilei group, $\Coad(g) \mu = (\boldsymbol{\ell}', g', p', E')$, with
\besub
\label{coadjoint action galilei}
\begin{align}
\boldsymbol{\ell}' & = R \boldsymbol{\ell} - \bb \times (R \bg) + \bc \times (R \bp)\,, \\
\bg' & = R (\bg - \bp e)\,, \\
\bp' & = R \bp\,, \\
E' & = E + \la \bb , R \bp \ra\,.
\end{align}
\esub

There are two important conclusions to draw from this example. The first one is that clearly, unlike a symplectic manifold, the dual $\fg$ is not (in general) homogeneous under the coadjoint action. For instance here, the norm of the momentum $\bp$ is invariant under the coadjoint action. So, starting from a moment with a given $\bp$, we cannot reach another physical state with a different norm of the momentum by a coadjoint action. This means that while, on a symplectic level, we can map one trajectory to any other possible trajectory with a suitable transformation in $G$, it is not possible to reach all moments (\ie physical states) from one moment. One then say that there are different \emph{orbits} of the coadjoint representation.

The \emph{Coadjoint orbits} of a moment $\mu_0 \in \fg^*$ are defined as
\begin{equation*}
\cO_{\mu_0} = \left\lbrace \mu = \Coad(g) \mu_0 \vert g \in G\right\rbrace \subset \fg^*\,.
\end{equation*}
In other words, an orbit $\cO_{\mu_0}$ is the set of all physical states a system can reach starting from a state $\mu_0$ by transformations in $G$. The orbits partition the dual $\fg^*$, which is depicted on the figure \ref{f:coadjoint orbits}.
\tikzset{middlearrow/.style={
        decoration={markings,
            mark= at position #1 with {\arrow{>}} ,
        },
        postaction={decorate}
    }
}
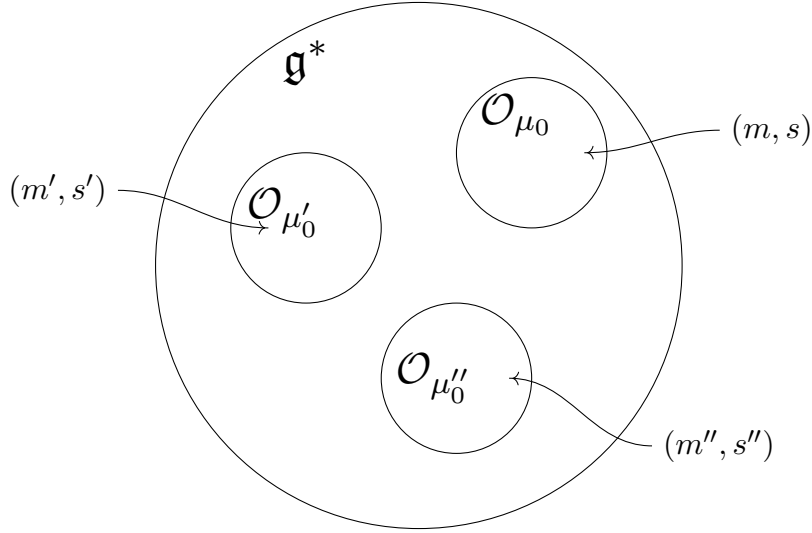
\begin{figure}[!ht]
\begin{tikzpicture}
  \draw (0,0) circle (3.5cm);
  \draw (1.5,1.5) circle (1cm);
  \draw (-1.5,0.5) circle (1cm);
  \draw (0.5,-1.5) circle (1cm);
  \node[scale=1.7] at (-1.5,2.7) {$\fg^*$};
  \node[scale=1.5] at (1.3,2) {$\cO_{\mu_0}$};
  \node[scale=1.5] at (-1.8,0.7) {$\cO_{\mu'_0}$};
  \node[scale=1.5] at (0.2,-1.5) {$\cO_{\mu''_0}$};
  \draw [middlearrow={0.98}] (-4,1) to[out=0,in=180] node[pos=0,left,scale=1.1]{$(m',s')$} (-2,0.5);
  \draw [middlearrow={0.98}] (4,1.8) to[out=180,in=0] node[pos=0,right,scale=1.1]{$(m,s)$} (2.2,1.5);
  \draw [middlearrow={0.98}] (3.1,-2.4) to[out=180,in=0] node[pos=0,right,scale=1.1]{$(m'',s'')$} (1.2,-1.5);
\end{tikzpicture}
\caption{Coadjoint orbits within the moment space $\fg^*$, and the Casimir invariants (here shown for a group which admits a mass $m$ and a spin $s$ as Casimir invariants, \textit{e.g.} the Poincar\'e group) that index them. Note that, unlike what the figure suggests, in general $\dim \cO_{\mu_0} < \dim \fg^*$.}
\label{f:coadjoint orbits}
\end{figure}

Now, as we have also seen from the Galilean example, there are invariants under the coadjoint action. These invariants then help index each coadjoint orbit, they are called \emph{Casimir invariants} \footnote{In the literature, it is more common to see Casimir invariants in the context of Lie algebras, not so much in the context of dual of Lie algebras. However due to the (linear) pairing between the two, these definitions are equivalent.}.

The Kirillov-Kostant-Souriau theorem then tells us that all coadjoints orbits $\cO_{\mu_0}$ are symplectic manifolds. In fact, an orbit will be the symplectic manifold that describe the dynamical system (\textit{e.g.} an elementary particle) corresponding to the Casimir invariants on this orbit.

\medskip

The second important observation from the coadjoint action of the Galilei group on Galilean moments, which is crucial to understand for section \ref{ExoCarrSec}, is that the transformations \eqref{coadjoint action galilei} do not match physical experiments. There is indeed a contradiction between experiment, which tells us that the action of a Galilean boost changes the momentum of the system, and the fact that this shift of momentum is missing in \eqref{coadjoint action galilei}. The solution to this is that the momentum should be shifted by $m \bb$, \ie the mass of the system times the boost. However, mass is not in the list of physical quantities appearing in the Galilei group (or rather, in the dual of its Lie algebra). As such, the coadjoint orbits, and symplectic manifolds derived from the Galilei group only correspond to massless dynamical systems. However, we know that massive Galilean systems exist, so mass must come from somewhere.

The above observation is linked to the very last part of the KKS theorem \ref{thm:kks}, which states that we may need to consider the \emph{central extension} of a given group. A central extension of a group $G$ is an extension by a group $E$ such that the extension group is in the center of the resulting group $H = G \ltimes E$. In practice, it is usually more convenient to consider central extensions at the level of Lie algebras. Given a Lie algebra $\fg$ with a set of generators $(A_i)$, computing a central extension of this algebra amounts to add a generator $B$ such that it commutes with all $A_i$ \footnote{And check that the resulting set of generators and their commutation relations do form indeed an algebra, \ie obey the Jacobi identity.}. If the resulting algebra is isomorphic to the original algebra $\fg$ then the extension is said to be trivial (and in group terms, one would then have $H = G \times E$), however if the resulting algebra is not isomorphic to $\fg$ then then central extension is said to be non-trivial.

Trivial central extensions have no physical meaning. However, non-trivial central extensions have an important impact on the physical parameters used to describe a system. Non-trivial central extensions display an accounting problem: a dynamical system may require more physical parameters to be described than the number of symmetries in the group. Also, such extensions display, in a sense, an obstruction of the possibility to represent a given group on certain spaces, typically on phase space and wavefunctions \cite{Barg54,LevyLeblond69}. This obstruction is then parametrized by a constant (or $n$ constants for an $n$ dimensional extension), which enters the list of physical parameters used to describe a system.

In particular, the Galilei group always features at least one\footnote{Non-trivial central extensions, which are related to the second cohomology group of $G$, $H^2(G)$, are a rather complicated topic and, when considering groups acting on spacetime, their number (the dimension of $H^2(G)$) typically depends on the dimension of spacetime.} non-trivial central extension, which yields the Bargmann group \cite{Barg54}, whose definition we recalled in \eqref{bargGroup}. It features an additional generator, whose dual we denote $m$: it corresponds to physical mass\footnote{One can think of this central extension situation for the Galilei group as follows. The Poincar\'e group has 10 generators (in $3+1$ dimensions), and dual to time and spatial translation generators are the energy and momentum. Now, in relativistic mechanics, the mass of the system is not a degree of freedom as it is computed from the energy and momentum. However, when taking the contraction $c \rightarrow \infty$ of the Poincar\'e group to obtain the Galilei group, the number of generators stays the same, but the relation between momentum, energy and mass is broken. The Galilean mass then appears as an additional degree of freedom, which can only be described by an extension of the Galilei group.}. The coadjoint action of the Bargmann group is then $\Coad(g) \mu = (l', g', p', E', m')$, with
\begin{align*}
\boldsymbol{\ell}' & = R \boldsymbol{\ell} - \bb \times (R \bg) + \bc \times (R \bp) + m \bc \times \bb\,, \\
\bg' & = R (\bg - \bp e) + m (\bc  - \bb e)\,, \\
\bp' & = R \bp + m \bb\,, \\
E' & = E + \la \bb , R \bp \ra + m \bb^2/2\,, \\
m' & = m\,,
\end{align*}
where now $\bp$ and the other physical quantities have the expected transformations under a group action. This coadjoint action has three Casimir invariants, provided $m \neq 0$,
\begin{align*}
C_1 & = S^2, \quad S = \boldsymbol{\ell} - \bg \times \bp / m, \\
C_2 & = E - \frac{\bp^2}{2m}, \\
C_3 & = m
\end{align*}

As the Galilean example shows, considering whether a group has non-trivial central extensions is crucially important in order not to miss complete classes of physical states (massive Galilean systems here).

\bigskip

Finally, after all the considerations explained above, one can start to build the symplectic structure on the orbits, in order to derive physical models for dynamical systems.

The Maurer-Cartan form (with a slight abuse of notation) $\Theta = g^{-1} dg$ is a one form on $G$ which takes value in the Lie algebra $\fg$. We can thus consider its pairing with an element $\mu_0$ in the dual $\fg^*$, which yields the Cartan 1-form $\varpi$ of the dynamical system which is able to reach the physical state $\mu_0$,
\begin{equation}
\varpi = \mu_0 \cdot \Theta\,.
\end{equation}
Note that the choice of basepoint $\mu_0$ within an orbit $\cO_{\mu_0}$ does not matter, as the (seemingly) different presymplectic structures obtained are equivalent. Hence, in practice, one chooses the simplest $\mu_0$ in the orbit of the system considered.

Let us give the example of a non-relativistic spinless massive particle with internal energy $E_0$ and mass $m$. Such a system is described ``at rest'' by the basepoint moment $\mu_0 = (0, 0, 0, E_0, m)$ in the dual of the Bargmann algebra. The contraction of $\mu_0$ with the Maurer-Cartan form $\Theta = g^{-1}dg$ of the Bargmann group $\wG$ displayed in \eqref{bargGroup} yields\footnote{After the reparametrization $h \rightarrow t$, $\bc \rightarrow \bx$, $\bb \rightarrow \bp/m$, $b_0 \rightarrow s$ on the group.} the Cartan resp. Souriau forms,
\besub
\begin{align}
& \varpi_0^{\mathrm{Gal}} = \mu_0 \cdot \Theta = -E_0 dt - m ds +  \bp\cdot d\bx -  \frac{\bp^2}{2m}\, dt\,,
\label{BCform}
\\
& \sigma_0^{\mathrm{Gal}} = d\varpi_0^{\mathrm{Gal}} = d\bp \wedge \left(d\bx-\frac{\bp}{m} dt\right)\,.
\label{BSform}
\end{align}
\label{BCSforms}
\esub
These forms are defined on the evolution space $\cE \simeq \wG / \SO(3)$ with coordinates $\big(\bx,\bp,t,s\big)$, where $t$ is non-relativistic time and $s$ is the ``vertical'' coordinate along the central extension.

We see that we recover the forms \eqref{varpi_galilean_mechanics} (up to irrelevant total differentials) and \eqref{sigma_galilean_mechanics}, for free particles.

The KKS construction is thus a powerful method to canonically obtain the dynamics of a system invariant under any Lie group $G$. However, it only yields models without interactions. There is no canonical way to obtain the model of a system with interactions, only heuristics.

In particular, we can introduce minimal coupling to an electromagnetic field represented by a closed 2-form on space-time, $\IF= \dfrac{1}{2} F_{\mu\nu}dx^\mu \wedge dx^\nu = E_i dx^i \wedge dt + \half F_{ij}dx^i\wedge dx^j$, where the spatial components of the field strength tensor encode the magnetic field, for instance $F_{ij}=\epsilon_{ijk}B_k$ in $3+1$ dimensions. Then Souriau's rule is to add $\IF$ to the free form $\sigma_0=d\varpi_0$ \footnote{Note that this minimal coupling rule is in general \emph{different} from the naive ``minimal coupling'' rule $p_\mu \to p_\mu - e A_\mu$ with $A^\mu = (A^0 \equiv V,{\boldsymbol{A}})$. An example where the two heuristics are different is, \textit{e.g.} an ``exotic'' Galilean particle \cite{Galexo,DH01} see Sec.~\ref{GalSec}.},
\beq
\sigma_0  \to
\sigma = \sigma_0 + e\IF \,,
\label{Smincoupling}
\eeq
where $e$ is the electric charge.

\section{Galilean particle dynamics}\label{GalSec}

While the main focus of this review is about Carroll dynamics, there are strong parallels with Galilean dynamics, and hence some review of the latter is warranted for future comparison, which is what this section will be used for.

As computed in the previous section \eqref{BCSforms}, the Souriau 2-form for a non-relativistic (Galilean) spinless particle with mass $m$ is
\begin{equation}
\sigma_0^{\mathrm{Gal}} = d\bp \wedge \left(d\bx-\frac{\bp}{m} dt\right)\,.
\end{equation}

Now, one of the equation of motion obtained when studying the kernel of this 2-form is the \emph{velocity/momentum relation},
\beq
\frac{d\bx}{dt}=\frac{\bp}{m}\,.
\label{velrel}
\eeq

While we usually take for granted that the momentum on phase space is related to the velocity of the particle, we will see with Carroll models that this is not always the case.

Note that massive non-relativistic systems can conveniently be studied in the ``Bargmann'' framework \cite{Eisenhart,DBKP,DGH91,Morand}, where dynamical systems are recast to move on a ``Bargmann'' spacetime, which has one more dimension than the usual non-relativistic spacetime. In practice, one adds a coordinate $s$, which is the same $s$ appearing in the 1-form \eqref{BCform}. The dynamics then project down to a non-relativistic spacetime, without $s$. Such considerations are particularly convenient since the isometries of a Bargmann spacetime span exactly the Bargmann group, which removes the need to consider central extensions.

The well-known conserved quantities associated with Galilean symmetry will not be reproduced here with the exception of the Galilean boost momentum,
 \beq
 \bd^{\mathrm{Gal}}=m\bx - t\,{\bp}\,,
 \label{Gboostmom}
 \eeq
 we record for later comparison  with \eqref{LLconsquant}, \eqref{mCboost}, \eqref{d0form0}, \eqref{dipolemom} and \eqref{exoCboost} and \eqref{Cfieldboostmom}.

Now, for the above free particle, we can write the 2-form as $\sigma_0^{\mathrm{Gal}} = \Omega_0-d\mathscr{H}_0\wedge dt$ with $\mathscr{H}_0=\bp^2/2m$. Considering electromagnetic (EM) interactions, in the static case, amounts to add $e\IF$ to $\sigma_0$ as in \eqref{Smincoupling}, which modifies both the symplectic form and the Hamiltonian, respectively,
\beq
\Omega_0  \to \Omega = \Omega_0 + \smallover{e}/{2} F_{ij} dx^i \wedge dx^j
\aand
\mathscr{H}_0 \to \mathscr{H}_{em} = \frac{\bp^2}{2m}+eV\,.
\label{OHincoup}
\eeq
The vector gauge field ${\boldsymbol A}$ and the electrostatic potential $V$ determine the magnetic and electric fields,
\beq
\bnabla \times \bA = \bB\,, \qquad \bnabla V = -\bE\,,
\label{eq_B_E}
\eeq
respectively.

We mention for completeness that by changing the basepoint $\mu_0$ (with different Casimir invariants, \ie considering another orbit $\cO_{\mu_0'} \neq \cO_{\mu_0}$), one can build models for: (i) spinning particles  \cite{LightHall,SSD,HPA79} and (ii) for \emph{massless particles}, in both the Relativistic and Galilean  frameworks \cite{SSD} including curved space.

Light which propagates in an optical medium can, in particular, be viewed as a massless particle which moves in a curved manifold \cite{DHFermat, SpinOptics, GravSpin,DFinsler}. The optical index  of the  medium, $n(\bx)$, allows us to define a metric
\beq
g_{ij} = n^2(\bx)\,\delta_{ij}
\eeq
on space such that the gradient, $\bnabla (1/n)$, plays the r\^ole of an effective ``electric'' force. When $n=\const$ the motion is free and instantaneous (as seen above). For light with spin $s$, the non-uniform optical index $n = n(\bx)$
yields an anomalous velocity term with approximate equations of motion eqn.~\# (14) of \cite{DHFermat},
\beqa
\dot{\br} \approx\bp-\frac{s}{k}\,\bnabla \left(\frac{1}{n}\right)\times\bp,
\qquad
\dot{\bp}\approx
-n^3k^2\bnabla\left (\frac{1}{n} \right)\,,
\label{LinDuv}
\eeqa
consistently with \cite{LightHall}.
\goodbreak

\kikezd{Exotic Galilean dynamics in the plane}%

In the plane the Galilei group has been known to admit a second,  central extension \cite{LLGal} which allows for an extended ``exotic'' dynamics \cite{BGGK,LSZ,Galexo,DH01,NCLandau,HPl}.
Now we spell out the Souriau framework for the ``exotic'' (i.e. doubly-centrally extended) Galilean particle with mass $m$ subjected to the background electromagnetic field. Setting
$
\theta={\kappa}/{m^2}\
$
as in \eqref{NCtheta}, the ``exotic'' Souriau  $2$-form
\beqa
\sigma_{\ex}^{\mathrm{Gal}}
&=&
\underbrace{
(dp_i-{e}E_i dt)\wedge(dx^i-\frac{p_i}{m\;}dt)
\;+\;
\half eB\,\epsilon_{ij}\,dx^i\wedge{}dx^j}_{\sigma^{\mathrm{Gal}}_{em}}
\;+\;\underbrace{\,\theta
\,\epsilon_{ij}\,dp^i\wedge{}dp^j
}_{exotic}\,\qquad
\label{GalexoS}
\eeqa
is defined on the evolution space $\cE=\IR^2\times\IR^2\times \IR = \Big\{\bx, \bp, t\Big\}$. It can be split as
$
\sigma_{\ex}^{\mathrm{Gal}}=\;\Omega_{\ex}^{\mathrm{Gal}}-d\cH\wedge dt\,,
$
where
\beq
\Omega_{\ex}^{\mathrm{Gal}}= \;\underbrace{d\bp\wedge d\bx+ \smallover{e}/{2} F_{ij} dx^i \wedge dx^j}_{\Omega_{em}} + \underbrace{\,
\frac{\kappa}{2m^2}
\,\epsilon_{ij}\,dp^i\wedge{}dp^j}_{exotic}\,\,,
\qquad
\mathscr{H}_{em} =\; \frac{\bp^2}{2m}+eV\,.
\label{GalexoSbis}
\eeq
The system is regular when the determinant of the symplectic form does not vanish,
\beq
\det(\Omega_{ij}^{\mathrm{Gal}})=\bigg(\frac{m^*}{m\;}\bigg)^2 \neq0\,,
\where
m^*\equiv m^*_G=m\big(1- \theta\,eB\big)\,.
\label{Galeffmass}
\eeq
Here
$m^*$ is an \emph{effective mass} \cite{Galexo,DH01}. Dropping the suffix $G$,
 the Poisson brackets associated with \eqref{GalexoS} are, for $m^*\neq0$,
\beq
\{x_{i},x_{j}\}=
\displaystyle\frac{m\,}{m^*}\,\theta \epsilon_{ij}
\qquad
\{x_{i},p_{j}\}=\displaystyle\frac{m\,}{m^*}\,\delta_{ij},
\qquad	
	\{p_{i},p_{j}\}=\displaystyle\frac{m\,}{m^*}\,eB
	\,\epsilon_{ij}\,,
\label{GalexoPBOld}
\eeq
or equivalently,
\beq
\{x_{i},x_{j}\}=
\left(\displaystyle\frac
{1}{1-\theta eB}\right)\theta\epsilon_{ij},
\;\;\;
\{x_{i},p_{j}\}=\left(\displaystyle\frac{1}{1-\theta eB}\right)\delta_{ij},
\;\;\;	
\{p_{i},p_{j}\}=\left(\displaystyle\frac{1}{1- \theta eB}\right)eB\epsilon_{ij}\,.\;
\label{GalexoPB}
\eeq
\! In particular, the coordinates do not commute when $\theta\neq0$.
The associated Hamilton equations  are
\besub
\begin{align}
\frac{m^*}{m\;}\dot{x}_{i}
\; =&\;
\frac{p_{i}}{m\,} - e \theta \epsilon_{ij}E_{j}\,,
\label{Gvelrel}
\\
\dot{p}_i = & \; e E_i +  e B \epsilon_{ij} \dot{x}^j\,.
\label{Gpdot}
\end{align}
\label{Galexoeqmot}
\esub
Note the velocity and the momentum cease to be parallel when $\theta\,\bE \neq0$. For $\theta=0$ we recover  the usual expressions for a charged particle in a constant magnetic field
 and for $B=0$ we get the exotic expression \#(2.3) in \cite{DH01}.
The system  becomes singular when the effective mass vanishes, $m^*=0$, and  Hamiltonian (alias ``Faddeev-Jackiw'') reduction \cite{FaJa,DH01} yields  \cite{Galexo,DH01,NCLandau}:

\medskip
\begin{thm}
\textit{eqn. \eqref{Galexoeqmot} implies that when its effective mass \eqref{Galeffmass} vanishes,}
\beq
m^*=0\,,
\label{Gm*0}
\eeq
\textit{and when the electric field is constant then an ``exotic'' Galilean particle follows the Hall law \eqref{Halllaw}. }
\label{ThmII.1}
\end{thm}

Indeed, for $m^* = 0$, one has $e \theta = 1/B$, and thus the first equation \eqref{Gvelrel} yields for the momentum $p_i = m \epsilon_{ij} \frac{E_j}{B}$. Now, if the electric field is constant, then $\dot{p}_i= 0$, and the second equation \eqref{Gpdot} reduces to the Hall law \eqref{Halllaw}.

The behavior  when the effective mass changes sign is studied conveniently using chiral decomposition \cite{PlChiral,Chiral11}. We have the Souriau form \eqref{GalexoS} is locally
$\sigma_{\ex}^{\mathrm{Gal}}= d \varpi_{\ex}^{\mathrm{Gal}}$\,, where the Cartan $1$-form
$\varpi_{\ex}^{\mathrm{Gal}}$ corresponds to the ``exotic'' phase space Lagrangian \cite{DH01}
\footnote{In $d=2$ spatial dimensions, the vector product is a pseudoscalar, $\ba\times\bb= \epsilon_{ij} a^i b^j$.},
\beq
\cL_{\ex}^{\mathrm{Gal}}=\underbrace{
(\bp-e\bA\,)\cdot \frac{d\bx}{dt}
-\mathscr{H}_{em}\,dt}_{\cL_{em}}
\;+\;\underbrace{
\frac{\theta}{2}
 \,\bp\times \frac{d\bp}{dt}}_{exotic}\;.
\label{GalexoLag}
\eeq
We record for further reference that in a constant $B$-field the \emph{conserved  angular momentum} is   \cite{DH01}
\begin{equation}
\ell=
\bx \times \bp+\half{eB}{\bx}{}^2+\half\theta \bp{}^2
+{\chi}\,,
\label{GalexAmom}
\end{equation}
where the real constant ${\chi}$ is identified as the \emph{anyonic spin} \cite{anyons,JNExospin, DHExoSpin}. The relation to planar vortex dynamics \cite{Bilbao,DoshiG} will be studied elsewhere.

\subsection{Galilean anyons with spin-field coupling}\label{sfcoup}

The planar model can  be further extended. In $d=2$ spatial dimensions, the anyonic spin is just a constant $\chi$, and a term involving the spin-field Hamiltonian\footnote{For an ``exotic photon'', the spin-field term \eqref{anymagmom} plays a particular role on the horizon, see Sec.~\ref{BHh}. Another
example is studied in \cite{WuZhuFeng}.},
\beq
\mathscr{H}_{any}=-\mu{\chi}B\,,
\label{anymagmom}
\eeq
where $\mu$ is a magnetic momentum can be added to the spinless expression \eqref{GalexoS}. Then we end up with the Souriau form of a massive, exotic anyon with spin in an electromagnetic (EM) field,
\beqa
\hskip-5mm\sigma^{\mathrm{Gal}}\hskip-4mm
&&=
\nn
\\ \hskip-6mm
&&\underbrace{(dp_i-{e}E_i dt)\wedge(dx^i-\frac{p^i}{m\;}dt)}_{electric}+
\underbrace{\half eB\,\epsilon_{ij}\,dx^i\wedge{}dx^j}_{magnetic}
+ \underbrace{\frac{\theta}{2}\,
\epsilon_{ij}\,dp^i\wedge{}dp^j}_{exotic}
+
\underbrace{\mu{\chi} d B \wedge dt}
_{spin-field} \,.\qquad\quad
\label{GalexoSspin}
\eeqa
The new term merely shifts the force,
\beq
eE_i \to E^*_i = e E_i + \mu{\chi} \partial_i B,
\label{Elshift}
\eeq
and the equations of motion  become again \eqref{Galexoeqmot} up to this shift.

Note the important role which is played by the magnetic field $B$ in Eq.~\eqref{Elshift}. The gradient of $B$ behaves as an effective electric field which adds up to the background electric field $\boldsymbol E$ in Eq.~\eqref{Elshift}. The effective electric field appears from the interaction of the anyonic spin with the background magnetic field described by the Hamiltonian~\eqref{anymagmom}. The spin-field interaction plays the same role as the electrostatic potential $e V$ in the Hamiltonian~\eqref{OHincoup} for a usual electrically charged particle. This analogy identifies the effective electrostatic potential for the anyons, $eV_{eff} = - \mu{\chi} B$. The inhomogeneities in the latter produce the force on the anyon which can be described, according to Eq.~\eqref{eq_B_E}, via an effective electric field  ${\boldsymbol E}_{\mathrm{eff}} = (\mu{\chi}/e) {\boldsymbol \nabla} B$.

When the effective mass vanishes, $m^* = 0$, the system becomes once again singular. However then the magnetic field,
  $B = B_{crit} = m^2/e\kappa$,  is necessarily \emph{constant}  and therefore the new, $\mu$-term drops out: the anyonic spin has \emph{no impact} on the (Hall) motions.

\goodbreak
Massless Galilean particle models can also be constructed by the KKS method by choosing appropriately the base point $\mu_0$ in the dual Lie algebra. They can be spinless (``Fermat particle''-- see sec.\ref{s:massless_carr}), or can also carry spin \cite{SSD,DHFermat,SpinOptics}. The same method works also in the relativistic case. Such a particle can be coupled to a background metric \cite{Saturnini,GravSpin,DMS18}.

The ``exotic'' model introduced in \cite{Galexo,DH01} has anomalous gyromagnetic factor $g=0$. More general cases are studied in \cite{AnAn}. An optical medium can  be viewed also as a special metric \cite{SpinOptics}; then for massless  particles we get a \emph{Spin-Hall effect for light} \cite{LightHall,DHFermat,SpinOptics,DHchiral}.

\subsection{Galilean/Bargmann symmetry in field theory
}\label{GalBfield}

So far we studied Galilean invariant classical models.
Now we present a field-theoretical exemple to be compared later with the discussion of fractons in sect.\ref{fractonSec}.

In \cite{LLDirac} L\'evy-Leblond proposed to describe a non-relativistic particle with spin $1/2$ in an electromagnetic field by a counterpart of the Dirac equation.
Although his equation can, of course, be studied in any dimension \cite{GomisNovell,DuvalLL,GGT,HPA92,Cho92}, here we restrict ourselves to $d=2$. Then it can be obtained  \cite{DHP95} by starting with the $4$-component Dirac equation in 1+2 dimensions,
\beq
\big(ic\gamma_{\pm}^\alpha D_\alpha-m\big)\psi_{\pm}=0\,,
\label{2dDirac}
\eeq
where $D_\alpha=\p_\alpha -ieA_\alpha$ is the gauge-covariant derivative, the two-component spinors $\psi_{\pm}$ are the chiral components and $\gamma_{\pm}^\alpha = \big(\pm(1/c)\sigma^3,i\sigma^1,i\sigma^2\big)$ are the 1+2 dimensional Dirac matrices. Setting
{\small\beq
\psi_+=e^{-imc^2t}\barray{c}\Psi_+\\{}\tilde{\chi}_+\earray
\aand
\psi_-=e^{-imc^2t}\barray{c}\tilde{\chi}_-\\{}\Psi_-\earray,
\eeq}
dropping $iD_t\tilde{\chi} \ll mc^2\tilde{\chi}$ ,
eqn. \eqref{2dDirac} becomes
\beq
\barraynb{lll}
iD_t\Phi-(\bsigma\cdot\bD){\chi}&=&0\,,
\\
(\bsigma\cdot\bD)\Phi+2mc\chi &=&0\,,
\earraynb
\label{LLeq}
\eeq
where $\Phi={\tiny\barray{c}\Psi_+\\{}\Psi_-\earray}$
and
$\chi=c{\tiny\barray{c}\tilde{\chi}_-\\{}\tilde{\chi}_+\earray}$.
Using  $(\bD\cdot\bsigma)^2 = \bD^2+eB\sigma_3$  allows us to deduce the 2nd-order equations
\beq
\barraynb{lll}
iD_t\Phi&=&-\frac{1}{2m}\big[\bD^2+eB\sigma_3\big]\Phi\,,
\\[6pt]
iD_t\chi &=&-\frac{1}{2m}\big[\bD^2+eB\sigma_3\big]\chi
-\frac{e}{2m}(\bsigma\cdot\bE)\Phi\,.
\earraynb
\label{Paulieq}
\eeq
The component $\Phi$ thus satisfies the two-component Pauli equation which further separates into two uncoupled scalar equations with opposite charges. The component $\chi$ loses in turn its proper dynamics and is determined by the large component, $\Phi$.

The free L\'evy-Leblond equation\footnote{The  discussion also holds true for the Schr\"odinger equation.}, \eqref{LLeq} with $A_\mu=0$, is, by construction, invariant w.r.t. the Bargmann group \cite{Barg54}--- a non-trivial central extension of the Galilei group \footnote{The full group of invariance of the L\'evy-Leblond equation (and of the Schr\"odinger equation) is the conformal extension of the Bargmann group, called the Schr\"odinger group \cite{DHPspinor}.}, cf. eqn. (8) and (26) in ref. \cite{LLDirac}, If one works with \emph{projective} representations, a Galilei boost by $\bb$ is implemented as,
\beq
\barray{c}
\Phi(\bx,t)\\ \chi(\bx,t)
\earray \to
e^{im(\bb\cdot\bx+\half\bb^2t)}
\barray{cc}
\II_2 & 0
\\
-\half \bsigma\cdot\bb &\II_2
\earray
\barray{c}
\Phi(\bx+\bb{t}, t)\\ \chi(\bx+\bb{t}, t)
\earray .
\label{BoostonLL}
\eeq
Thus boosts act on the component $\Phi$ as
a usual Galilei boost does on a scalar,
\beq
\Phi(\bx,t) \to e^{im(\bb\cdot\bx+\half\bb^2t)}\,\Phi(\bx+\bb t, t),
\label{Boostonscalarfield}
\eeq
however it mixes the $\Phi$ and $\chi$ components.
The well-known associated conserved quantities are listed in \cite{LLDirac}.

Studying Galilean field theories on a Newton-Cartan (N-C) spacetime (def. \ref{NCSdef}) is not convenient: the central extension  does not belong to the isometries of the N-C spacetime, therefore we  have no true, only projective representations of the Galilei group.

The natural context to study Galilean field theories is  provided by the Bargmann space  \cite{Eisenhart,DBKP,DGH91}, see def. \ref{BargSdef}. Bargmann structures have an additional spacetime coordinate $s$; the isometries of a flat Bargmann structure are given by the Bargmann group, see \eqref{Bargaction}. Bargmann wavefunctions are $\Psi(\bx, t, s)$ and are equivariant with respect to the new coordinate,
\beq
L_\xi \Psi = i m \Psi
\label{equivariance}
\eeq
($\xi = \partial_s$).
 This means that ``ordinary'' wavefunctions $\Phi$ are lifted to Bargmann space, where their phase freedom is incorporated into the $s$ coordinate.
\begin{equation}
\Psi(\bx, t, s) = e^{ims} \Phi(\bx, t)\,.
\end{equation}
It is this form which explains how wave functions (or scalar fields) behave under Bargmann/Galilei transformations, see \eqref{Bargaction}: the projective factor in \eqref{BoostonLL} and \eqref{Boostonscalarfield} follows, for example, from the Bargmann lift of the symmetry. Hence, these wavefunctions $\Psi(\bx, t, s)$ allow the definition of a true representation of the Bargmann group, without projective factors.

Lastly, the Bargmann structure is also natural for Galilean field theories in the sense of covariance. For instance, the Schr\"odinger equation on a Newton-Cartan spacetime is equivalent to a massless Klein-Gordon equation on a Bargmann structure, while the L\'evy-Leblond equation is equivalent to a massless Dirac equation on Bargmann space which has 1 higher dimension \cite{DHPspinor}.

These properties will play an important r\^ole for fractons, see sec.\ref{fractonSec}.

Applying Noether's theorem \ref{fieldNoether} to the spin $1/2$ L\'evy-Leblond system, we get \cite{LLDirac,DHPspinor}
\beq
J^0=|\Phi|^2,
\qquad
J^i =\frac{1}{2im}\Big(\Phi^{\dagger}\bnabla\Phi-(\bnabla\Phi)^{\dagger}\Phi\Big)+\bnabla\times \left(
\frac{1}{2m}\Phi^{\dagger}\sigma_3\Phi\ \right)\,.
\label{LLconscurr}
\eeq
The conserved quantities of a free particle, eqn. \# (5.6) of \cite{DHPspinor},
\beq
\barraynb{llll}
\bP &=& \displaystyle\int\underbrace{\left\{\frac{1}{2i}\big(\Phi^{\dagger}\bnabla\Phi-(\bnabla\Phi)^{\dagger}\Phi\big)\right\}}_{{\bm{\cP}}}d^2\bx
&\text{linear momentum}
\\[12pt]
\ell &=& \displaystyle\int\!\bx\times {\bm{\cP}}\,d^2\bx +\displaystyle\int\half\Phi^{\dagger}\sigma_3\Phi\,d^2\bx
&\text{angular momentum}
\\[10pt]
\bd^{\mathrm{Gal}}&=& m\! \displaystyle\int\!\Phi\Phi^{\dagger}\,\bx
\,d^2\bx -t\bP
&\text{boost momentum}
\\[10pt]
H &=& \displaystyle\int\left\{\frac{1}{2m}\bnabla\Phi(\bnabla\Phi)^{\dagger}\right\}\,d^2\bx
&\text{energy}
\\[14pt]
M &=& m\! \displaystyle\int\Phi\Phi^{\dagger}\,d^2\bx
&\text{mass}
\earraynb
\label{LLconsquant}
\eeq
span under Poisson bracket the Bargmann algebra.
 The angular momentum $\ell$ confirms  that the components of $\Phi$ have spins $\pm\half$. The conserved quantity associated with Galilean boosts, $\bd$,
should be compared with the boost momentum of a point particle, \eqref{Gboostmom}.

\section{Carroll dynamics} \label{genCarrollSec}

After recalling the Galilean case,
henceforth we focus our attention at Carroll particles which are indeed our main objects of interest.
The label $\{\cdot\}^{\mathrm{Car}}$ analogous to $\{\cdot\}^{\mathrm{Gal}}$ will  be understood and omitted.

The dynamics of a Carroll particle in an electromagnetic
field can be derived in two steps as outlined in section \ref{GalSec}: first a free model is constructed by the KKS method applied to the Carroll group \cite{SSD,Carrollvs,ConfCarr,Ancille,Marsot21}, which is then coupled to the em field by the rule \eqref{Smincoupling}.

An element $g=g(R, \bb, \bc, b_0)$ of the Carroll group $\gC$ is parametrized by $\frac{d(d+3)}{2} + 1$ numbers. Dual to them  are the ``moments'' \cite{SSD}, which are the physical parameters that describe the state of a particle at a point. A moment is $\mu=\mu(\ell, \bd, \bpi, m) \in \carr^*(d)$, where $\ell \in \rso(d)^*$ is the angular momentum, $\bd$ the boost momentum [center of mass], $\bpi$ the conserved momentum, and $m$ is an arbitrary real number.
The Carroll group acts on these moments with its coadjoint action,
\besub
\begin{align}
\ell' & = \ell + \bb \times R \bd - \bc \times R \bpi + m \bb \times \bc\\
\bd' & = R \bd + m \bc \\
\bpi' & = R \bpi + m \bb
\label{carrcoadpi} \\
m' & = m
\end{align}
\label{carrcoad}
\esub
The invariants under the coadjoint action are the  \emph{Casimir invariants}.
Notice that $m$ is \emph{not} an externally given quantity but a \emph{moment} and a \emph{Casimir invariant} associated now \emph{not} with a central extension as for Galilei/Bargmann theory, but with (Carrollian) \emph{``time'' translations} along Carrollian time $s$.
 It is a sort of ``Carrollian energy'' cf. \eqref{CarrollinB}.

The KKS  orbit calculation we follow here provides us with a mathematical framework but does \emph{not} give a physical interpretation for the moments $\ell, \bd, \bpi, m$.  In view of the physical applications which will come, for now we make the choice to interpret $m$ as the mass of a particle, due to the analogy of the Carroll and the Galilei groups.
 This interpretation is not fixed, though:  in fracton theory reviewed in sec.\ref{fractonSec} for instance,  it can be viewed instead as an electric charge, which can plainly take both positive or negative values. The mathematical framework applies to both cases.

The word ``momentum'' has indeed  three related but somewhat different meanings here. Firstly, one has  $\bp$, which we define to be always $m \bv$. Secondly there is the \emph{canonical} or \emph{generalized} momentum, defined as $\bP = \partial \cL / \partial \bx'$.
For a non relativistic particle in a magnetic field, for instance, $\bP = m \bv - e \bA$.
At last, ``momentum'' is also the \emph{conserved quantity} dual to space translation symmetry (a ``moment'' in Souriau's terminology \cite{SSD}), which we shall call here \emph{``impulsion"} and denote by $\bpi$. The three definitions coincide in some simple cases but not always, see, e.g., \eqref{consexoCarr}.

The dynamical description for Carroll particles is readily found by embedding  the Carroll group $\gC$ \eqref{BCmatrix} into the Bargmann group $\wG$ \eqref{bargGroup} as a subgroup  by \emph{``freezing out''} (Galilean) \emph{time translations} \cite{Carrollvs,Carroll4GW}  in $\gB$,
\beq
h=0\,.
\label{CarrollinB}
\eeq

The moments allow us to define Casimir invariants. As said above, the first one is $m$, the \emph{mass} of the particle.
Carroll particles can also carry spin, defined as an additional Casimir invariant. However the planar case discussed in Sec.~\ref{ExoCarrSec}  will be sufficient for our purposes here.

\subsection{Massive Carroll particles}\label{mCpart}

We consider first  the free massive case $m\neq0$ with no spin.
The $dt$ -terms in \eqref{BCSforms}  are switched off  and  we end up (consistently with  \# (A.9) and (A.10) in  \cite{Carrollvs}), with,
\besub
\begin{align}
\varpi_0 =& \; \;\bp\cdot d\bx - mds\,,
\label{Carr0Cart}
\\
\sigma_0  =& \;  d\varpi_0 = d\bp \wedge d\bx\,.
\label{Carr0Sou}
\end{align}
\label{genCarrollCS}
\esub
These forms are defined on  ``Carroll evolution space''
\beq
\cE=\IR^d\times\IR^d\times\IR = \Big\{\bx, \bp, s\Big\}\,.
\label{Carev}
\eeq
Note that Carrollian time, $s$, does not appear in the Souriau 2-form.

Compared to the Galilean theory of sec. \ref{s:presymplectic} and sec.\ref{GalSec},
not only  $E_0$, the internal energy is lost but, even more importantly, the usual \emph{kinetic term} $(\bp^2/2m)\,dt$  in \eqref{OHincoup} of the Galilean theory is \emph{missing} from the Cartan form \eqref{Carr0Cart} (or equivalently, from the Hamiltonian $\mathscr{H}$  \eqref{Carroll2sigma} and thus also from the $\cL$ Lagrangian \eqref{0CarrLag}).

The usual $-(\bp/m) ds$ is missing from behind the $d\bx$ in $\sigma_0$,
   \emph{leaving $\bp$ and $d{\bx}/ds$ unrelated}: the usual Galilean velocity relation \eqref{velrel} is lost.
This fundamental difference between Galilei and Carroll systems  would not be altered by
adding further (either electromagnetic or central extension) terms.

Splitting the Souriau form into symplectic  form and Hamiltonian as in \eqref{sigmaform},
\beq\medbox{
\sigma_0=\Omega_0 - d\mathscr{H}_0\wedge d s \with
\Omega_0 = d\bp \wedge d\bx
\aand
\mathscr{H}_0 \equiv 0\,,}
\label{Carroll2sigma}
\eeq
shows that a free massive Carroll particle
has an \emph{identically zero Hamiltonian}.
The corresponding Hamilton equations read as follows:
\beq
({x^i})^{\prime}=\{x^i,\mathscr{H}_0\}=0,\qquad {p}_i^{\prime}=\{p_i,\mathscr{H}_0\}=0\,,
\label{primeds}
\eeq
where the prime,
$
\{\,\cdot\,\}^{\prime} = d/ds\,,
$
denotes a derivative with respect to Carrollian time.
Equations~\eqref{primeds}
imply therefore:
\goodbreak

\begin{thm}
\textit{A free massive Carroll particle \underline{does not move}},
\beq
\bx(s)=\bx_0,
\qquad
\bp(s)=\bp_0\,,
\label{nofreemotion}
\eeq
\label{ThmFreeNoMotion}
\end{thm} \vskip-10mm
cf. (A.11) in \cite{Carrollvs}.
\goodbreak

\smallskip
A free massive particle is  Carroll-symmetric by construction and is also manifest by looking at \eqref{genCarrollCS}. The associated conserved quantities (Souriau's ``moment'' \cite{SSD}) in $d=2$ spatial dimensions are as follows:
\besub
\begin{align}
\ell =& \; \bx \times \bp &\text{angular momentum},
 \label{mCangmom}
 \\
\bpi =& \; \bp\ &\text{linear momentum},
\label{mCmom}
\\
\bd^{\mathrm{Car}} =& \; m \bx\, &\text{C-boost momentum}.
\label{mCboost}
\end{align}
\label{consCarr}
\esub
Comparison with the Galilean boosts in \eqref{Gboostmom} shows that the difference comes, once again, from the decoupling of position and momentum. These expressions will later
 be generalized  \eqref{consexoCarr}. The superscipt $\{\,\cdot\,\}^{\mathrm{Car}}$ will henceforth be dropped and we'll write simply $\bd^{\mathrm{Car}}\equiv\bd$.

\begin{corollary}
\textit{\small The conservation of the C-boost momentum $\bd$ in \eqref{mCboost} implies, consistently with Theorem \ref{ThmFreeNoMotion}, that a massive particle with Carroll boost symmetry \underline{can not move}.}
\label{nomotCoroll}
\end{corollary}
\goodbreak

\kikezd{Lagrangian formulation}.
The  Carrollian dynamics can also be described by a Lagrangian. In the free  massive case, replacing $\bp$ by $\bv=\bp/m$ and viewing $\bv$ and $\bx^{\prime}$ as independent  variables, the
Euler-Lagrange equation of
\beq
\cL_0 = m\bv\cdot \bx^{\prime}\,,
\label{0CarrLag}
\eeq
confirm the ``no motion'' conclusion \eqref{nofreemotion}:
 $\bv$ is indeed a Lagrange multiplier, yielding the free equation of motion $\bx^{\prime}=0$ in \eqref{primeds},
supplemented with $\bv=\bv_0=\const$
We note also \textit{en passant} that eqn. \eqref{0CarrLag} confirms the C-boost symmetry: it is enough to consider the Hamiltonian action
\beq
\cS = \int \!\cL_0 ds = \int\! m\bv\cdot d\bx\,,
\label{CarrHaction}
\eeq
which does \emph{not} involve $s$ and is therefore trivially invariant under C-boosts \eqref{Cboost}.
The Noether moment is $\bd$ in \eqref{mCboost}.

Coupling a Carroll particle to an electromagnetic (EM) field by the rule \eqref{Smincoupling} amounts to
\beq
m\bv \to m\bv-e\bA \;\aand\; 0\equiv \mathscr{H}_{0} \to \mathscr{H}_{em}= eV
\label{emHam}
\eeq
 which yields, in Lagrangian form,
 \beq
\cL_{em}=(m\bv - e\bA)\cdot\bx^{\prime}
 - eV\,.
\label{emCLag}
\eeq
Variation w.r.t. $\bp$ and $\bx$ viewed as independent variables implies  the equations of motion \# (3.21a-b) of Ref.~\cite{Marsot21},%
\beq
\bx^{\prime}= 0\,,
\qquad
\bv^{\prime} = \smallover{e}/{m} \bE\,.
\label{emCeqmot}
\eeq
Note that the magnetic field is not involved
here.
To understand how this comes about, note that the variational equations are,
\beq
(x^i)^{\prime}=0
\aand
mv_i^{\prime}=eE_i+{eB}\epsilon_{ij}(x^j)^{\prime}\,.
\label{mEMeq}
\eeq
The first equation here switches off the Lorentz-force in the second one: ``no motion'' implies ``no Lorentz force'', allowing us to  generalise Theorem \ref{ThmFreeNoMotion}: %

\begin{thm}
\textit{\small A \emph{massive} charged}  \textit{Carroll particle in an electromagnetic background field \underline{does not move}:
$
\bx = \bx_0=\const
$  }
\label{ThmEMNoMotion}
\end{thm}

This statement will be modified for centrally extended Carroll particles, see Theorem~\ref{ThmVI.4}.

We notice that while the presence of the electric field $\bE\neq0$  changes  $\bv$, this does not affect the position, $\bx$, in its  evolution w.r.t. Carrollean time $s$.\\
For completeness we record also the Souriau form obtained by the rule \eqref{Smincoupling},
\beq
\sigma_{em}  = \sigma_{0}+ e\IF = md\bv \wedge d\bx
+ \smallover{e}/{2} F_{ij} dx^i \wedge dx^j + eE_i dx^i \wedge ds\,,
\label{emCarrSou}
\eeq
whose kernel yields again \eqref{emCeqmot}\footnote{We also get
$\bE\cdot\bx^{\prime}=0$ i.e., that the velocity is perpendicular to the electric field. However this is reduced to an identity because of the first relation in \eqref{emCeqmot}. Spinning  Carroll equations in $d=3$ spatial dimensions were  presented in \cite{Marsot21}.}.
\goodbreak

\kikezd{Immobility and Carroll boost symmetry}

The relation between immobility and Carroll boost symmetry emphasized in Corollary \ref{nomotCoroll} can also be understood as follows.
Let us consider a massive particle and assume that it is invariant under Carroll boosts and time translations \emph{only}. Let us emphasis that since we do not consider rotation and spatial translation symmetries, the model considered here is more general than that of usual Carroll particles.
 This particle is described by
a $2d + 1$ dimensional evolution space $\cE = \big\{(\bx, \bv, s)\big\}$. Its Cartan 1-form is,
\begin{equation}
\varpi = A_i(\bx, \bv, s) dx^i + B_i(\bx, \bv, s) dv^i - \mathscr{H}(\bx, \bv, s) ds\,,
\label{CCartan}
\end{equation}
which is the most general 1-form on $\cE$. Here $\mathscr{H}$ is the Hamiltonian.

The equations of motion for this most-general dynamical system are then given by,
\begin{equation}
\label{general_eom}
\left\lbrace
\begin{array}{l}
\displaystyle \frac{dx^j}{ds}(\partial_j A_i - \partial_i A_j) + \frac{dv^j}{ds}(\partial_{v^j} A_i - \partial_i B_j) = - \partial_i \mathscr{H} - \partial_s A_i \\[0.5em]
\displaystyle \frac{dx^j}{ds}(\partial_{v_i} A_j - \partial_j B_i) - \frac{dv^j}{ds}(\partial_{v_j} B_i - \partial_{v_i} B_j) = \partial_{v_i} \mathscr{H} + \partial_s B_i
\end{array}
\right.
\end{equation}

The next step is to require Carroll boost-invariance and time translation invariance. This requires the Cartan 1-form obeys $L_X \varpi = df$  (cf. \ref{sympNoether}), for
\beq
X = \beta_i \partial_{v_i} - \beta_i x^i \partial_s,
\label{infCboost}
\eeq
\and for some function $f$, implying
$d L_X \varpi=0$ (cf. \ref{sympNoether}). Likewise for the time symmetry, $d L_T \varpi = 0$, with $T = \partial_s$.

These symmetry considerations require the following conditions,
\besub
\begin{align}
\label{conditions_dlxvarpi=0}
& \beta_k \partial_{v_k} (\partial_j A_i - \partial_i A_j) + \beta_i (\partial_s A_j + \partial_j \mathscr{H}) - \beta_j (\partial_s A_i + \partial_i \mathscr{H}) = 0 \\
& \partial_s(\partial_j A_i - \partial_i A_j) = 0 \\
& \beta_k \partial_{v_k} (\partial_{v_j} A_i - \partial_i B_j) + \beta_i (\partial_s B_j + \partial_{v_j} \mathscr{H}) = 0 \label{sym_link_a_H} \\
& \partial_s (\partial_{v_j} A_i - \partial_i B_j) = 0 \\
& \beta_k \partial_{v_k} (\partial_{v_j} B_i - \partial_{v_i} B_j) = 0 \\
& \partial_s (\partial_{v_j} B_i - \partial_{v_i} B_j) = 0 \\
& \beta_k \partial_{v_k} (\partial_s A_i + \partial_i \mathscr{H}) = 0 \\
& \partial_s(\partial_s A_i + \partial_i \mathscr{H}) = 0 \\
& \beta_k \partial_{v_k} (\partial_s B_i + \partial_{v_i} \mathscr{H}) = 0 \\
& \partial_s(\partial_s B_i + \partial_{v_j} \mathscr{H}) = 0
\end{align}
\esub

At this point, let us consider two possibilities. First, we will consider Cartan 1-forms/Lagrangians which do not depend on acceleration (\ie $dv^i$) which means that $B_i = 0$. In a latter case, we will consider general systems.

\paragraph{Acceleration-independent Lagrangians}

The first natural set of dynamical systems to look at are those whose Lagrangian/Cartan 1-form does not depend on the acceleration, \ie $B_i = 0$. In that case, the equations of motion and the constraints obtained above simply considerably. In particular, the symmetries requirements imply that $\partial_s \partial_{v_j} A_i = 0$, which in turns means that $\partial_{v_k} \partial_i H = 0$. With the other equations, one finds that $\partial_{v_i} H$ is a constant.

Hence, looking at the second equation of motion \eqref{general_eom}, for systems with accelerations-independent Lagrangians, non-trivial motions require the Hamiltonian to be of the form $H = c_i v^i + g(x^i)$, for some constants $c_i$ and a function $g$.
The second equation of motion is thus of the form,
\begin{equation}
\frac{dx^j}{ds} \partial_{v_i} A_j = c_n\,.
\end{equation}
Note that since we want the system to be regular (in the sense that there must be a unique solution to the equations of motion), we have $\partial_{v_i} A_j \neq 0$. This term however, cannot depend on $s$, must be so that $\partial_i \partial_{v_k} A_j = \partial_j \partial_{v_k} A_i$, and must be so that $c_i = - \frac{1}{n} \partial_{v_j} \partial_{v_i} A_j$ per \eqref{sym_link_a_H}.

Solving all the above conditions, the most general acceleration independent Cartan 1-form invariant under Carroll boosts and time translations, with non-trivial motion is described by,
\begin{equation}
\varpi = \left(\alpha v_i - c_k v^k v_i + \half v^2 c_i + \sum_i f_i + \gamma_i s\right) dx^i - \left(c_k v^k +g(x^i)\right)ds
\end{equation}
for $\alpha, \gamma_i$ constants, and the functions $f_i = f_i(v_k - v_i, x_j)$, with $k \neq i$, \textit{e.g.} for $n = 3$, $f_1 = f_1(v_2 - v_1, v_3 - v_1, x_j)$, satisfying $\partial_i \partial_{v_k} f_j = \partial_j \partial_{v_k} f_i$. At least one of the $c_i$ must be non-zero in order for the system to move.

Assuming that the functions $f_i$ vanish, the equations of motion are,
\begin{equation}
\label{general_eom_no_accel}
\left\lbrace
\begin{array}{l}
\displaystyle \left(\left(\alpha - c_k v^k\right) \delta_{ij} + c_i v_j - c_j v_i \right)\frac{dv^j}{ds}  = - \partial_i g(x^j) - \gamma_i \\[0.5em]
\displaystyle \left(\left(\alpha - c_k v^k\right) \delta_{ij} + c_i v_j - c_j v_i \right) \frac{dx^j}{ds} = c_i
\end{array}
\right.
\end{equation}

For standard Carroll dynamics \eqref{Carr0Cart}, which features no motion, $\alpha = 1$, $c_i = 0$, $\gamma_i = 0$, and $f_i = 0$.

If one asks at this point that the Lagrangian must not depend on terms higher than the square of the velocity, which is the case for most physical systems, then $c_i = 0$\footnote{$v^2 c_i dx^i/ds$ being a third power of velocity term.} and the system does not move.

\begin{thm}
\textit{\small
A regular massive dynamical system, described by an acceleration-independent Lagrangian, containing no higher terms than velocity square ones, invariant under Carroll boosts and time translations do not move.}
\end{thm}

Note that the equations restricting drastically the form of the Hamiltonian, and forbidden motion for ``physical'' systems, were $\beta_k \partial_{v_k} \partial_i \mathscr{H} = 0$, and $\beta_k \partial_{v_k} \partial_{v_i} \mathscr{H} = 0$. If one of the boost symmetry were to be broken, for instance in $d = 2$, $\bbeta = (\beta_x, 0)$, the Hamiltonian is only constrained to be $H = c_x v_x + g(v_y, x, y)$. The arbitrary dependency on $v_y$ will then allow non-trivial motion. Hence, for ``reasonable physical systems'', breaking the Carroll boost invariance, even only partially, is necessary in order to allow motion.

\medskip

\paragraph{Acceleration dependent Lagrangians}

A more general class of dynamical systems are those whose Lagrangian depend on the acceleration, see \textit{e.g.} \cite{LSZ,acceldep} for physical examples. This is actually the case of some systems in this review, due to the ``exotic'' terms of ``extended'' dynamical systems in $d = 2$, for instance \eqref{GalexoLag} for Galilean systems, and later with \eqref{exoCarrollC} for Carroll systems.

For systems with acceleration, \ie when $B_i \neq 0$, the constraints \eqref{conditions_dlxvarpi=0} imposed by the Carroll boost symmetry and the time translation symmetry are not enough to restrict motion in a meaningful manner. We get that the right hand sides of the equations of motion \eqref{general_eom} do not depend on the velocity nor on time, but they may still depend on the position.

\medskip

As a conclusion, we see that ``physically reasonable'' dynamical systems, which do not have acceleration terms, or very specific high order velocity terms, in their Lagrangian, and which are invariant under Carroll boosts (and time translation), cannot move. There are two ways to find motion for these systems:
\begin{itemize}
\item Include acceleration terms in their Lagrangian description, which is effectively what we will study in sections \ref{ExoCarrSec}, \ref{BHh}.
\item Or, consider systems with only partial Carroll boost symmetry, which is what we will study in some following sections, for instance section \ref{noparticlemot}, and also \ref{exoCcoup}.
\end{itemize}

Things are even more subtle for massless particles, as it will be explain in sec. \ref{s:massless_carr}.

\subsection{Multiparticle Carollian systems}\label{MultiCarroll}

Now we discuss briefly multiparticle systems, which are discussed in \cite{Bergshoeff14,CDG23,ZHZ23}.
Consider first a collection of $N$ isolated free massive Carroll particles in $d$ dimensions which do not interact between themselves. We described them by the generalization of \eqref{CarrHaction},
\beq
\cS_0= \int\left(\sum_{a=1}^N m_a\bv_a\cdot \bx_a^{\prime}\right)ds=
\int\sum_{a=1}^Nm_a\bv_a\cdot d\bx_a\,.
\label{NCLag}
\eeq

Each $\bv_a$ here can be viewed as an independent Lagrange multiplier, and therefore we get \emph{decoupled} equations of motion
\beq
m_a\bx_a^{\prime}=0 \;\Rightarrow\; \bx_a = \big(\bx_a\big)_0
\aand
\bv_a^{\prime}= 0
\;\Rightarrow\; \bv_a =\big(\bv_a\big)_0\,,
\label{NCfreeeqmot}
\eeq
$a= 1, \dots ,N$.
Recalling that C-boosts \eqref{Cboost} act on Carroll time $s$, but not on $\bx$ implies the $s$-independence of the  action \eqref{NCLag}, from which we infer the C-boost symmetry of the system. The associated conserved  boost momentum is the sum of the individual expressions,
\besub
\begin{align}
\bd = &\; \sum_a \bd_a = \sum_{a}m_a\bx_a
= M \bX_0\,,
\label{Ndipole}
\\
M=&\, \sum_{a}m_a\,,
\qquad
\bX_0 = \frac{\sum_{a}m_a (\bx_a)_0}{M}\,.
\label{NCoM}
\end{align}
\label{Nboostmom}
\esub\vspace{-5mm}

\begin{thm}\textit{\small
A system composed of non-interacting massive Carroll  particles is C-boost symmetric. None of the individual particles which compose the system can move, implying obviously that its center of motion \eqref{NCoM} is fixed, consistently with the conservation of its Carroll momentum \eqref{Ndipole}.}
\label{ThmIV.4}
\end{thm}

Similarly, the conserved total linear momentum is the sum of separately conserved particle momenta,
$
\bP = \sum_a m_a\bv_a.
$
 This expression contains a curious ambiguity reminiscent of the ``internal energy'' and ``internal angular momentum'' of Galilean systems \cite{SSD,Zhang:2011du,Zhang:2012cr}: by \eqref{NCfreeeqmot}
each of the $\bv_a$ is defined \emph{up to an arbitrary constant vector} $\bc_a$.
\goodbreak

Lagrangian \eqref{NCLag} of $N$ uncoupled massive particles in $d$ dimensions is formally equivalent to a single particle in the dimension $Nd$. Therefore, most of the conclusions we made in the previous section for a single Carroll particle still apply.

Moreover, recall the peculiar causality of Carroll systems. The light-cone degenerates into a ``vertical line'' \cite{Leblond}, therefore the particles at different positions are causally disconnected and cannot interact with each other in a way  consistent with the Carroll symmetry. Hence, a collection of free Carroll particles is strictly the sum of its parts, not more.

What happens when we put such a system of massive, not interacting particles into an external electromagnetic field? Position and momenta, $\bx_a$ and $\bv_b$ are uncoupled, and each $\bx_a$ remains fixed, as we state
 in Theorem \ref{ThmEMNoMotion}. Therefore:

\begin{corollary}\textit{\small
The center of mass $\bX$ \eqref{NCoM} of $N$ non-interacting massive Carroll particles remains motionless even the system is put into an electromagnetic field.}
\label{NCorollIV.2}
\end{corollary}

What happens when an interaction is introduced by adding some, \textit{a priori} arbitrary, potential? Can such a system move? To see that , we consider  coupled Carroll particles through an arbitrary position and momentum-dependent $N$-body potential $V(\bx_a, \bv_a, s)$ \footnote{Remember that $\bv_a$ is (linear momentum)/mass, and is {\rm a priori} unrelated to the velocity, $\bx^{\prime}_a$.},
\begin{equation}
\cS =
\int \Big(\sum_a m\bv_a \cdot \bx_a' - V\Big) ds = \cS_0 - \int V ds \,.
\end{equation}
The equations of motion still separate,
\besub
\begin{align}
m_a \bx_a' & = \frac{\partial V}{\partial \bv_a}\,,
\label{xprime}
\\[6pt]
m_a \bv_a' & = - \frac{\partial V}{\partial \bx_a}\,.
\end{align}
\label{xvprime}
\esub
It follows that these particles \emph{do} move whenever the potential depends on $\bv_a$, $\frac{\partial V}{\partial \bv_a} \neq 0$.  Their center of mass $\bX_0$ in \eqref{NCoM} remains however fixed when the ($\bv$-dependent) forces on the r.h.s. of \eqref{xprime} add up to zero,
\begin{equation}
\sum_a \frac{\partial V}{\partial \bv_a} = 0\,.
\label{vNIII}
\end{equation}
This condition is indeed a sort of \emph{``velocity dual'' to Newton's Third law}, obtained by changing $\bx_a$ to $\bv_a$ \footnote{
The relation \eqref{vNIII} should be compared to the discussion in
\cite{SSD} (pp. 141-142 and on pp. 152-153) where Souriau states  ``Galilean relativity + the Maxwell Principle imply Newton's 3rd law''. Remember, though, that boost symmetry is broken in the Carroll context as seen above.
See also also \cite{Zhang:2011du,Zhang:2012cr,AnkerZiegler} for related developments.}.

Condition \eqref{vNIII} is satisfied in particular by the curious momentum-dependent potential of the 2-particle system
$
V=\big(\half(\bv_1-\bv_2)^2\big)
$
(we took for simplicity $m_1=m_2=1$)
considered in \cite{Bergshoeff14}, for  which \eqref{xvprime} requires
$
\bx_1^{\prime}=-\bx_2^{\prime}=\bv_1-\bv_2
$
and
$
\bv_a^{\prime}=0\,
$
so that $\bv_a(s)=(\bv_a)_0=\const$ cf. for non-interacting systems in \eqref{NCfreeeqmot}. Thus the position coordinates \emph{move} with constant but opposite Carrollian velocities,
whereas their center of mass remains fixed.

More generally, \eqref{vNIII} holds whenever the potential is the momentum-equivalent of the usual two-body potential,
\beq
V = \sum_{a > b = 1}^NV_{ab}(|\bv_a - \bv_b|)\,.
\label{Vvpot}
\eeq
\kikezd{Relation to the dipole moment}

 The conservation of the center of mass is reminiscent of that of the dipole moment in fractonic systems, discussed in the next section. Can we incorporate into the multi-Carrollian picture  dipole momentum conservation? The answer is positive provided we notice that the quantities $m_a$ play only a formal role of ``mass'' and can in fact also be viewed as \emph{charge}. This association is natural, especially if we notice the similarity between the operator of time translations in Carrollian systems which is the Hamiltonian $\mathscr{H}$, which is the counterpart of the charge operator $Q$ in the fracton models (see, for example, \cite{Pena-Benitez:2021ipo,Gromov:2022cxa}). Therefore we view the $m_a$ as a charge, $m_a \to q_a$ and allow the latter to take both positive and negative values, say $q_a = \pm q$. Then the definition of the center of mass~\eqref{Ndipole} reduces to the familiar expression for the conserved total dipole moment:
\begin{align}
\bd = \sum_{a} q_a\bx_a\,.
\label{NdipoleQ}
\end{align}

\begin{thm}\textit{
The motion of interacting Carroll particles conserves the total dipole moment~\eqref{NdipoleQ} as for fractons, provided the interaction is realized via the momentum-dependent two-particle potential~\eqref{Vvpot}.}
\label{ThmIV.5}
\end{thm}

\subsection{Massless Carroll particles}\label{s:massless_carr}

Massless Carroll particles constructed along the KKS lines \cite{ConfCarr} behave differently: unlike their massive counterparts, they \emph{can move} (as observed on the black hole horizon \cite{MZHLett}).
 The difference is highlighted by their momentum: we readily see from \eqref{carrcoadpi} that when the mass vanishes, $m=0$, the conserved norm of their momentum, $|\bpi|$, is promoted to a Casimir invariant.
Massless particles with zero or with non-zero momentum  behave differently as we show, below, providing us with two different classes of massless Carroll particles.

Consider first \emph{massless particles with non-zero conserved momentum},
\beq
\bpi\neq0\,.
\label{pinon0}
\eeq
Then the KKS algorithm yields the spinless ``\emph{Fermat particles}'' of geometrical optics \cite{SSD,DHFermat,ConfCarr,SpinOptics}   --- which are  \emph{both} Galilean \emph{and} relativistic \emph{and} Carrollian.
This comes from the coadjoint orbits for all three of the \textit{Galilei, Poincar\'e} and \textit{Carroll} groups
$\gG,\,\gP,\gC$ we are interested in are in fact those of their common Euclidean subgroup, $\gE$, obtained when boosts and time translations are dropped \cite{SpinOptics}. \emph{Geometric Optics is Euclidean} --- both in direct and historical sense \cite{Euklides}.

``Spinless light'' [alias a Fermat  particle] \emph{propagates instantaneously} along the light rays of geometrical optics, see Theorem \ref{ThmFermat} below.
In detail, the geometrical model of Ref.~\cite{SSD} describes a light ray by a pair $(\bx, {\bu})$, where ${\bx}$ is an arbitrary point on the ray and $\bu$ is a unit vector \footnote{This is akin to special relativity where the spatial velocity of massless particles also has unit norm, with units where $c = 1$. This highlights, once again, that these orbits lie in Poincar\'e orbits.}
such that $\bpi = k \bu$ is oriented along the ray.
The norm of the momentum $\vert \bpi \vert = k$ is a Casimir invariant which is assumed not to vanish.
\goodbreak

\kikezd{Coadjoint action of the euclidean group $\gE$}.

An element $(A,\bc)$ of the euclidean group $\gE$ acts on
$\mu_0 = (\boldsymbol{\ell}, \bpi)$ in the dual euclidean algebra $\mathfrak{e}^*$ labeled as,
\beq
\boldsymbol{\ell} \to A \boldsymbol{\ell} - \bc \times A\bpi\,,
\qquad
\bpi \to  A\bpi\,.
\label{euklgroup}
\eeq
We immediately have that the above Euclidean orbits lie within the Carroll orbits \eqref{carrcoad} with $m = 0$ and $\bd = 0$.
 The Casimir invariants are,
\begin{equation}
\vert \bpi \vert = k,
\qquad
\boldsymbol{\ell} \cdot \frac{\bpi}{k} = j\,,
\label{colj}
\end{equation}
where the non-zero constant $k$  is Souriau's ``color" \cite{SSD}, which corresponds to the absolute value of the conserved momentum. $j$ is the spin of the particle.
Choosing the basepoint $\mu_0=(0,\bpi) \in \mathfrak{e}^*$, the KKS algorithm \cite{SSD,Kost,Kir} endows $\cE$ with Cartan resp. Souriau forms
\beq
\varpi_0 = k\,\bu\cdot d\bx\,,
\qquad
\sigma_0= k \, d\bu\wedge d\bx\,.
\label{free0Sour}
\eeq

Equivalently, the dynamics is given by the Lagrangian
\beq
\cL_0ds= \bpi\cdot \bx^{\prime}ds = k \bu\cdot \bx^{\prime}ds= \bpi\cdot d\bx
\label{FermatLag}
\eeq
cf. \eqref{0CarrLag} in the massive case, which
makes  the Carroll symmetry manifest, since C-boost \eqref{Cboost} act only (by  reparametrization) on $s$.
In fact,  $s$ can even be dropped from the evolution space \eqref{Carev}, leaving us with
\begin{equation}
\cE= \IR^3 \times \IS^2 = \Big\{(\bx,\bu)\Big\}\,
\label{Carrevm0}
\end{equation}
which is thus 5-dimensional.
While the dynamics can be found within the Carroll group,  $s$ has effectively dropped out from the Cartan 1-form \eqref{free0Sour} and thus also from the evolution space, \eqref{Carrevm0}. Indeed, the  contribution proportional to $ds$ to $\varpi_0$  in \eqref{Carr0Cart} valid in the massive case is suppressed by letting the mass to go to zero.

We emphasise  that while $s$ is not needed to describe the evolution of the system, we could (trivially) add it by hand and return to the Carroll the evolution space \eqref{Carev}. This would restore the action of Carroll boosts on \eqref{Carev} --- but its action  would have no effect on the dynamics.

A light ray is determined by the characteristic foliation of $\sigma_0$ in \eqref{free0Sour}, $\delta\bu=0$ and $\delta\bx\propto\bu\,,$
which is indeed a straight line oriented along $\bu$.
The description by $({\bu, \bx})$ is however redundant: $\bx_1$ and $\bx_2$ lie on same ray if they differ by a multiple of $\bu$. A light ray is thus labeled ultimately by the  two orthogonal vectors,
\beq
{\bu}
\aand
\bq=\bx-(\bu\cdot\bx)\bu\,,
\qquad
\bu\cdot\bq=0.
\label{Fermalight}
\eeq
The ``space of motions'' of a Fermat  particle is thus
$\cM= T\IS^2$,  the tangent bundle of the 2-sphere, depicted in Fig.~\ref{FermatPic}. In conclusion:

\begin{figure}[ht]
\includegraphics[scale=.57]{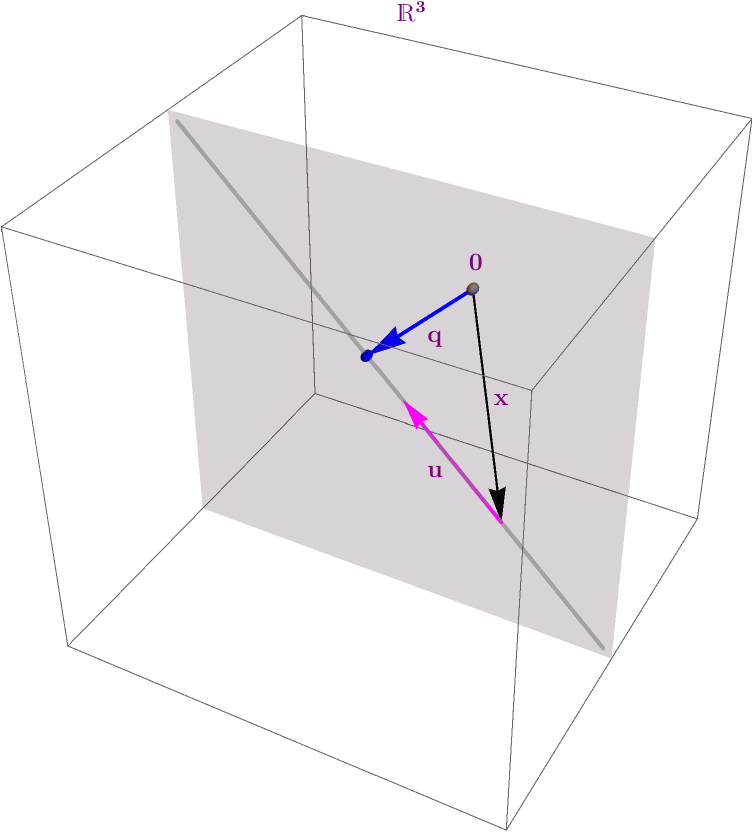}
\qquad\qquad
\includegraphics[scale=.57]{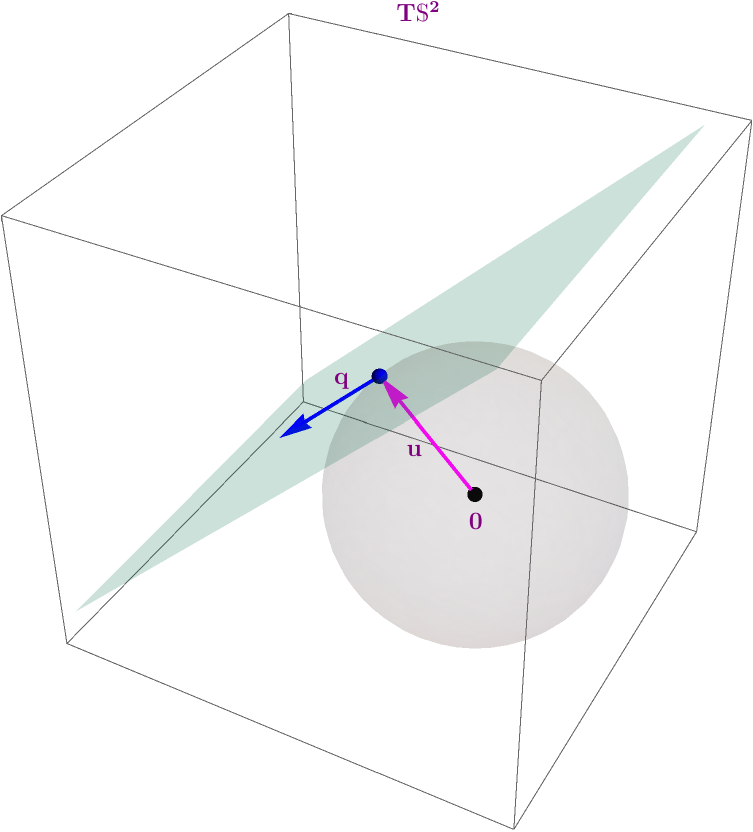}
\\\vskip-8mm
\null\hskip-8mm
(a) \hskip70mm (b)\\\vskip-4mm
\caption{\textit{\small (a) A Fermat particle in flat space moves along oriented straight lines. The unit vector \red{$\bu$} gives its direction  and the vector \blue{$\bq$} points to its closest point to a (chosen) origin.  $\bx$ is any point on the ray. (b) The space of motions is thus the tangent bundle of the unit 2-sphere, endowed with $k$-times its canonical symplectic form.
}
\label{FermatPic}
}
\end{figure}

\begin{thm}
\textit{ \small A {free, mass, spin and  charge-less} Carroll particle is equivalent \strut{}\footnote{Up to interchanging Galilean and Carrollian times, $t \leftrightarrow s$.}  to the  euclidean ``Fermat particle'' in refs. \cite{SSD,DHFermat,SpinOptics} and  \emph{moves} instantaneously, along the oriented straight lines of geometrical optics.}
\label{ThmFermat}
\end{thm}

An intuitive explanation is that there is no canonical quantity with the dimension of velocity in that theory, so either they do not move at all, or they move instantaneously. Since we know that they do move (the trajectory is a ray), it has to be instantaneous.

Recalling that Carroll boosts, \eqref{Cboost},
 leave the  free dynamics  \eqref{free0Sour} - \eqref{FermatLag}
  invariant, the Noether theorem yields the conserved bost momentum,
\beq
{\bd} \equiv 0\,,
\label{d0form0}
\eeq
cf. eqn. \# (VI.17) of \cite{ConfCarr}, which is again
 consistent with letting $m\to0$ in the massive boost momentum \eqref{mCboost}.
However, unlike in the massive case, its conservation is  does \emph{not} allow us to infer any conclusion about the dynamics --- consistently with those instantaneous \emph{motions} we had found.

\goodbreak
Extension to spin in $d\geq3$ space dimensions (which allows us to study the ``spin-Hall effect for light", see \cite{LightHall, DHFermat, SpinOptics}) is not considered in this paper.
Anyonic spin in $d=2$ will be studied in sec. \ref{ExoCarrSec}.
The description of Carroll particles can be extended to curved space \cite{Bergshoeff14}.

\kikezd{Coupling to a background electromagnetic field}.

A free particle  constructed by the KKS method can be coupled to an electromagnetic field by the rule \eqref{Smincoupling}.
The Carrollian version can be obtained by taking the Carrollian limit  \cite{LBLL}.
One starts with the relativistic electromagnetic tensor
\begin{equation}
\IF = F_{ij} dx^i \wedge dx^j + \frac{E_i}{c} dx^i \wedge dx^0\,.
\end{equation}
 However, rescaling the electric
 field  as $E_i \mapsto \widetilde{E}_i = E_i / (c C)$ one gets, after introducing Carrollian time by  $x^0 = s / C$ (so that $C \rightarrow \infty$ is the Carrollian limit),
 \begin{equation}
\IF = F_{ij} dx^i \wedge dx^j + \widetilde{E_i} dx^i \wedge ds\,
\label{contrem}
\end{equation}
where Carroll time, $s$, has replaced  the usual time coordinate\footnote{The same result is obtained \cite{Carrollvs,Morand} by switching to light-cone coordinates $u$ and $v$. Then the restriction to
the  $u = \const$ null hypersurface  is again \eqref{contrem}
 with $\widetilde{E}_i = E_i - F_{iz}$.}.

The electric and magnetic fields defined above satisfy, in the vacuum, the Carroll-Maxwell laws \cite{Carrollvs}, either in their electric version,
\begin{equation}
\left\{\begin{array}{lll}
\epsilon_{ijk} \partial_j E^k +  \partial_s B^i = 0\,, && \partial_i B^i = 0
\\[3pt]
\partial_s E^i = 0\,, && \partial_i E^i = 0
\end{array}\right.
\qquad \text{electric}
\end{equation}
or in their magnetic version,
\begin{equation}
\left\{\begin{array}{ll}
\partial_s B^i = 0\,, & \partial_i B^i = 0
\\[2pt]
\epsilon_{ijk} \partial_j B^k - \partial_s E^i = 0\,, & \partial_i E^i = 0
\end{array}\right.
\qquad \text{magnetic}
\end{equation}
respectively \footnote{When compared to Galilean electromagnetism \cite{LBLL,Carrollvs} the words ``electric'' and ``magnetic'' are interchanged, underlining the dual nature of Carrollian vs Galilean physics. These equations are written in $d=3$ spatial dimensions.}.
\goodbreak

One may wonder then what would happen when we couple a  \emph{massless} Carroll particle with non-zero ``color'' $k\neq0$ in \eqref{colj}, endowed also with an electric charge to electromagnetism. Such particles would be described by the Souriau 2-form
\begin{equation}
\sigma = k\, du_i \wedge dx^i + \frac{e}{2} F_{ij} dx^i \wedge dx^j + e E_i dx^i \wedge ds\,.
\end{equation}
The model looks similar to a massive, non-relativistic charged particle in a background electromagnetic field, --- but with the color $k$ playing the role of a mass, and $\bu$ playing that of  momentum (except that $\bu$ is a unit vector).
The motions are the integral leaves
of $\ker \sigma$,  given by,
\begin{equation}
{\bE}\cdot{\bu}=0\,,
\qquad
\frac{d\bx}{ds}\times\bu=0\,,
\qquad
k\frac{d\bu}{ds}-e\bE+e{\bB}\times\frac{d\bx}{ds}=0\,.
\label{colormodelbis}
\end{equation}

Note that the third equation forbids the existence of massless Carroll particle with color $k$ in a pure electric field background. However, if the magnetic field does not vanish, the system interestingly admits Hall-type solutions.
Assume indeed for simplicity that electric and magnetic fields are
 $\bE=(E^1,E^2,0)$ and $\bB=(0,0,B)$ with $E^{i}=\const\neq 0,\,B=\const\neq0$.
 Then some calculation yields,
\begin{equation}
\frac{d\bx}{ds}= \left(\frac{{|\bE|}^2}{|\bE\times \bB|}\right)\,\bu\,,
\qquad
\vec{\bu}=\frac{\bE\times \bB}{|\bE\times \bB|}\,.
\end{equation}%
When $\bE=0$, we find instead helical motions.

\null\hskip-6mm
\begin{table}[thp]
\begin{tabular}{|l|c|c|c|}
\hline
&free Carroll no charge $e=0$
& charged $e\neq0$ in EM field
& in curved metric
\\
\hline\hline
$m \neq 0
$
& no motion
& no motion
& no motion
\\
\hline
$m = 0, k = 0$
& no motion
&  Hall motions
&  no motion
\\
\hline
$m=0, k \neq 0
$
&instantaneous motion & helical/Hall motions  &geometric optics
\\
Fermat particle
&  straight lines
&
&
\\
\hline
\end{tabular}\vskip3mm
\caption{\textit{\small Motion of an unextended Carroll particle with
vanishing
spin, $s=0$.}
}
\label{CarrollTable1}
\end{table}

\subsection{``No-particle" motion}\label{noparticlemot}

A curious degenerate case arises, though, for
\beq
\bpi=0,
\label{pi0}
\eeq
for which the KKS algorithm would yield \emph{identically zero} Cartan (and thus Souriau) forms,
\beq
\varpi_{m = 0} \equiv 0 \aand \sigma_{m = 0} \equiv 0\,,
\label{noparticle}
\eeq
respectively, --- and thus \emph{no dynamics} at all!
 However coupling ``it'' naively to electromagnetism  by the rule \eqref{Smincoupling} yields \dots a \emph{Carroll   dynamics with a ``purely  electromagnetic Souriau form}\,,
\begin{equation}
\sigma^{(0)} = \frac{e}{2} F_{ij} dx^i \wedge dx^j + e E_i dx^i \wedge ds,
\label{p0sigma}
\end{equation}
 whose characteristic foliation yields neverthess non-trivial equations of motion (in $2+1$ dimensions).

The weird-looking ``naked'' electromagnetic Souriau form \eqref{p0sigma} can actually be derived from the massive 2-form \eqref{emCarrSou}. Replacing $\bp$ by $m\bv$ we get,
\beq
\sigma^{(m)} = md\bv \wedge d\bx
+ \smallover{e}/{2} F_{ij} dx^i \wedge dx^j + eE_i dx^i \wedge ds\,.
\label{vemCarrSou}
\eeq
Then letting  here $m\to0$ switches off the first term, leaving us just with \eqref{p0sigma}. Remarkably, the electric charge drops out from the equations of motion (as long as it does not vanish), leaving us with:

\begin{corollary}
Coupling minimally a charged degenerate ``no-particle'' with vanishing conserved momentum,  $\bpi=0$,  having
 therefore no free dynamics, \eqref{noparticle},
the system will  move by following the Hall law~ \eqref{Halllaw} \ie,
\begin{equation}
\label{Hall_pure_carroll}
(x^i)^{\prime} = \epsilon_{ij} \frac{E^j}{B}\,.
\end{equation}
\label{corollIV.2}
\end{corollary}\vskip-8mm
How can a ``no-particle'' have non-trivial motion? The mystery is clarified
by observing that belonging to the kernel of
 $\sigma^{(0)}$ means  that the  \emph{combined forces vanish along the trajectories}. Perpendicularly to the electric direction, the motion is thus ``free''!
\goodbreak

\kikezd{Partially broken C-boost symmetry}

From the symmetry point of view, Carroll boosts are generated by the 2-parameter vector field
$
X = \beta_i \partial_{v_i} -\beta_i x^i \, \partial_s
$,  \eqref{infCboost}
with  $\bbeta = (\beta_1, \beta_2)$, implemented on the Souriau 2-form \eqref{p0sigma} as,
\begin{equation}
L_X \sigma^{(0)} = -(\bE \times \bbeta) dx^1 \wedge dx^2
\,.
\label{sigmaBchange}
\end{equation}
Since the above should vanish for a symmetry of the system, we see that the Carroll boosts symmetry requires \emph{$\bbeta$ to be parallel to the electric field, $\bE$}. Thus the electromagnetic field breaks the full Carroll boost invariance, only the boost in the direction of $\bE$ remains. However, no C-boost symmetry in the direction perpendicular to $\bE$ implies no obstruction against motion. Thus we get:

\begin{thm}\textit{ \small The electromagnetic field breaks partially the C-boost symmetry : C-boosts perpendicular to the direction of the electric field $\bE$ are broken, while those parallel  to $\bE$ remain  symmetries. C-boost symmetry is consistent with the Hall law \eqref{Hall_pure_carroll}: motion is possible in perpendicularly to the electric field but not parallel to it.}
\label{partboostthm}
\end{thm}

C-boost symmetry thus implies immobility
however not conversely:  $\bd$ is conserved in the direction parallel to the electric field $\bE$. However its momentum vanishes because the particle is massless. No conclusion can be drawn from its conservation therefore. In the broken  (perpendicular) direction we have in turn no conserved momentum at all -- and thus neither garantie nor obstruction against motion.
In conclusion, the immobility  derived from the conserved boost momentum does not work for massless particles.


Using a more conventional language, putting $\bp = m\bv$ into the massive Carroll Lagrangian with EM coupling \eqref{emCLag} to get
 \beq
\cL_{em}=(m\bv - e\bA)\cdot\bx^{\prime}
 - eV\,.
\label{mvCLag}
\eeq
This Lagrangian is gauge-invariant in that it changes by a time derivative under $\bA \to \bnabla f$, $\cL_{m}\to \cL_{m}-\p_s\big(e\p_if x^i\big)$. Its variational equations are, instead of \eqref{mEMeq}
\beq
m(x^i)^{\prime}=0
\aand
mv_i^{\prime}=eE_i+{eB}\epsilon_{ij}(x^j)^{\prime}\,.
\label{m0EMeq}
\eeq
Then letting  $m\to0$ switches off the first equation \emph{leaving $(x^i)^{\prime}$ undetermined}, while the second one in \eqref{m0EMeq} reduces, when $e\neq0$, to the no-force-along-the-trajectory condition
\beq
E_i+{B}\epsilon_{ij}(x^j)^{\prime}=0
\quad\ie\quad
\IF_{\mu\nu}(x^{\nu})^{\prime}=0,
\
\label{Fxprime}
\eeq
which is the Hall law, \eqref{Hall_pure_carroll}. Note that  the electric charge has dropped out.
We note for completeness that this first-order equation comes from the first-order purely electromagnetic Lagrangian
\beq
\cL^{0}_{em}= -\bA\cdot\bx^{\prime} - V\,,
\label{m0CLag}
\eeq
which is indeed the $m\to0$ limit of \eqref{mvCLag}.
\goodbreak

The  behaviour of a ``no-particle'' under C-boosts can conveniently be studied by looking at the classical action,
\beq
\cA = \int \cL^{0}_{em}ds = \int\!(\bA\cdot d\bx + Vds)\,.
\label{m0Caction}
\eeq
Assuming for simplicity that the fields are constant, the action
changes under a C-boost \eqref{Cboost} as,
$$
\cA \to \cA + \int\!({surface\ term})
- \int \!\Delta\,
\where
\Delta = \big(E_1b_2 dx^1 x^2 + E_2b_1 dx^1 x^2\big).
$$
A C-boost is  a symmetry if $\Delta$ is also a surface term. Having assumed $\bE=\const$, we must have
$
d\Delta = -(\bE \times \bb)\, dx^1 \wedge dx^2 = 0,
$
which implies :

\begin{thm}
\textit{A C-boost \eqref{infCboost} oriented along the electric field, $\bE$, is a symmetry for a no-particle in a constant planar electromagnetic field $(\bE,B)$ but is broken otherwise.
}
\label{brokenCboosts}
\end{thm}

For $\bE=(0,E_y)$, for example, boosts are symmetries in the $y$ direction but broken in that of $x$. The first one induces the conserved Noether quantity
$
D_y =  yV(y),
$
whose conservation implies, for $E_y=-\p_yV\neq 0$, no motion in the $y$ direction, but does not exclude motion in the direction perpendicular to the electric field.

The correlation of ``allowed" and ``forbidden'' directions with the C-boost symmetry is analogous that of dipole symmetry  for fractons, whose immobility can be concentrated to lines or even surfaces \cite{Bidussi}.

\subsection{Carroll ``no motion'': interpretation through General Relativity}\label{CarrGR}

Carroll structures implying  Carroll dynamics are plentiful in General Relativity: in fact, any null hypersurface of a Lorentzian spacetime carries a Carroll structure \cite{Carrollvs,Morand,CiambelliNull}.

\tikzset{middlearrow/.style={
       decoration={markings,
           mark= at position #1 with {\arrow{>}} ,
       },
       postaction={decorate}
   }
}
\tikzmath{\lc = 0.5;}

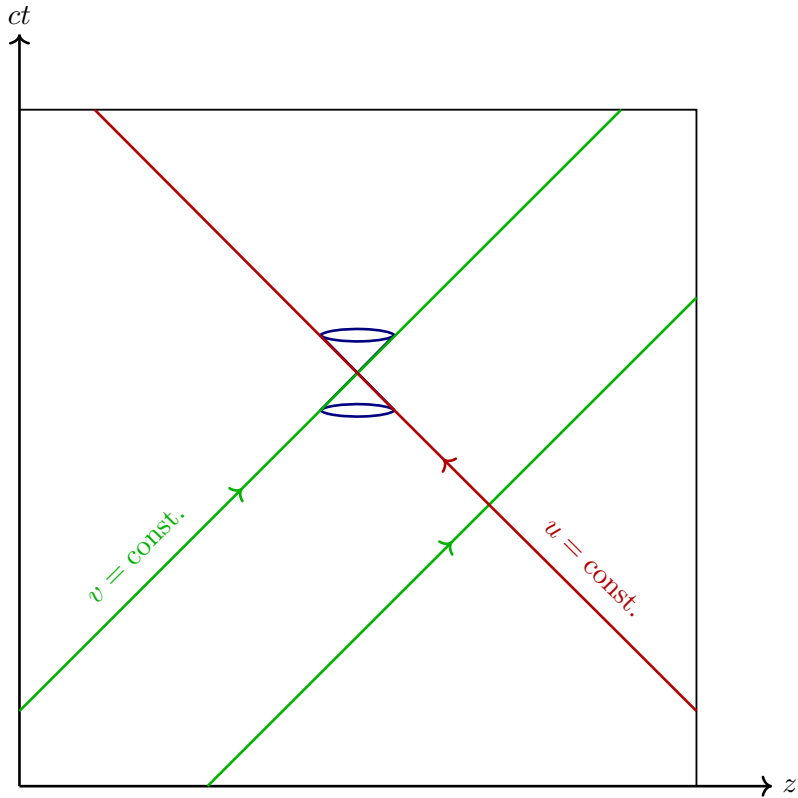
\begin{figure}[ht]
\begin{tikzpicture}[line width=1pt,scale=1, every node/.style={transform shape}]
 \draw [line width=1pt,middlearrow={1}] (0,0) -- node[pos=1,right]{$z$} (10,0);
 \draw [line width=1pt,middlearrow={1}] (0,0) -- node[pos=1,above]{$ct$} (0,10);
 \draw [line width=0.7pt] (0,9) -- (9,9) -- (9,0);

 \draw [black!50!blue](4,5) arc (170:10:{1*\lc} and {0.2*\lc}) coordinate[pos=0] (a);
 \draw [black!50!blue](a) arc (-170:-10:{1*\lc} and {0.2*\lc}) coordinate (b);
 \draw [black!50!blue](a) -- ([yshift={2cm*\lc}]$(b)$) coordinate (c);
 \draw [black!50!blue](b) -- ([yshift={2cm*\lc}]$(a)$) coordinate (d);
 \draw [black!50!blue](d) arc (170:10:{1*\lc} and {0.2*\lc});
 \draw [black!50!blue](d) arc (-170:-10:{1*\lc} and {0.2*\lc});

 \draw [black!30!green,middlearrow={0.5}] (2.5,0) to[out=45,in=-135] (9,6.5);
 \draw [black!30!green,middlearrow={0.37}] (0,1) to[out=45,in=-135] node[pos=0.45,rotate=45,yshift=0.4cm]{$v = \const$} (a) --  ([xshift=4cm,yshift=4cm]a);

 \draw [black!30!red,middlearrow={0.42}] (9,1) to[out=135,in=-45] node[pos=0.40,rotate=-45,yshift=0.4cm]{$u = \const$} (b) --  (1,9);

\end{tikzpicture}
\caption{\textit{\small Spacetime diagram. The light-cone is in \blue{\bf blue}. Photon trajectories are either \dgreen{\bf\emph{outgoing}} ($\dot{z} > 0$) depicted in \dgreen{\bf green} or \red{\bf \emph{ingoing}} depicted in \red{\bf red}.}}
\label{f:spacetime}
\end{figure}

If one wants to induce the motion of a particle on a light-like hypersurface, then the class of particles to consider is that of massless particles.
 Let us  look at  such particles in flat spacetime. Their geodesic motion is given by the set of first order equations,
\begin{align}
\label{geodesics}
\dot{X}^\mu & = P^\mu\,, \\
\dot{P}^\mu & = 0\,,
\end{align}
where the dot denotes the derivative with respect to some affine parameter.
Pick coordinates $(x, y, z, t)$ such that the momentum for outgoing massless particles be
$
P = (0, 0, 1, 1)\,,
$
yielding the spacetime diagram in figure \ref{f:spacetime}.

In- and outgoing null geodesics define new coordinates that are well suited for the study of null trajectories. Let us then introduce null coordinates,
$ u = \frac{ct + z}{2}\,, \,
v = \frac{ct - z}{2} \,$
and compute the geodesic equation of motion \eqref{geodesics} for a massless particle, for an ``outgoing''  particle with momentum $P$ chosen as above,
 for instance. We readily see that $\dot{u} = 1$ and $\dot{v} = 0$, meaning that outgoing massless particles have their trajectories along a $v = \const$ hypersurface, while $u$ is a natural evolution parameter. Conversely, an ingoing massless particle  would follow $u = \const$ hypersurfaces, with $v$ the evolution parameter. The situation is depicted in figure \ref{f:spacetime}.
The other equations of an outgoing particle  are, from \eqref{geodesics}, given by,
\begin{equation}
\label{induced_carroll_eom}
\frac{d x}{du} = 0\,, \qquad \frac{d y}{du} = 0\,,
\end{equation}
where we used $u$ as the evolution parameter.

What is said above allows  us to understand why the ``no-motion'' statement for Carroll particles is so natural: the $v = \const$ hypersurfaces define the Carroll structure.  Such a structure would be described by 2 spatial coordinates $x$ and $y$ and by another one ``along'' the null direction, $u$. Geometrically, $u$ would be what we called the ``Carroll time''. The outgoing massless particles stay on these Carroll structures, and their equations of motion are thus induced from the ambient spacetime \eqref{induced_carroll_eom}.
Thus there is ``no motion'' in the $(x, y)$ plane, as the induced momentum vanishes in the spatial directions, as expected for Carroll dynamics.

For the Carroll structures that are found as null hypersurfaces in lorentzian spacetimes, the intrinsic statement that Carroll particles do not move mean simply  that particles move in the null direction that defines the hypersurface, \ie, that it follows a null geodesic in the ambient spacetime. The massless particle has all its momentum directed toward the null direction, and has no spare momentum left to deviate in the other directions. In that sense, the statement of the Red Queen \footnote{According to \cite{LewisC}: ``A slow sort of country! \dots it takes all the running you can do, to keep in the same place.''}
makes perfect sense for Carrollian physics.

So in a sense, we can think of Carroll structures/dynamics as an effective description of phenomena which follow null geodesics. This also implies that finding Carrollian motion means finding potential deviations from null geodesics.


Comparison of the behaviors in the massive/massless cases yields further insight: eqn.~\eqref{emCeqmot} implies that the mass $m\neq0$ ``glues" the particle to a fixed position.  However in the massless limit the particle is ``liberated" and moves as in \eqref{Hall_pure_carroll}.
This non-analytical behavior of the mass-induced spatial ``confinement''  of the Carroll particle has similar features with the conventional Hall effect~\eqref{Halllaw} in two spatial dimensions.

The motion of a particle with a nonvanishing (even infinitesimally small, but nonzero) electric charge $e \neq 0$ is radically different from that of a particle with an exactly zero charge $e = 0$. In the first case, the electromagnetic field generates the Hall motion of the particle~\eqref{Halllaw}, while in the second case, the background field does not affect the motion at all. Moreover, despite the paramount requirement of the nonvanishing electric charge $e \neq 0$, the Hall motion does not depend explicitly on the actual value of the electric charge $e$~\eqref{Halllaw}. In the Hall conditions (with perpendicular electric and magnetic fields), the motion is non-analytic in $e$ as $e=0$ and $e \neq 0$ cases are qualitatively different.

The ``no-particle" motion of the Carroll particle is a non-analytical phenomenon but now in the particle mass rather than in the electric charge: the motion is radically different for massive ($m \neq 0$) and massless ($m = 0$) cases. The second similarity appears because the actual value of the mass does not enter the solution of equations of motion for the Carroll particle, similarly to the electric charge in the Hall effect.
\goodbreak

\subsection{Carroll field theories}\label{fieldCarrSec}

Carrollian field theory can also be considered. As already pointed out some time ago by
Henneaux  \cite{Henneaux}, such theories involve only time derivatives; spatial derivatives are absent. They will be useful to understand fractons, discussed in the following section \ref{fractonSec}.

As explained in section \ref{GalBfield},
the L\'evy-Leblond equations  \eqref{LLeq} can be obtained from the relativistic Dirac theory by letting the velocity of light go to infinity \cite{LLDirac}. Now we show that  letting the velocity of light go instead to zero yields Carroll field theory~\cite{Marsot21}.

The Carrollian limit of the Dirac equation
is otained by writing it in terms of $s = x^0 / C$ coordinate \cite{Carrollvs,Marsot21} (instead of $x^0 = c t$ as for Galilean contraction),
\begin{equation}
C\left(i\gamma^0 \partial_s + \frac{i}{C}\, \gamma^j \partial_j - m \right) \Psi = 0\,.
\end{equation}
Choosing $\gamma^0\propto\sigma^3$ we get, in the $C \rightarrow \infty$ limit \footnote{Carroll contraction can also be seen as the limit $c \rightarrow 0$, where $c$ is the usual speed of light, --- a fact which lead to L\'evy-Leblond to his initial \textit{clin d'oeil} \cite{Leblond}.},
\beq
\gamma^0 \p_s \Psi = - i m \Psi\,,
\qquad \Psi = \left(\begin{matrix}\psi_+ \\ \psi_-\end{matrix}\right)\,.
\label{DiracCarr}
\eeq
The Dirac matrix $\gamma^0$ has been choosen diagonal, therefore it does not mix the spinor components and
the first order field equations split into two uncoupled equations for the components,
\beq
\medbox{
\p_s \psi_\pm = \mp im\psi_\pm\,.}
\label{Carrfeq1}
\eeq
This equation is the Carroll counterpart to the L\'evy-Leblond equation \eqref{LLeq}.
It can also be derived from its Lagrangian which is Carroll limit of the Dirac Lagrangian,
\begin{equation}
\cL_C = \frac{i}{2}\big(\Psi\gamma^0\p_s\Psi^*-\Psi^*\gamma^0\p_s\Psi\big) - m |\Psi|^2\,.
\label{LLLag}
\end{equation}

While the above first-order Carroll field equations \eqref{DiracCarr} describe spinors, it is also possible to have the same equation but for scalar fields. For instance, quantization of the Carroll model \eqref{genCarrollCS} yields the same equation \eqref{Carrfeq1} but implemented on a scalar field $\Phi$ \cite{Marsot21},
\begin{equation}
\p_s \Phi = im\Phi\,.
\label{scalarCarr1}
\end{equation}

That the same equation holds for spinors and scalar fields comes from their rather trivial form. Indeed, differences between spinor and scalar field equations typically involve terms with spatial derivatives, such as $\bsigma\cdot\bD$ in \eqref{LLeq}, --- but spatial derivatives are missing in Carroll theories.
While in the free case these equations coincide, their behavior may be different, though, when coupled to a background field. For instance, in Galilean quantum mechanics, the free Schr\"odinger and the Pauli equations are formally identical, despite one acting on scalar fields, and the other one acting on spinors but they become different when coupled to an electromagnetic field.

Similar statements holds for the second-order system. Let us indeed consider the Klein-Gordon equation for a free scalar field of mass $m$ in $1+d$ dimensions with coordinates $x^0, x^1 \dots x^d$ and a metric of signature $(-1, \underbrace{1,\dots, 1}_{d}\,)$,
\beq
\p_\mu \p^\mu\Phi-m^2C^2\Phi=0\,,
\label{KGd+1}
\eeq
where the parameter $C$ has the dimension of velocity and is, in general, different from the velocity of light $c$. Introducing Carroll time, $s$, by putting $x^0 = s/C$ , \eqref{KGd+1} becomes,
$$
\Big[\frac{1}{C^{2}} \bigtriangleup - \p_s^2 - m^2\Big]\Phi=0\,.
$$
Then letting $C\to \infty$ we end up with the second-order-in-$s$ equation with no spatial derivatives,
\begin{equation}
\medbox{
\p_s^2 \Phi + m^2 \Phi = 0\,.}
\label{Carrfeq2}
\end{equation}
In the same spirit, the Klein-Gordon Lagrangian is, in terms of Carroll time $s$,
\beqa
\cL_{KG}=
 C^2 \Big(\p_s\Phi\p_s\Phi^*+\frac{1}{C^2}|\bnabla\Phi|^2
-m^2\Phi\Phi^*\Big)\,.
\label{KGLag}
\eeqa
The overall factor $C^2$ can be dropped and the
$C\to\infty$ limit of the bracketed quantity is
\beqa
\cL^{(2)}_{C} = \vert \partial_s \Phi \vert^2 - m^2 \vert \Phi \vert^2.
\label{CarrLag2}
\eeqa

\medskip
We underline here a characteristic property of Carrollian field theories: space derivatives are missing  from the equations of motion and Lagrangians.

Let us then look at the symmetries. While these theories are built using the Carroll limit, or directly, by quantizing the Carroll group \cite{Marsot21},
 their symmetry groups turn out to be {much larger} than the Carroll group alone: due to the absence of spatial derivatives they  are invariant under the transformation
\beq
\Phi\left(\bx,s\right) \to \widehat{\Phi}\left(\bx,s\right) =e^{imT(\bx)}\Phi \left(\bx,s\right)
\label{supertrslcarr}
\eeq
for an arbitrary function $T(\bx)$.
The absence of spatial derivatives implies also that the  solutions of the free first-order theory \eqref{Carrfeq1} are of the form,
\begin{equation}
\Phi(\bx,s) = e^{ims}\phi(\bx)\,,
\label{solcarr1}
\end{equation}
where $\phi(\bx)$ is an arbitrary function of $\bx$ alone. The symmetry \eqref{supertrslcarr} then corresponds to shifting the time coordinate by an arbitrary function $T(\bx)$ of the position,
\begin{equation}
s \mapsto s + T(\bx)
\label{ssupertransl}
\end{equation}
called, in the BMS context \cite{BMS} a \emph{supertranslation}. Augmented with rotations and spatial translations, we then get  an infinite dimensional symmetry,
\beq
\Phi\left(\bx,s\right) \to \widehat{\Phi}\left(\bx,s\right) =e^{imT(\bx)}\,\Phi(R \bx + \bc,s)\,,
\label{fullsymcarr}
\eeq
where we recognize the ``super-Aristotle'' group $\gA^\infty$  (item \# 9. in sec.\ref{GCAgeo}), implemented as in \eqref{Supercarrollaction}. Note that, unlike for Galilean field theories \ref{GalBfield}, in the present, Carroll context, all symmetries of the field equations are  isometries of the underlying spacetime, as $\gA^\infty$ is actually the isometry group of flat weak Carroll structures, see def. \ref{CarrWSdef}. We recall that
Aristotle $\gA$ is a subgroup of Carroll $\gC$, itself being a subgroup of ``super-Aristotle'' $\gA^{\infty}$, cf. \eqref{ACAinf}.
Choosing $T(x)=b_0$ yields the Aristotle group $\gA$ in $d+1$ dimensions, composed of the Euclidean group $\gE$ augmented by   translations of Carroll time, $s$.
The next (i.e. first order in $\bx$) term,
$
T(\bx) = -\bb_1 \cdot \bx\,,
$
yields Carroll (C-) boosts.

An irreducible representation of the Carroll group on the general solutions \eqref{solcarr1} can be defined  \cite{Marsot21}. One has, for a Carroll transformation $\rC$ in \eqref{BCmatrix}, $\Phi \to \Phi \circ \rC^{-1}$, \ie,
\begin{equation}
\Phi(\bx, s)  \to
 e^{i[-b_{0}+\bb_{1}\cdot(\bx-\bc)]}\,\Phi(R^{-1}(\bx - \bc),s) \,.
\label{carrrep}
\end{equation}

We underline that, even though Carroll boosts (and time translation) are \emph{kinematical} in that it comes from the action of the group on a spacetime -- more precisely, on \emph{Carroll spacetime} with coordinates $\bx$ and $s$.

Additional insight into Carroll symmetry can be gained by looking at the Lagrangian
for the Klein-Gordon equation \eqref{KGd+1}, which is invariant under Lorentz boosts,
\beq
\bx \to \bx' = \bx + \frac{\gamma^2}{\gamma+1}(\bbeta\cdot\bx)\bbeta + \gamma\bbeta\, x_0\,,\qquad
x_0 \to x_0' = \gamma(x_0+\bbeta\cdot\bx),
\label{Lboost}
\eeq
where $\gamma=(1-\bbeta^2)^{-1/2}$ . Then putting
$s = Cx_0,\,\bb=-C\bbeta$ and letting $C\to \infty$
\eqref{Lboost} becomes the C-boost \cite{Leblond}
\beq
\bx' = \bx \aand
s' = s - \bb\cdot\bx\,,
\label{Cboostbis}
\eeq
which is precisely the Carroll boost \eqref{Cboost}.
The  conserved quantity
\beq
{\bd}=m \displaystyle\int\!\Phi\Phi^{\dagger}\,\bx\,d^2\bx\,
\label{Cfieldboostmom}
\eeq
generated by C-boosts we call Carroll field boost momentum
will reappear again  in the fracton section as dipole momentum, \eqref{dipolemom}.
 Comparison with
the Galilean boost momentum $\bd^{\mathrm{Gal}}$ in \eqref{LLconsquant}
indicates the absence of the momentum term $-t\bP$.
the conserved angular momentum for the first-order system  \eqref{DiracCarr} is in turn
\beq
\ell = \int \Big\{\bx\times\frac{1}{2i}\big(\Phi^{\dagger}\bnabla\Phi-(\bnabla\Phi)^{\dagger}\Phi)\Big\}d^2\bx
+
\half\!
\int\!\Big\{|\psi_+|^2-|\psi_-|^2\Big\} d^2\bx\,,
\label{fracangmom}
\eeq
consistently with having spins $\pm 1/2$, cf. the L\'evy-Leblond expression  in \eqref{LLconsquant}.

\goodbreak
\section{Fractons and Carroll particles}\label{fractonSec}

\subsection{The fracton model}\label{fmodel}

Hypothetical quasiparticles in condensed matter systems with limited mobility called \emph{fractons}  attract much current attention \cite{Pretkofractons,PretkoCY,Gromov:2018nbv,Seiberg,Bidussi,JainJensen,Grosvenor:2021hkn,Surowka,Zoo}.
As it is understood now, the existence of restricted-mobility quasiparticles is ultimately a consequences of exotic spacetime symmetries. The Carroll particles constitute an important class of fractons as we show below.

The fracton phase of matter can be described by {complex scalar field theories}.
A Lagrangian has been put forward (e.g. in  \cite{Bidussi}, eqn. \# (2.50)),
\beq
\cL = \underbrace{\frac{i}{2}\big(\Phi\p_t\Phi^*-\Phi^*\p_t\Phi\big) - m |\Phi|^2}_{\cL_0}
\;-\;\lambda\underbrace{\Big(\p_i\Phi\p_j\Phi-\Phi\p_i\p_j\Phi)
(\p_i\Phi^*\p_j\Phi^*-\Phi^*\p_i\p_j\Phi^*\Big)}_{self-interaction \ \cL_{int}}\,,
\label{PhiLag}
\eeq
where $\Phi(\bx, t)$ is a complex scalar field.
A second-order Lagrangian could also be considered by replacing the first-order-in-time derivative terms by $\vert \partial_t \Phi \vert^2$ \cite{Bidussi,JainJensen},
\beq
\cL^{(2)} = \underbrace{\vert\partial_t\Phi\vert^2 -m^2|\Phi^2}_{\cL_0^{(2)}}
\;-\;\lambda \Big(\p_i\Phi\p_j\Phi-\Phi\p_i\p_j\Phi)
(\p_i\Phi^*\p_j\Phi^*-\Phi^*\p_i\p_j\Phi^*\Big)\,.
\label{PhiLag2}
\eeq

The simple model~\eqref{PhiLag2} incorporates a very nontrivial non-perturbative physics that features strongly coupled phases. This fact could be guessed from a nontrivial dimensionality of the coupling $\lambda$ in Eq.~\eqref{PhiLag2}. In the absence of mass, $m=0$, the coupling $\lambda$ sets a single spatial scale and thus can be absorbed into the redefinition of spatial coordinates. The resulting theory contains no small parameters and, thus, is intrinsically non-perturbative.
(The physical situation is very similar to the three-dimensional Yang-Mills theory, which has a single dimensionful coupling constant $g$ that can also be absorbed by a scaling transformation to a redefinition of coordinates. This fact has important physical consequences implying, for example, that quark-gluon plasma in four spacetime dimensions does not allow a perturbative treatment even at high temperatures where the gauge coupling should become small due to asymptotic freedom. The reason is that the system in thermal equilibrium experiences dimensional reduction to its three-dimensional counterpart, which has no perturbative treatment~\cite{Linde:1980ts}).

In the scalar models of type~\eqref{PhiLag}, fractons can be treated in a large-$N$ limit, where the field space is extended to multiple fields $\Phi \to \Phi^a$ with $a = 1, \dots, N$. Treating $1/N$ as a small parameter, the system can be considered in a standard large-$N$ perturbation theory. The solution for the ground state reveals that this class of models is indeed characterized by strongly correlated phases~\cite{Jensen:2022iww}. Moreover, the classical dipole symmetry of the model~\eqref{PhiLag}, which we discuss later, is spontaneously broken in the whole phase diagram of these models even in the absence of a scalar potential $V(\Phi)$. As a result, the ground state of such a model is characterized by dipole superfluids of various types~\cite{Jensen:2022iww}. Despite the fact that the true ground state cannot be revealed in the classical approach, the classical symmetries of the fracton model~\eqref{PhiLag} are important for the quantum behavior through a low-energy Goldstone description.

We study first the free classical dynamics given by the Lagrangian  $\cL_0$, given by the first term in Eq.~\eqref{PhiLag}.
It is reminiscent of the usual Galilei Lagrangian in that it has a first order derivative in time $t$. However, it is also different from the Galilean expression as it has no space derivatives: it is, in fact  Carroll field theory.

Variation of $\cL_0$ in \eqref{PhiLag} yields the truncated first-order wave equation
\beq
\medbox{
\p_t \Phi(\bx, t) = i m \Phi(\bx, t)\,}
\label{freefrac1}
\eeq
whose solutions are,
\begin{equation}
\Phi(\bx,t) = e^{imt}\phi(\bx)\,,
\label{solmpos}
\end{equation}
where $\phi(\bx)$ is an arbitrary function of $\bx$ alone.
The second-order Lagrangian $\cL_0^{(2)}$ in \eqref{PhiLag2}  yields instead a truncated \emph{Klein-Gordon} equation  with no space derivatives,
\begin{equation}
\p_t^2 \Phi(\bx, t) + m^2 \Phi(\bx, t) = 0\,,
\label{freefrac2}
\end{equation}
which is indeed the ``square'' (iteration) of \eqref{freefrac1}. The  missing of spatial derivative terms from both equations  \eqref{freefrac1} and \eqref{freefrac2}
 is analogous to the absence of the kinetic term $\bp^2/2m$ in  the classical Carroll dynamics, studied in sec.\ref{mCpart}.

We note for further reference that eqn. \eqref{freefrac1}
 is indeed \emph{identical} (up to $s \leftrightarrow t$) to
eqn. \eqref{scalarCarr1} satisfied by a Carroll scalar field. It is also reminiscent of a Dirac equation except for missing spatial derivatives.
The second order equation, \eqref{freefrac2}, is in turn the Carroll field equation \eqref{Carrfeq2}. Thus we conclude: \emph{free fracton field theory is a Carroll field theory}.

We highlight that this statement works only at the classical level, as the interaction term in ~\eqref{PhiLag2} drives towards the non-perturbative domain with strongly interacting phases and broken symmetries~\cite{Jensen:2022iww}. While strongly coupled effects cannot be captured by a classical theory (especially by a free classical theory), our analysis below provides us with a relation between the classical models which shows that the fractonic and Carrollian models are indeed related.

The theory given by the Lagrangian \eqref{PhiLag} with $\lambda = 0$ has a trivial energy spectrum:
\beq
\varepsilon_{\boldsymbol k} = m\,,
\eeq
implying that the group velocity of the excitations is zero:
\beq
{\boldsymbol{v}}_{\boldsymbol{k}} = \frac{\partial \varepsilon_{\boldsymbol k}}{\partial \boldsymbol{k}} = 0\,,
\eeq
confirming that the quasiparticles described by the free Lagrangian $\cL_0$  \emph{do not move} \cite{Pretkofractons,PretkoCY,Gromov:2018nbv,Seiberg,Bidussi,JainJensen,Grosvenor:2021hkn,Zoo} --- which is indeed their ``{\rm raison d'\^etre}''. This  hints at the similarity of fractons with massive Carrollian particles we studied in sec.\ref{mCpart}.

The independence of the particle energy level from its momentum is  characteristic of the so-called flat bands, the subject which is under staring interest in condensed matter physics. Flat bands appear in the magic-angle bilayer graphene~\cite{Tarnopolsky:2018mxs} which hosts unconventional superconductivity at low temperatures~\cite{magic_angle_superconductivity}. Fermionic models and the relation to Carroll physics of flat bands has been discussed very recently Ref.~\cite{Bagchi:2022eui}.

Let us now see how Carrollian concepts provide a key to study fractons.

\subsection{Dipole vs Carroll boost symmetry
}\label{fracsymmSec}

Condensed matter physicists  argue that the manifest  Aristotle symmetry of fractons can be augmented by a somewhat mysterious  \emph{dipole} (or even multipole) \emph{symmetry} implemented through a position-dependent but time independent phase factor,
\beq
\Phi\left(\bx,t\right) \to \widehat{\Phi}\left(\bx,t\right) =e^{im(b_0- \bb_{1}\cdot \bx)}\Phi \left(\bx,t\right)\,,
\label{dipolesym}
\eeq
where $\bb_1 \in \IR^2$ and where a constant phase rotation by $b_0$ has also been added \cite{Pretkofractons,PretkoCY,Gromov:2018nbv,Seiberg,Bidussi,JainJensen,Ozkan}.
 Infinitesimally we have,
\besub
\begin{align}
\delta\Phi(\bx, t) = & \quad \;i\beta_0\, \Phi(\bx, t),
\label{Cmass}
\\
\delta\Phi(\bx, t) = & -i(\bbeta_1\cdot\bx)\, \Phi(\bx, t)\,
\label{indipact}
\end{align}
\label{fracsymm}
\esub
\!($\beta_0= \delta b_0,\,\bbeta= \delta\bb$).
\emph{Dipole symmetry} is reminiscent of Carroll symmetry: it implies \emph{immobility} --- the characteristic feature of fractons    \cite{Seiberg,JainJensen,Bidussi,Grosvenor:2021hkn,Surowka}.

Fractons and Carroll particles are in fact closely related  \footnote{Remarkably, the word ``Carroll'' does not even  appear in the cond-mat literature (with the notable exception of  \cite{Bidussi}).}. Below we show that their \emph{dipole symmetry is in fact the remnant of a curious double Carroll boosts symmetry of the free system broken to a single, ``internal'' symmetry by the interaction}.

We start with the free first-order Lagrangian $\cL_0$ in \eqref{PhiLag}, which is manifestly invariant under Aristotle transformations (see sec.\ref{GCAgeo}).
However the absence of spatial derivatives implies  that it is also invariant by the ``internal phase shift''
in \eqref{dipolesym} \cite{Seiberg,Bidussi,JainJensen}.
We note also that $\cL_0$ is in fact invariant under
\beq
\Phi\left(\bx,t\right) \to \widehat{\Phi}\left(\bx,t\right) =e^{imT(\bx)}\Phi \left(\bx,t\right)
\label{multipole}
\eeq
for any function $T(\bx)$,
called, in cond-mat, a multipole transformation \cite{Bidussi}.
 This follows  from \eqref{solmpos} and the absence of spatial derivatives by insertion into \eqref{freefrac1}.

The expressions \eqref{multipole} (and \eqref{dipolesym} in particular) are reminiscent of a Galilei boost acting on a scalar field, eqn. \eqref{Boostonscalarfield} ---  except that it does \emph{not} shift the position $\bx$,
 hinting rather at an intriguing similarity with Carroll field theory (sec.\ref{fieldCarrSec}).
 Bidussi et al. \cite{Bidussi} recognized indeed that dipole (resp. multipole)  symmetry is in fact a Carroll transformation which shifts the \emph{time} $t$ but not the position,
\begin{equation}
\bx \to \bx,\qquad
t \mapsto t + T(\bx)\,,
\label{tsupertransl}
\end{equation}
called, in the BMS context, a \emph{supertranslation} \cite{BMS}.

The dipole (multipole) symmetries of free fractons belong to the ``super-Aristotle'' group $\gA^{\infty}$ (sec. \ref{GCAgeo})  implemented as in \eqref{fullsymcarr} \emph{after interchanging} $s$ and $t$.
Now the Carroll group $\gC$ is a subgroup of $\gA^{\infty}$.
 Carroll time translations and Carroll boosts are just the first two lowest-order supertranslations.
 Dipole symmetry \eqref{dipolesym} is thus recovered as Carroll-boost (C-boost) symmetry \emph{built on shifting the time, $t$}.
The Noether theorem (sec.\ref{fieldNoether}) applied to the free Lagrangian $\cL_0$ in  \eqref{PhiLag} then yields the conserved quantities,
\besub
\begin{align}
q =& \;\;\;\; \int \vert \Phi(\bx,t) \vert^2 \, d^3\bx \,,
\label{fracecharge}
\\
\bd = & - \int \vert\Phi(\bx,t)\vert^2 \bx\, d^3 \bx\,,
 \label{dipolemom}
\end{align}
\label{fracCQ}
\esub
\!interpreted as  conserved \emph{electric charge} \footnote{The charge and mass densities are assumed to be up to sign proportional.} and \emph{dipole moment} \footnote{To be compared to \eqref{Cfieldboostmom}}.

While the conservation of charge~\eqref{fracecharge} does not affect how the excitations move, the immobility of a single particle follows from the conservation of the dipole moment~\eqref{dipolemom}. Indeed, the motion of a particle would immediately change the dipole moment of the system so that the particle cannot move if the dipole moment is conserved. However, the motion of two oppositely charged particles is allowed because such a motion conserves not only the electric charge~\eqref{fracecharge} but also the dipole moment~\eqref{dipolemom}, provided the distance between the particles stays unchanged.

For any solution of the wave equations \eqref{freefrac1} (or \eqref{freefrac2}),
the particle density is manifestly independent of time \cite{Marsot21,deBoer20},
\beq
|\Phi(\bx,t)|^2= |{\phi}(\bx)|^2\,,
\label{eq_particle_density}
\eeq
providing us with the field-theoretical analog of \emph{immobility}:
the conserved charges \eqref{fracCQ},
\beq
q = -{m}\!\int\!
|\phi({\boldsymbol{x}})|^2d\bx\,
\aand
\bd=m\int\!|{\phi}(\bx)|^2\bx\, d^2\bx\,,
\label{fractoncharges}
\eeq
are manifestly $t$-independent. Thus:

\begin{thm}
\textit{The free fracton Lagrangian  $\cL_0$
 and thus the equation of motion \eqref{freefrac1} are  identical to the free Carroll expressions \eqref{Carrfeq1} and \eqref{CarrLag2}, up to interchanging $t$ and $s$.
The dipole (resp. multipole) symmetry of free fractons
  can be understood as  symmetry under Carroll boosts \eqref{Cboostbis} (resp. supertranslations).
  The conserved dipole moment in \eqref{fractoncharges} and the conserved C-boost moment in \eqref{Cfieldboostmom} are the same, $\bd=\bd^{\mathrm{Car}}$.}
\label{Carrolfraction}
\end{thm}
Free fractons can thus be seen as Carroll field theories. However they can be seen also from another, complementary point of view.

Recall first certain similarities shared with the boost symmetry in Galilean quantum mechanics, \eqref{Boostonscalarfield} \cite{DBKP,DGH91} \footnote{
To incorporate the phase degree of freedom by lifting the wave function to a fiber bundle with a new, ``internal'' coordinate has first appeared in Geometric Quantization
\cite{Souriau67} and  then in ``monopole without strings'' discussions  \cite{WuYang76,Yang77,Balachandran79}. The latter are discussed also in the Prequantization Chap.~V. written around 1975 for the planned  but never completed revised edition  of Souriau's book \cite{SSD}.
}.
Galilean boosts are represented projectively, i.e., up to a phase on the wavefunctions. The phase freedom can be incorporated by lifting the latter from  Galilean (or Newton-Cartan) spacetime to one dimension higher, to Bargmann spacetime $\cB$ \cite{DBKP,DGH91}  with coordinates $(\bx, t, s)$,
where the variable $s$ corresponds to the phase. The lifted wavefunctions are equivariant,
\begin{equation}
\Psi(\bx, t, s) = e^{ims}\Phi(\bx, t)\,,
\label{solequiv}
\end{equation}
for some function $\Phi(\bx, t)$.
 The $1$-parameter central extension of the Galilei group (called the Bargmann group) acts on $\cB$ and the phase factor in the Galilean phase symmetry, \eqref{Boostonscalarfield} is recovered by equivariance \eqref{solequiv} with respect to the ``vertical'' variable, $s$.

Now we argue that
the same procedure could be applied also for fractons. We could indeed consider fractons as wavefunctions lifted at some (for now unknown), fiber bundle $\cQ$ with coordinates $(\bx, t, s)$ whith a ``vertical'' coordinate $s$. $\cQ$  is analogous to Bargmann space  --- except that its  base is an Aristotle, instead of Newton-Cartan  manifold. Such wavefunctions would still be  \eqref{solequiv}. Given the form \eqref{solmpos} of  free fractons, the latters lift to $\cQ$ by equivariance, as,
\begin{equation}
\Psi_{\mathrm{free}}(\bx, t, s) = e^{ims} \left(e^{imt} \phi(\bx)\right)\,.
\label{solfreeQ}
\end{equation}

Let us emphasise that for free fractons the time, $t$, and the internal variable, $s$, play an equivalent role: what previously looked as a Carroll transformations acting on the time $t$, could  be seen as Carroll (or even BMS-type) transformations acting on the submanifold defined by fixing
$t$, shifting on the internal (phase) coordinate $s$,
$s \mapsto s + T(\bx)\,$ cf. \eqref{ssupertransl}
implying, by  \eqref{solfreeQ},
\beq
\Phi\left(\bx,t\right) \to \widehat{\Phi}\left(\bx,t\right) =e^{imT(\bx)}\Phi \left(\bx,t\right)\,.
\label{multipolebis}
\eeq

The remarkable result is that C-boosts involving $t$ or $s$ yield the same implementation, \eqref{dipolesym}, and thus the same conserved quantity. This follows from the \emph{missing} of both $s$ and $t$ and having just $\bx$ in the exponent.
The clue is the double role of \eqref{solmpos}: both equivariance and solution.

 Thus conservation laws can be derived in \emph{two different ways}: either  ``horizontally'', by identifying fracton wavefunctions \eqref{solmpos} as defined on a Carroll structure, or ``vertically'' by lifting the wavefunctions through \eqref{solequiv} to a fiber-bundle in one dimension higher.

Free fractons can be coupled to a background electromagnetic field using the standard minimal coupling procedure,  $\partial_{\mu} \to \partial_{\mu} - i e A_{\mu}({\boldsymbol{x}})$.
But the free Lagrangian $\cL_0$ has no spatial derivatives $\boldsymbol\nabla$, therefore
 the vector part of the gauge potential, $\boldsymbol{A}$, does not enter into the Lagrangian.

The electric potential $V = V(\bx,t)$ does appear via $\partial_t \to \partial_t - i e V ({\bx,t})$, but it only modifies the phase of the solution,
\beq
\Phi_V(\bx,t) = e^{i[mt + \int^t\!eV(\bx,t) dt]}\phi(\bx)\,,
\label{fraclift2}
\eeq
cf. in~\eqref{solmpos},
without affecting the particle density~\eqref{eq_particle_density},
$|\Phi(\bx,t)|= |\phi(\bx)|$. Therefore free fractons
 do \emph{not} move even in an e.m. field: the immobility statement is extended to fractons in external fields.
The same conclusion can be  drawn also from dipole conservation:
$\Phi_V$ in \eqref{fraclift2} solves the coupled field equation
\beq
\big(\partial_t - i e V ({\bx},t)\big)\Phi_V=im\Phi_V\,,
\label{vfraceq}
\eeq
which implies in turn the conservation of $\bd$ in \eqref{dipolemom},  which boils down again to \eqref{fractoncharges}.

\smallskip
That Aristotle symmetry generates a Carrollian model is not surprising: the same happens at the classical level.
Aristotelian particles can be constructed, for instance, by the KKS algorithm \cite{AncilleAristotle}. Denoting an element of the dual by $\mu = (\ell, \bp, \mE)$, where we have as usual $\ell$ the (anyonic) spin of the particle, $\bp$ its momentum and $\mE$ its energy, the coadjoint action is
\beq
\Coad(g) \mu = (\ell - \bc \times A \bp, A \bp, \mE).
\eeq
 Thus the coadjoint orbits are just the Euclidean orbits shifted by $\mE$.
The Casimirs are the norm of the momentum, $|\bp|$, and the energy $\mE$ \footnote{In 3 dimensions, we would have also the scalar spin $\boldsymbol{\ell} \cdot \bp$.}.
For $|\bp| \neq 0$ and $\mE \neq 0$, the Cartan 1-form is \footnote{The Lagrangian resp. Souriau form are, accordingly, $\cL_{Arist}=\bp\cdot\frac{d\bx}{dt} + \cE$ (notice, however, that $\cE$ is merely a constant here) and $\sigma = d\bp\wedge d\bx$.},
\beq
\varpi_{Aris} = \bp\cdot d\bx + \mE dt\,.
\label{ArisCartan}
\eeq

However in \eqref{ArisCartan} we recognize that we formally have the same model as the Carroll one in \eqref{genCarrollCS}, in the interchange $s\leftrightarrow t$ and $m \leftrightarrow \mE$.

The Aristotle group $\gA$ is by construction (and also manifestly) a symmetry of \eqref{ArisCartan}. However, owing to the fact that the time translation in the Aristotle group appears as a trivial central extension of the Euclidean group, the coupling between space and time coordinates are uncoupled explaining why the Aristotle  models may have position-dependent time translations, \ie, supertranslations. Then, as mentioned above, the Carroll group is a subgroup of Aristotle + supertranslations.


Turning now to the full Lagrangian \eqref{PhiLag}, we note first  the self-interaction term  $\cL_{int}$ is invariant under the phase transformations \eqref{dipolesym},
\beq
\cL_{int}(\Phi)=\cL_{int}(\widehat{\Phi})\,
\eeq
which are thus symmetries for the interacting system \eqref{PhiLag}~\cite{Bidussi}.
Then the Noether theorem implies that the free expressions \eqref{fracCQ} remain correct for the full dynamics.

What about the Carroll approaches?
One sees readily that while the lifting condition \eqref{equivariance} remains valid, the field equations for interacting fractons do change. Thus the ``horizontal'' Carroll transformation involving $t$, \eqref{tsupertransl} with $T(x)=-\bb_1\cdot \bx$, is not more a symmetry for the full expression \eqref{PhiLag}. However the ``vertical'' one with $s$, \eqref{Cboost}, is still a symmetry , confirming the dipole symmetry for  \eqref{PhiLag}.
 Their curious accidental identity is not more valid  for $\lambda \neq 0$.

\smallskip
One can wonder if the BMS-type higher symmetry survives the addition of the interaction term $\cL_{int}$ in \eqref{PhiLag}. Plugging the Ansatz \eqref{multipole}   into \eqref{PhiLag}, we readily see that
\beq
\cL \rightarrow \widehat{\cL} = \cL + \lambda \Phi^4 (\partial_i \partial_j T) (\partial_i\partial_j T)\,.
\eeq
Not being a total derivative, the shift breaks the symmetry, unless
\beq
\partial_i\partial_j T=0
\quad\text{for all}\quad i,j\,.
\label{intsymm}
\eeq
It follows that only terms up to linear order  in $\bx$,  $T(\bx) =  \beta_{0}-\vec{\bbeta}_{1}\cdot \vec{\bx}$ survive in general, as in \eqref{dipolesym}.

Similarly, a dilation
$\Phi \to \widehat{\Phi} = \Lambda \Phi,\, \Lambda=\const$   breaks the scale invariance of the free Lagrangian,
\beq
\cL_0 \rightarrow \widehat{\cL}_0 = \Lambda^2\cL_0
\qquad\text{but}\qquad
\cL_{int} \rightarrow \Lambda^4 \cL_{int}\,.
\eeq
In conclusion,

\begin{thm}
\textit{The  BMS-type infinite multipole symmetry \eqref{multipole} of the free  Lagrangian $\cL_0$  is broken by the interaction term $\cL_{int}$ to the first two (constant and linear) terms, as in eqns \eqref{dipolesym}. The first of these generates conserved mass/electric charge, while the linear one, which is indeed an  internal Carroll boost \eqref{Cboost}, generates the conserved dipole moment $\bd$ in \eqref{fractoncharges} which implies {immobility} even for interactions considered in \eqref{PhiLag}.}
\end{thm}
\goodbreak

With hindsight to the Galilean case, it would be expected that the boost symmetry is broken by the interaction. It is therefore remarkable that \emph{one  of two Carroll boost}
--- namely the $s$-based ``internal one'' -- does survive the addition of $\cL_{int}$ which thus keeps its \emph{dipole symmetry} \footnote{Bidussi et al. \cite{Bidussi} call the ``horizontal'' ($t$-based) transformation ``Carroll boost'' and the ``vertical'' one based on $s$ ``internal''.}.

The link between Carroll symmetry and fractons is also investigated in \cite{Figueroa-OFarrill:2023vbj}.

\section{Doubly-extended Carroll dynamics}\label{ExoCarrSec}

\subsection{Free massive doubly-extended particle}\label{freeCarroll}

A  fact which escaped attention until recently is that in $d=2$ space dimensions the Carroll group admits a \emph{two-parameter  central extension}
\cite{Ancille,Azcarraga,Marsot21}. Its central charges shall be denoted by $\kappa_{\ex}$ and $\kappa_{\ma}$, respectively,
whose non-vanishing will henceforth be assumed.
Such additional charges are allowed by the theory and they may play an important role for planar systems with Carroll symmetry.

The doubly-extended group, $\widehat{\gC}$, can be represented by $6\times6$ matrices (Eqn \# (2.9) in \cite{Marsot21}) to be compared with \eqref{BCmatrix},
\beq
\barray{clccc}
A^i{}_j & 0 & c^i &0 & \epsilon^i{}_k b^k
\\
-b_k A^k{}_j &1 &f &0 &A_{\ex}
\\
0 & 0 &1 &0 &0
\\
c_k \epsilon^k{}_l A^l{}_j &0 &A_{\ma} &1 &- (f + \bb\cdot\bc )
\\
0 & 0& 0 & 0 &1
\earray\,,
\label{exoCmatrix}
\eeq
where $A_{\ma}$ and $A_{\ex}$ are new parameters associated with the double extension. The extended algebra
has a 8-parameter moment map,
  $\mu(\ell, \bd, \bpi , m, \kappa_{\ma}, \kappa_{\ex})$
with commutators,
\beq
\barraynb{llll}
[J, K_i] = \epsilon_{ij} K_j\,, \quad
&[K_i, K_j] = \epsilon_{ij} A_{\ex}\,, \quad
&[J, P_i] = \epsilon_{ij} P_j\,, \quad
&[K_i, P_j] = \delta_{ij} P_0\,,
\\[6pt]
[J, P_0] = 0\,, \quad
&[K_i, P_0] = 0\,, \quad
&[P_i, P_j] = \epsilon_{ij} A_{\ma}\,, \quad
&[P_i, P_0] = 0\,,
\earraynb
\label{dCarrollCR}
\eeq
where $J$ is the rotation, the $(K_i)$ are  boosts, the $(P_i)$ spatial translations, $(P_0)$ is time translation, and $A_{\ex}$ and $A_{\ma}$ are the exotic and magnetic extensions, respectively.

More insight into the extended algebra structure
is obtained by comparison with previously considered
 extensions. The one with $A_{\ex}$ is plainly the same
 as the ``exotic'' extension considered for the planar Galilean algebra \cite{LLGal,Galexo,DH01}.
The extension $A_{\ma}$ in the doubly extended Carroll algebra \eqref{dCarrollCR} implies in turn the non-commutativity of spatial translations,
consistently with its interpretion as an additional, ``internal'' magnetic field see sec.\ref{GeoZooSec}.

The extended Carroll group acts on these moments by \emph{coadjoint action},
\besub
\begin{align}
\ell' & = \ell + \bb \times A \bd - \bc \times A \bpi + m \bb \times \bc - \half \kappa_{\ma} \bc^2 - \half \kappa_{\ex} \bb^2\\
\bd' & = A \bd + m \bc + \half \kappa_{\ex} \epsilon \bb \\
\bpi' & = A \bp + m \bb - \half \kappa_{\ma} \epsilon \bc \\
m' & = m \\
\kappa_{\ma}' & = \kappa_{\ma} \\
\kappa_{\ex}' & = \kappa_{\ex}
\end{align}
\label{ecarrcoad}
\esub
The free doubly-extended model obtained by the KKS algorithm is, by construction, symmetric w.r.t. the doubly-extended Carroll group $\widehat{\gC}$, with associated conserved quantities,
\besub
\begin{align}
\ell =& \; \bx \times \bp +  \displaystyle\half{\theta}\bp^2
 + \displaystyle\half{\kappa_{\ma}} \bx^2 + \chi\,,
 &\text{angular \ momentum }
 \label{exoCangmom}
\\[3pt]
\pi_i =& \; p_i -\kappa_{\ma}\, \epsilon_{ij} x^j\,,
&\text{``impulsion''}
\label{exoCmom}
\\[3pt]
D^i =& \; m\Big(x^i +\theta\,\epsilon^{ij} p_j\Big)\,,
&\text{C-boost \ momentum}
\label{exoCboost}
\end{align}
\label{consexoCarr}
\esub
completed  with
$ m\,, \, \kappa_{\ex}\,,\,\kappa_{\ma}
$, cf. \# (3.18) in \cite{Marsot21} and sec.\ref{genCarrollSec} for the terminology.

These quantities are  analogous to the ``exotic'' Galilean expressions in a constant magnetic field \cite{Galexo,DH01}.
The real constant $\chi$ in \eqref{exoCangmom} is Carrollian/anyonic spin, defined in \eqref{2CCasi}.
The boost momentum $\bd$ \eqref{exoCboost} plays a particularly important role, as it will be explained below.

The unusual form of the ``impulsion'' $\bpi$  in \eqref{exoCmom} deserves a comment.
First we note that $A_i = - \half (\kappa_{\ma}/e) \epsilon_{ij} x^j\,$ is a vector potential for $\kappa_{\ma}$  viewed as a (constant) ``internal magnetic field".
Then $\bpi$  has the unexpected form
$\pi_i = p_i + 2\,e\, A_i$, where $\bp$ is just a coordinate. However in terms of the \emph{canonical momentum}  $P_i = p_i - \half \kappa_{\ma} \epsilon_{ij} x^j$
obtained from the Lagrangian \eqref{exoCLag},
$\pi_i = P_i - \half \kappa_{\ma} \epsilon_{ij} x^j=
P_i+e\, A_i$.

\kikezd{Casimirs}.
When the mass does not vanish, $m\neq0$, we have four Casimirs \cite{Marsot21}: $\kappa_{\ma}$, $\kappa_{\ex}$, the mass, $m$, and the \emph{anyonic spin}, $\chi$, defined by,
\begin{equation}
m^*\chi = m^* \ell -  \bd \times \bpi  + \frac{\kappa_{\ex}}{2 m}  \bpi^2 +  \frac{\kappa_{\ma}}{2 m} \bd^2\,,
\label{2CCasi}
\end{equation}
where
\beq
m^* = m\left(1{-}\frac{\kappa_{\ex}}{m^2}\kappa_{\ma}\right)\,
\label{Ceffcharge}
\eeq
is an effective mass, analogous to $m^*_G$ in  \eqref{Galeffmass} in the Galilean case.

Eqn. \eqref{2CCasi} is valid also for $m^*=0$ when both the spin $\chi$ and the total angular momentum $\ell$ drop out.
In this case, realized for a special relation between the Casimir invariants $\kappa_{\ma}\cdot\kappa_{\ex}=m^{2}$, we get:
\beq
\frac{1}{m} \bd \times \bpi
- \frac{1}{2\kappa_{\ma}}\bpi^2
- \frac{1}{2 \kappa_{\ex}} \bd^2 = 0\,.
\label{Cmstar0}
\eeq

Multiplying \eqref{2CCasi} by $m^2\neq0$ and then letting $m\to 0$ yields in turn the expression valid in the massless case
\begin{equation}
\chi = \ell - \frac{\bpi^2}{2\kappa_{\ma}} - \frac{\bd^2}{2\kappa_{\ex}}\,,
\label{m0Casi}
\end{equation}
where the Casimir invariants $\kappa_\ma$ and $\kappa_\ex$ are assumed not to vanish. Equation~\eqref{m0Casi} implies that the Carrollian anyonic spin $\chi$ is given by the angular momentum $\ell$ with  contributions coming from the impulsion $\bpi$ and the C-boost  (alias generalized dipole) momentum.
\goodbreak

The KKS construction applied to  $\widehat{\gC}$ yields, for a particle with mass $m\neq0$, the Cartan  resp. Souriau-forms
 with two additional terms \cite{Marsot21},
\besub
\begin{align}
\varpi_{\ex} = &\;\; \bp\cdot d\bx\;\, + \frac{\theta}{2}\bp\times\d\bp
\qquad \;\;\;\, +\;\frac{\kappa_{\ma}}{2}\bx\times\d\bx\,,
\label{exoCarrollC}
\\[8pt]
\sigma_{\ex} = &\, d\bp \wedge d\bx
 +
\underbrace{\;\half\theta\,
 \epsilon_{ij}dp^i\wedge dp^j}_{exotic}
\;\,+ \;
\underbrace{\;\frac{\kappa_{\ma}}{2}\,\epsilon_{ij}dx^i\wedge dx^j}_{``internal \ magnetic''}\,\,,
\label{exoCarrollS}
\end{align}
\label{exoCarrollCS}
\esub
defined on Carroll evolution space $\cE$ \eqref{Carev},
cf. \# (3.10) in \cite{Marsot21}. The constant
\beq
\theta=\frac{\kappa_{\ex}}{m^2}\,
\label{exotheta}
\eeq
here is interpreted as  the {non-commutativity parameter}, see \eqref{exoCcommrel} below.

The terms with $\theta \sim \kappa_{\ex}$ in \eqref{exoCarrollCS} are those of exotic Galilean dynamics, \eqref{NCtheta}, and are in fact present also for
 \emph{relativistic anyons} from which the Galilean model can be deduced as it will be shown in sec. \ref{CarrPoinSec}.
The terms with $\kappa_{\ma}$ are  reminiscent of a (constant) ``internal \ magnetic'' field, as mentioned above. Cf. also \eqref{GalexoS}.

The Souriau form $\sigma_{\ex}$ in \eqref{exoCarrollCS} is formally $\Omega_{\ex} -{d\mathscr{H}_0}\wedge ds$  with \emph{zero} Hamiltonian,
\beq\bigbox{
\Omega_{\ex} = d\bp\wedge d\bx
+
\frac{\theta}{2}\,\epsilon_{ij}dp^i\wedge dp^j
+
\frac{\kappa_{\ma}}{2}\,\epsilon_{ij}dx^i\wedge dx^j\,
\aand \mathscr{H}_0 \equiv 0\,,}
\label{CESympHam}
\eeq
 cf. \eqref{Carroll2sigma}.
$\Omega_{\ex}$ is a closed and regular (i.e. symplectic) $2$-form on the $s=s_0=\const$ submanifold $\IR^2\times\IR^2=\Big\{\bx, \bp, s_0\Big\}$ of the Carroll evolution space $\cE$.

When the \emph{effective mass} $m^*$ is non-zero, $m^* \neq 0\,,$ i.e.
$\kappa_{\ex} \cdot \kappa_{\ma} \neq m^2$, the associated Poisson brackets  \cite{Ancille}
\begin{align}
    \big\{x^i,x^j\big\} & = \left(\dfrac
{1}{1-\theta \kappa_{\ma}} \right)\,\theta
\epsilon^{ij}\,,
\; \nonumber
\\[4pt]
\big\{x^j,p_i\big\} & =
  \left(\dfrac{1}{1-\theta \kappa_{\ma}
 }\right) \,\delta_i^{\,j}\,,
\;
\label{exoCcommrel}
\\[4pt]
\big\{p_i,p_j\big\} & = \left(\dfrac{1}{1-\theta \kappa_{\ma}
 }\right)\,\kappa_{\ma}\,\epsilon_{ij}\,
\nonumber
\end{align}
 are \emph{identical} with eqn. \eqref{GalexoPB} of the ``exotic'' Galilei case under the replacement $\kappa_{\ma} \sim eB$.
The first relation  implies that the coordinates do not commute
when $\kappa_{\ex}\neq0$
and the last one implies noncommuting momenta for $\kappa_{\ma}\neq0$.

\smallskip
Alternatively (and equivalently when $m^*\neq0$) we can consider the doubly-extended Carroll Lagrangian defined on the evolution space $\cE$ in \eqref{Carev} which has two more terms  added to the free unextended Carroll Lagrangian $\cL_0$  \eqref{0CarrLag},
\beq
\cL_{\ex} = \underbrace{\;\bp\cdot{\bx}^{\prime}\;}_{\cL_0}
\;+\;
\underbrace{\,\half\theta\,\epsilon_{ij}p^i(p^j)^{\prime}}_{exotic}
\;+\;
\underbrace{\,\half\kappa_{\ma}\,\epsilon_{ij}x^i({x}^j)^{\prime}}_{magnetic}\,\,\; .
\label{exoCLag}
\eeq
The associated variational equations (which are valid without the assumption $m^*\neq0$)
\beq
({x}^k)^{\prime}+\theta\,\epsilon^{kl}({p}_l)^{\prime}=0\,,
\qquad
({p}_i)^{\prime}-{\kappa_{\ma}}\,\epsilon_{ij}({x}^j)^{\prime}=0\,,
\eeq
 are integrated at once to yield the conserved vectors
\besub
\begin{align}
x^k  +  \theta\epsilon^{kl} p_l=&\const  = Q^k\,,
\label{Qcoord}
\\[2pt]
p_i - {\kappa_{\ma}}\epsilon_{ij}x^j =&\const = \pi_i\,.
\label{pii}
\end{align}
\label{freeCmot}
\esub
In particular, the 2-vector $\bQ = (Q^k)$ in \eqref{Qcoord}
is identified as the \emph{guiding center} \cite{Ezawa,Chiral11}.

The conservation of the boost momentum $\bd$ in \eqref{consexoCarr} alone does \emph{not} now imply  immobility for $\bx$ -- however it \emph{does imply} that of  $\bQ$ in \eqref{Qcoord}. Combining  with the conservation of $\bpi$ \eqref{pii} then yields the pair of decoupled equations
 which involve the {effective mass} \eqref{Ceffcharge},
\beq
m^*\,({x}_i)^{\prime}=0\,,
\qquad
m^*\,({p}_i)^{\prime}=0\,
\label{effmass}
\eeq
which allow us to infer:
\goodbreak
\begin{thm}
\textit{
A free doubly-extended Carroll particle with non-vanishing effective mass, $m^*\neq0$ in \eqref{Ceffcharge}, \emph{does not move}.}
\label{ThmFreeNoExtendedMotion}
\end{thm}
\goodbreak

However when the \emph{effective mass vanishes},
\beq
m^*= 0\quad \ie \quad \theta\,
 \kappa_{\ma}= 1\,
\quad\ie\quad
\kappa_{\ex}\kappa_{\ma}=m^2\,,
\label{0efm}
\eeq
then \emph{no conclusion} can be drawn from \eqref{effmass}.
For $m^*= 0$ the determinant of the symplectic form\footnote{Compare with \eqref{Galeffmass}.},
\beq
\det \big(\Omega_{ij}\big)=
\left(\dfrac{m^*}{m\;}\right)^2
= \left(1-\theta \kappa_{\ma}\right)^2
\label{Carrexodet}
\eeq
 becomes indeed zero, highlighting the fact that the variational system is singular, requiring ``Faddeev-Jackiw'' reduction
  \cite{FaJa,Galexo,DH01}.
The dimension of the phase space drops from 4 to 2.
The  variables $\bx$ and $\bp$ become redundant however
their (fixed) combination $\bQ = (Q^k)$ in \eqref{Qcoord}
is still a physical quantity.
In its terms the reduced  free symplectic form is simply
\begin{equation}
\Omega_{red} = \kappa_{\ma}\, dQ^1 \wedge dQ^2 =
\frac{1}{\theta}\,dQ^1 \wedge dQ^2\,.
\end{equation}
The reduced Poisson brackets and variational forms are thus
\beq
\big\{Q_1,Q_2\big\}_{red}= -\theta\,,
\qquad
\varpi_{red}= \frac{1}{2\theta}\,
\epsilon_{ij}\,Q^idQ^j\,,
\qquad
\cL_{red}=\frac{1}{2\theta}\,
\bQ\times{\bQ}^{\prime}\,.
\label{redCLag}
\eeq
The Hamiltonian $\mathscr{H}$ is  identically zero and the equations of motion,
${Q}_i^{\prime} = \{Q^i,\mathscr{H}\}$,
 are therefore,
\begin{equation}
Q_i^{\prime} = 0 \Rarrow \bQ = \bQ_0 = \const
\label{DotQ0}
\end{equation}
consistently with  \eqref{Qcoord}. In conclusion,

\begin{thm}
\textit{
When the effective mass of a massive system vanishes, $m^*=0$, the free massive doubly-extended particle model becomes singular. The dynamics of $\bx$ and of $\bp$ \emph{can not be} \emph{separately determined}, however  the guiding center $\bQ$ in  \eqref{Qcoord}, \emph{remains fixed}.}
The conserved quantities in \eqref{consexoCarr} are expressed in terms of the guiding center alone,
\besub
\begin{align}
\ell =\; &\frac{1}{2\theta} \bQ^2 & \mbox{\rm{angular \ momentum}}
\\[4pt]
\pi_i =& \,
-\frac{1}{\theta}\epsilon_{ik} Q^k &\mbox{\rm{``impulsion"}}
\\[4pt]
\bd =\; & m \bQ &\mbox{\rm{C-boost \ momentum}}
\end{align}
\label{QCarrCas}
\esub
consistently with the loss of two physical degrees of freedom.
\label{ThmMotionVanishingEffMass}
\end{thm}

For $m^*=0$ the (no-)motion of the guiding center, \eqref{DotQ0}, follows directly from the conservation of $\bd$ or of $\bpi$,
themselves proportional to $Q^k$.

We conclude that extended Carroll symmetry plays for the extended model a role analogous to that of Carroll in the unextended case, with the effective mass $m^*$ \eqref{Ceffcharge} replacing the ``naked'' mass, $m$.

\subsection{Massless doubly-extended particles}\label{exom0C}

Let us now study the double extension of the
massless Carroll particles discussed in section \ref{s:massless_carr}.
The coadjoint action of the Euclidean group \eqref{euklgroup} extends to one of the doubly extended Carroll group.
The group element $\big(A,\bb,\bc,f,A_{\ex}, A_{\ma}\big) \in \widetilde{C}$
is implemented on $\tilde{\mu}_0=(\ell,\bd,\bpi, m=0,\kappa_{\ex},\kappa_{\ma}) \in {\widetilde{\mathfrak{c}}\,}^*$ as,
\besub
\begin{align}
\ell \to &\; \ell- \bc\times A\bpi + \bb\times A\bd
- \half\kappa_{\ex}\, \bb^2   -\half\kappa_{\ma}\,\bc^2 \,,
\label{exoellC}
\\
\pi_i \to &\; (A\bpi)_i-\kappa_{\ma}\,\epsilon_{ij} c^j\,,
\label{exogC}
\\
D^i \to &\; (A\bd)^i+\kappa_{\ex}\,\epsilon_{ij}b^j
\label{exopC}
\end{align}
\esub
while $m=0, \kappa_{\ex}, \kappa_{\ma}$ are left invariant,
 \cite{Marsot21}. Having $\kappa_{\ma}\neq 0$ in \eqref{exopC} then implies that  the norm of the momentum is \emph{not} a Casimir invariant anymore  \cite{Marsot21}.
 We thus have only one kind of massless doubly-extended particle  --- namely the trivial one associated with $\bpi=0$. Put in another way, the ``color'' $k$ in a chosen  basepoint  can be eliminated by a translation which carries $\tilde{\mu}_0$  to the origin of the dual algebra, whose contribution to the KKS symplectic form is therefore obviously zero.
Thus the first term in \eqref{exoCarrollC} disappears, leaving us with the ``truncated'' or ``purely exotic'' Cartan and Souriau forms,
 \besub
\begin{align}
\varpi_{\ex} = &\;\; \frac{\kappa_{\ex}}{2}\bv\times\d\bv
\qquad \,\, \, +\;\,\half{\kappa_{\ma}}\bx\times\d\bx\,,
\\[4pt]
\sigma_{\ex} = &  \underbrace{\half\kappa_{\ex}\,
 \epsilon_{ij}dv^i\wedge dv^j}_{exotic}
\;\; + \;\;
\underbrace{\half{\kappa_{\ma}}\,\epsilon_{ij}dx^i\wedge dx^j}_{magnetic}\,\,,
\label{Csigmav}
\end{align}
\label{exoCarrollCSv}
\esub
\!where we reverted to the $\bv$-notation since $\bp = m \bv$ is not well-defined anymore\footnote{The same result is obtained alternatively by putting $\bp = m \bv$  in \eqref{exoCarrollC} and then letting $m\to0$.} which further extends the ``no-particle" model discussed in subsec. \ref{noparticlemot}.
Had we set the extension parameters to zero we would find vanishing forms and \emph{thus no particle trajectories}, as we had observed before in section \ref{s:massless_carr}. However keeping the terms induced by the central extension, we \emph{do find ``trajectories''}: the equations of motion become trivial and yield \emph{fixed positions}\footnote{The vector $\bu$ in the Fermat model of sec. \ref{s:massless_carr} is now ill-defined: null vectors have no direction.},
\beq
({x}^i)^{\prime}=0\,,
\quad
({p}_i)^{\prime}=0
\Rarrow
\bx(s)=\bx_0=\const,
\quad
\bp(s)=\bp_0=\const
\eeq

\begin{thm} \textit{
A ``purely exotic'' free massless doubly-extended Carroll particle \eqref{exoCarrollCSv} which has $\bpi=0$ \emph{does not move}: the double
extension eliminates the free motions we  studied in sec.\ref{s:massless_carr}.
}
\label{ThmIV.2+half}
\end{thm}

This no-motion conclusion will be modified when the particle is coupled to  external fields, see the section below.
\goodbreak

\subsection{Coupling to a gauge field: eppur si muove}\label{exoCcoup}

The free model can be coupled to an em field by  the rule \eqref{Smincoupling}. The external and ``internal''  magnetic fields  combine into the generalized magnetic field:
\beq
 {B^*} = eB + \kappa_{\ma}\,,
\label{combCmag}
\eeq
and yield non-commuting coordinates and momenta \cite{Marsot21},
\besub
\begin{align}
&\sigma_{\ex}=
dp_i\wedge dx_i
 + \frac{\theta}{2}\,\epsilon_{ij}dp^i\wedge dp^j
  +\half B^*\epsilon_{ij}dx^i\wedge dx^j
 + eE_idx_i\wedge ds\,,
 \label{exoCemS}
\\[6pt]
&\{x_{i},x_{j}\}=\displaystyle\frac{m\,}{m^*}\theta\,\epsilon_{ij}\,,
\qquad\{x_{i},p_{j}\}=\displaystyle\frac{m\,}{m^*}\,\delta_{ij}\,,
\qquad	
\{p_{i},p_{j}\}=\displaystyle\frac{m\,}{m^*}\,B^*
	\,\epsilon_{ij}\,,
\label{CexoPB}
\end{align}
\label{Cexodyn}
\esub
where the Carrollian effective mass which generalizes \eqref{Ceffcharge} now involves the combined field  \eqref{combCmag},
\beq
m^*=
m \left(1 - \theta B^*\right)\,,
\label{m**}
\eeq
assumed not to vanish\footnote{Compare with \eqref{GalexoS}, \eqref{Galeffmass} and \eqref{GalexoPB} in the Galilean case.}.
 For a static scalar potential $V = V(\bx)$, $\bE = -\bnabla V$, for example, the Hamilton equations are \cite{Marsot21}),
\besub
\begin{align}
({x}^i)^{\prime} = & - \frac{\theta}{1- \theta B^*} \epsilon^{ij}eE_j\,,
\label{exoCHallx}
\\[4pt]
\,({p}_i)^{\prime} = &\;\quad\frac{1}{1-\theta B^*}\,eE_i\;.
\label{exoCHallp}
\end{align}
\label{exoCHameq}
\esub
For completeness, we mention that the symplectic form associated with \eqref{exoCemS} is
\beq
\Omega_{\alpha\beta} =\left(
\begin{array}{cccc}
0 & B^{\ast } & -1 & 0 \\
-B^{\ast } & 0 & 0 & -1 \\
1 & 0 & 0 & \theta \\
0 & 1 & -\theta & 0%
\end{array}%
\right)
\Rarrow \Omega^{\alpha\beta}=\frac{1}{(1-\theta B^{\ast })}\left(
\begin{array}{cccc}
0 & \theta & 1 & 0 \\
-\theta & 0 & 0 & 1 \\
-1 & 0 & 0 & B^{\ast } \\
0 & -1 & -B^{\ast } & 0%
\end{array}%
\right)
\label{dexosymp}
\eeq
so that the Hamilton equations for $\mH = eV\,$ are indeed \eqref{exoCHameq}.
The position $\bx$ splits off from $\bp$ which is coupled to $\bx$ through the electric field, --- but its dynamics does not effect the motion of $\bx$.

The equations \eqref{exoCHameq} are reminiscent of but different from their Galilean counterparts in \eqref{Galexoeqmot}: $p_i$ is missing from the  ``velocity'' relation \eqref{exoCHallx} which is purely anomalous, whereas the Lorentz force is missing from the purely electric $\bp$-equation \eqref{exoCHallp}. The external magnetic field is hidden in $m^*$ through $B^*$.

\begin{thm} \textit{The position $\bx$ of a massive doubly-extended Carroll particle in a static  EM  field
with  non-vanishing effective mass $m^*\neq0$
 \emph{moves} by following the (anomalous) Hall law, \eqref{exoCHallx}.
}
\label{ThmVI.4}
\end{thm}
\noindent
However combining the equations \eqref{exoCHameq} implies,

\begin{thm}
\textit{The guiding center $Q^i = x^i + \theta\,\epsilon^{ij}p_j$ in \eqref{Qcoord} of a doubly-extended particle with $m^*\neq0$, coupled to an electromagnetic field {does not move},}
\beq
Q_i^{\prime} = 0 \Rarrow \bQ = \bQ_0 = \const
\label{Qfix}
\eeq
\label{ThmVI.5}
\end{thm}

\vskip-11mm
When $m^*\neq0$ the real-space motion is recovered by solving the $\bp$ equation and using \eqref{Qcoord},
\beq
x^i(s)=Q_0^i - \theta \epsilon^{ij} p_j(s)\,.
\label{xQp}
\eeq
For constant  EM  fields s.t. $m^*\neq0$, for example, the motion is perpendicular to the electric field,
\beq
x^i(s)=Q_0^i - \epsilon^{ij} \Big(\frac{e\theta}{1- \theta B^*}\Big)E_j s\,.
\label{xQpConst}
\eeq
For vanishing effective mass, $m^*=0$ the $\bx$-motion can not be recovered as said before (and consistently with \eqref{exoCHameq}
and \eqref{xQpConst}).

Another example is obtained by a radial electric field $\bE = f(|\bx|)\,\hat{\bx}$, where $f(|\bx|)$ is a function of the radius in the $d=2$ plane. Multiplying \eqref{exoCHallx} by $\hat{\boldsymbol x}$ shows that  the motion which is  circular.
In terms of the complex variables
$
\zeta= x^1+ ix^2$ and $\Pi = p_1+ip_2$,
\besub
\begin{align}
\zeta(s) &= |\bx_0| \exp[{-i\Omega s}]\,
\where
\Omega =\frac{e\theta}{1-\theta B^*(|\bx_0|)}\frac{f(|\bx_0|)}{|\bx_0|}\,,
\label{zetamotion}
\\[4pt]
\Pi(s) &= i\frac{|\bx_0|}{\theta}\,\exp[{-i\Omega s}] +\Pi_0
=
\frac{i}{\theta}\,\zeta(s) +\Pi_0\,.
\label{pimotion}
\end{align}
\label{zetapimotion}
\esub
Thus the momentum  rotates around $\Pi_0=\const$ (which can be absorbed into the guiding center) with the same angular velocity $\Omega$ as $\zeta$, but with a phase advance by $\pi/2$.

\begin{figure}[ht]
\hskip-5mm
\includegraphics[scale=0.25]{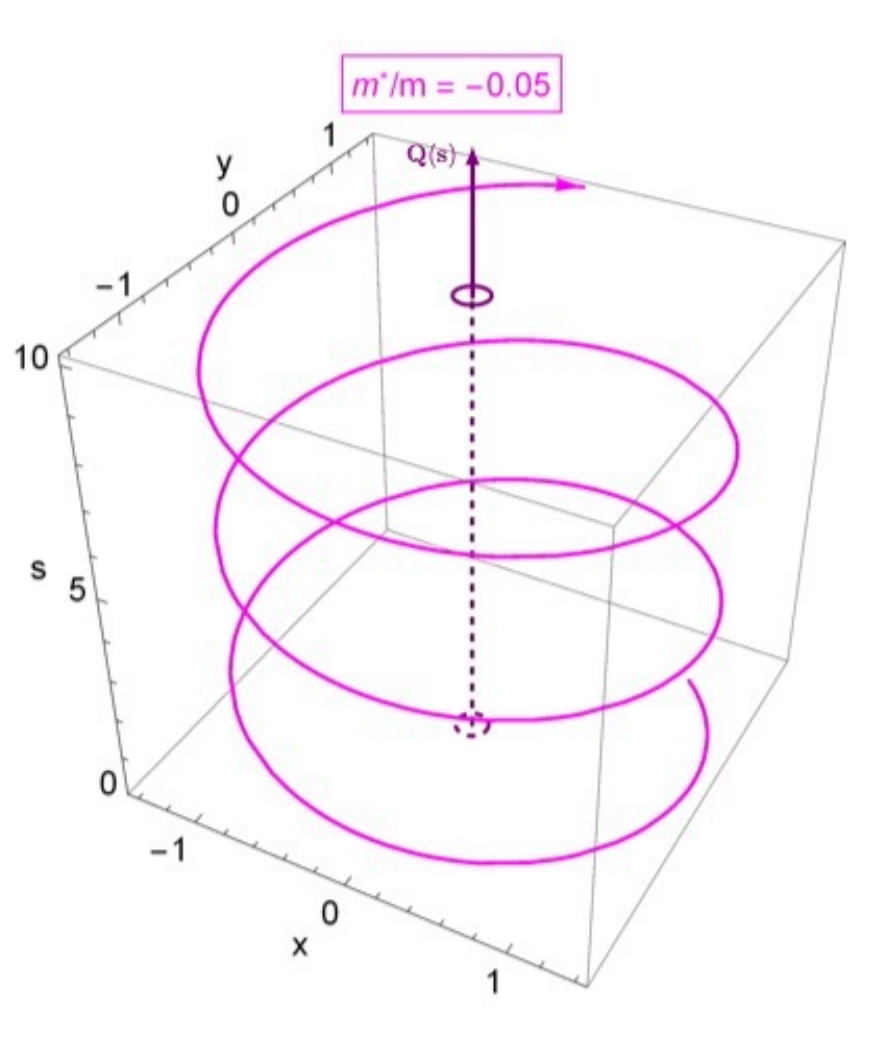}
\,
\includegraphics[scale=0.25]{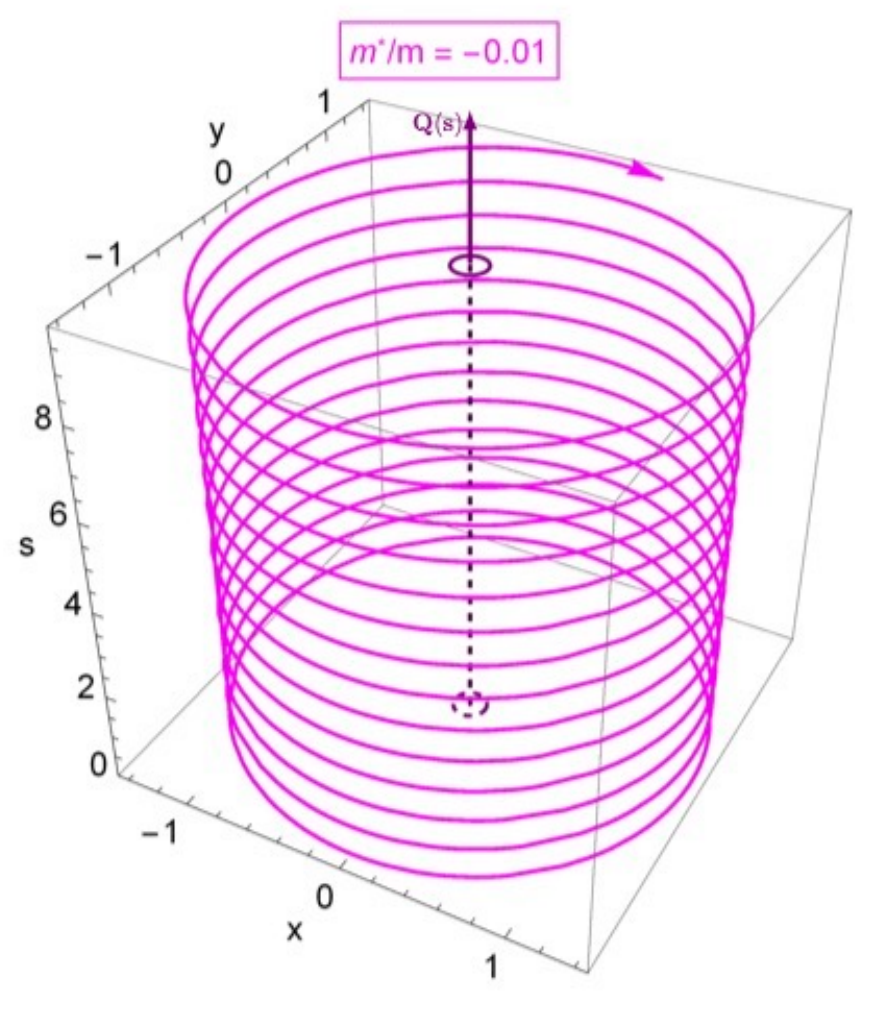}
\,
\includegraphics[scale=0.25]{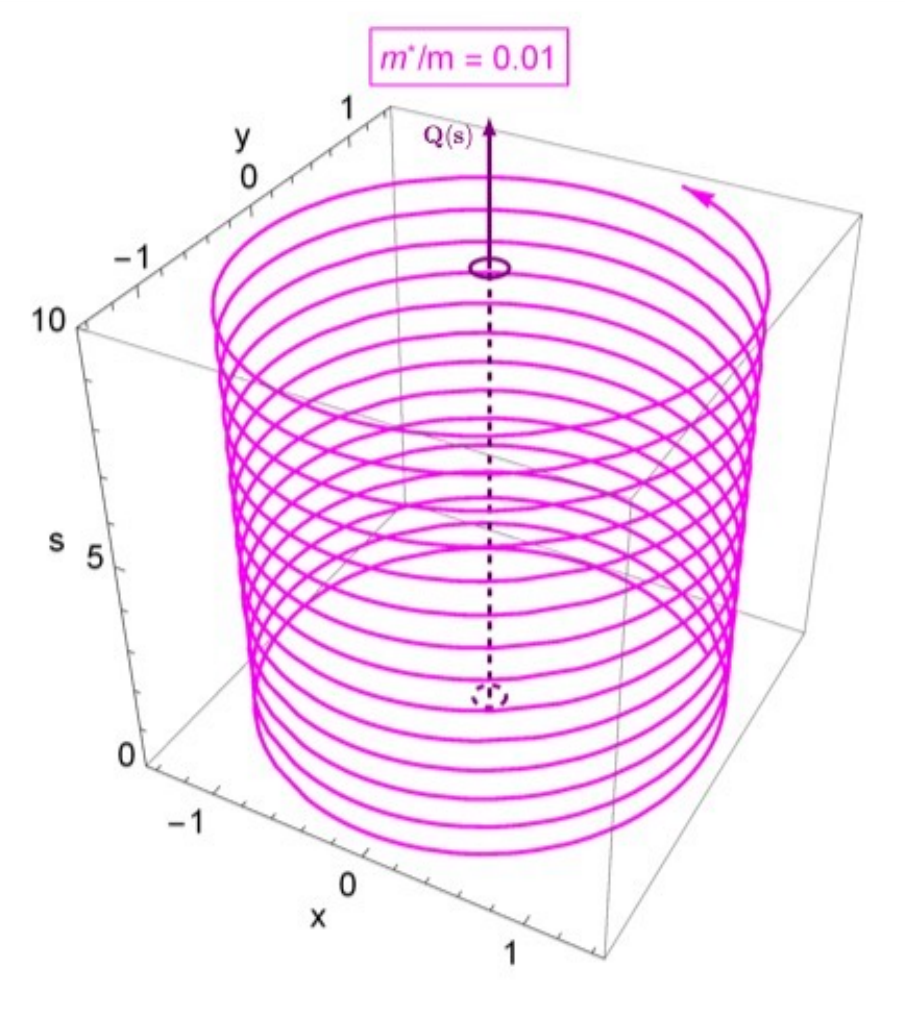}
\,
\includegraphics[scale=0.25]{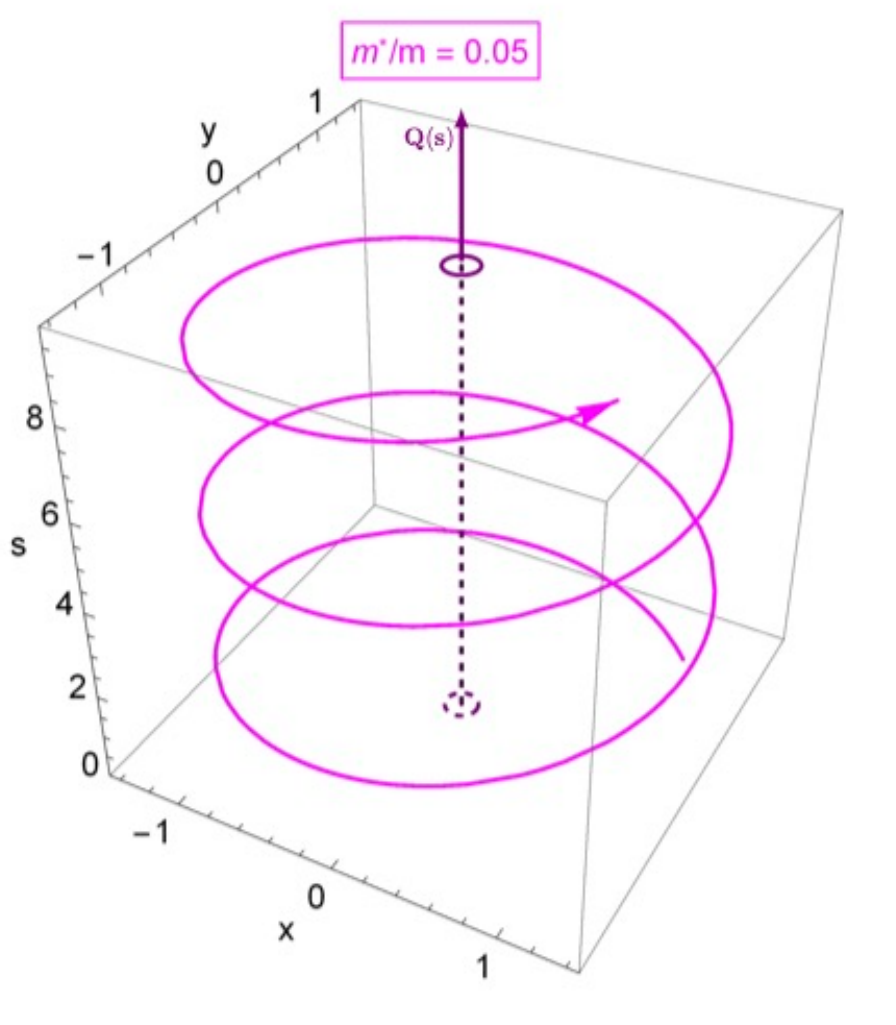}
\\\null\vskip-9mm
\caption{\textit{\small Trajectories for unit charge and exotic parameter in a radial electric field
$ \bE = {\bx}/{|\bx|^3}$ in the plane for
 effective masses passing from negative to positive,
 unfolded in Carroll time $s$.
The rotation around the fixed guiding center $\bQ(s)=\bQ_0$ speeds up when the effective mass increases from $m^*<0$. The motion is instantaneous for $m^*=0$ and changes orientation for $m^*>0$.}
\label{phasetransit}
}
\end{figure}

\goodbreak
For $m^*=0$ but $\bE\neq0$
 the eqns \eqref{exoCHameq} are ill-defined, however  their guiding center, i.e. their combination $\bQ$ in \eqref{Qcoord} remains fixed  when  $m^*$ changes sign by sweeping through zero,
 \beq
\frac{m^*}{m\;}({Q}^i)^{\prime} = \frac{m^*}{m\;} \left(({x}^i)^{\prime} + \theta
\,\epsilon^{ij}{p}_j^{\prime}\right) = 0
\eeq
by \eqref{Qfix}, including at $m^* = 0$. See
Fig.~\ref{phasetransit} for an insight  into the ``phase transition''.
The external forces vary the frequency $\Omega=\Omega(|\bx|)$ but the trajectories remain circular with fixed radius $|\bx_0|/\theta$ for all values of $m^*$, cf. \cite{Chiral11}.

\subsection{Extension to vanishing Carroll mass, m=0}\label{m0Csec}

The above results can be extended from massive to \emph{ massless} particles.
The clue is to rewrite \eqref{exoCHameq} for $m\neq0$ as,
\beq
({x}^i)^{\prime} = \frac{1}{B^* - m^2/\kappa_{\ex}} \epsilon^{ij} eE_j\,,
\qquad
{p}_i^{\prime} = m^2\frac{eE_i}{m^2-\kappa_{\ex} B^*}\;.
\label{IV.22BIS}
\eeq
Then letting $m\to 0$ the exotic charge $\kappa_{\ex}\neq0$ drops out and we get:
\begin{thm}
\textit{A massless, charged doubly-extended Carroll particle  in  combined electric and effective magnetic fields $\bE$ and $B^*$ \emph{moves} according to the generalized \emph{Hall law},}
\beq
({x}^i)^{\prime} = \,\epsilon^{ij}\frac{eE_{j}}{B^*}
\equiv \,\epsilon^{ij}\frac{eE_{j}}{eB + \kappa_{\ma}}
\;\aand\;
{p}_i^{\prime}=0\,,
\label{Emotionm0}
\eeq
\label{ThmIV.6}
\end{thm}
The uncoupled first-order eqns. \eqref{Emotionm0} and the ``interesting'' one for $\bx$ can in fact be derived form the first-order Lagrangian resp.  Souriau form defined on the $2+1$ dimensional Carroll manifold itself,
\besub
\begin{align}
\cL_{m=0}^{\ma}=&\half {B^*} \epsilon_{ij} x^i (x^j)^{\prime} - eV(\bx)\,,
\label{2+1CLag}
\\[4pt]
\sigma_{m=0}^{\ma}=&\half {B^*}\epsilon_{ij} dx^i\wedge dx^j
+
e E_i dx^i \wedge ds\,.
\label{2+1CSou}
\end{align}
\label{2+1CLS}
\esub
These expressions do not involve the exotic charge $\kappa_{\ex}$. Its magnetic counterpart, $\kappa_{\ma}$, is harmless, as it merely shifts the magnetic field by a constant and can even be turned off if we wish and then  \eqref{Emotionm0} reduces  to the usual Hall law \eqref{Halllaw}.
The expressions \eqref{2+1CLS} are consistent  also with the degenerate ``no-particle'' expressions \eqref{m0CLag} in sec. \ref{noparticlemot} associated with $\bpi=0$, written in the gauge $A_i=-\frac{B}{2}\epsilon_{ij}x^j$.

Returning to the Souriau framework, coupling the free massless particle with ``purely exotic dynamics'', \eqref{exoCarrollCSv}, to an external  EM  field by the rule \eqref{Smincoupling} yields,
 \beq
\sigma_{m=0} = \underbrace{
\half\kappa_{\ex}\,
 \epsilon_{ij}dv^i\wedge dv^j
 +
\half B^* \epsilon_{ij}dx^i\wedge dx^j}_{\Omega_{m=0}}
\;+\;
\underbrace{eE_i dx^i \wedge ds}_{-d\mH_{m=0}\wedge ds}\,.
\label{e0m0Sform}
\eeq
Rewriting the Souriau form $\sigma_{\ex}$  \eqref{exoCemS}
in terms of $\bv=\bp/m$ and then letting $m\to 0$ suppresses the first term and  yields once again \eqref{e0m0Sform}.
 The determinant of the symplectic form $\Omega_{m=0}$ in \eqref{e0m0Sform} is,
\beq
\det\big(\Omega_{m=0}\big)= (\kappa_{\ex})^2 (B^*)^2\,.
\label{m0det}
\eeq
Thus  the conditions $\kappa_{\ex}\neq0$ and $B^*\neq0$  guarantee that $\Omega_{m=0}$ is a regular symplectic form. However position $\bx$ and  $\bv$ are decoupled and the dynamics  \emph{projects} to Carroll spacetime
$\cC=\left\{\big(\bx,s\big)\right\}$: inuitively, the first term in \eqref{e0m0Sform} is switched off, and ``forgetting" about the ``irrelevant'' dynamics $\bv'=0$ we are left  with \eqref{2+1CLS}.

For $\Omega_{m=0}=0$ the system becomes singular and requires regularization called  ``Faddeev-Jackiw" [or Hamiltonian] reduction \cite{FaJa}. The rigorous statement is that in the singular case the 5-dimensional evolution space $\cE=\left\{\big(\bv,\bx,s\big)\right\}$
is reduced to 2+1-dimensional Carroll space $\cC$, endowed with the reduced dynamics \eqref{2+1CLS}. The exotic charge $\kappa_{\ex}$ drops out in the process. The reduced symplectic form $\sigma_{m=0}^{\ma}$ in \eqref{2+1CSou} is regular provided $B^*\neq0$.

The reduced dynamics in eqns \eqref{Emotionm0} -- \eqref{2+1CLS} yields Hall motion also for a massive, charged Carroll particle with no central extension, $\kappa_{\ex}=\kappa_{\ma}=0$ as long as $B\neq0$.

What can be the physical realization of these particles? Rigorously speaking, there are no strictly massless charged fields in the Standard Model of fundamental interactions in its low-temperature phase, which is relevant to present-day phenomenology~\cite{ParticleDataGroup:2022pth}. While genuinely massive constituents of the model can be treated as effectively massless particles at very high energies (such as leptons and quarks in heavy ion collisions~\cite{STAR:2005gfr}), they propagate in three spatial dimensions and, therefore, are not relevant for the peculiar features of two-dimensional physics considered in this section.

However, one cannot exclude that distant analogs could be found in the condensed matter context. As a promising  example of quasiparticles possessing massless and, simultaneously, electrically-charged excitations in two spatial dimensions has been realized in graphene, a two-dimensional monolayer crystal made of carbon atoms,  discovered experimentally in 2004~\cite{Novoselov:2005kj}.
In this material, the dynamics of the low-energy electronic excitations is described by the $2+1$ dimensional massless Dirac Hamiltonian,
\begin{equation}
{\mathscr H} = {\bar \psi} i {\boldsymbol \alpha}({\boldsymbol \nabla} - i e {\boldsymbol A}) \psi + e V\,,
\label{eq_H_graphene}
\end{equation}
where ${\boldsymbol \alpha} = (\gamma^0 \gamma^1, \gamma^0 \gamma^2)$ and the $\gamma^\mu$ are spatial Dirac matrices in $2+1$ dimensions. Being charged, the massless electronic quasiparticles interact with the background electromagnetic field $A^\mu = (V,{\boldsymbol{A}})$, thus providing us with a system that exhibits equivalences with quantum electrodynamics,
in particular in terms of the (quantum analogue of) Hall effect~\cite{Zhang:2005zz}.

In the absence of background electromagnetic fields, the Dirac particle described by the Hamiltonian~\eqref{eq_H_graphene} possesses the `relativistic' Dirac particle-like dispersion, $\varepsilon_{{\boldsymbol p}} = v_F |{\boldsymbol p}|$, where ${\boldsymbol p} = (p_1,p_2)$ corresponds to a quasimomentum of the particle with the origin at the tip of a Dirac cone. The particles propagate relativistically with the Fermi velocity, $v_F$, which is typically hundreds of times smaller than the speed of light. The massless nature of the excitations in free-standing graphene is protected by the symmetries of the underlying carbon crystal lattice, thus making the massless nature ($m=0$) of these charged excitations ($e \neq 0$) a robust feature of the material~\cite{CastroNeto:2007fxn}.

Strictly speaking, the quasiparticles in graphene should be distinguished from the Carroll particles because the former do propagate in the absence of background electromagnetic field contrary to the latter. Still, this example -- which has no analogues in terms of fundamental particles -- gives us some hope to find charged massless (quasi)particles in two spatial dimensions behaving similarly to charged, doubly-extended Carroll particles which, by Theorem \ref{ThmIV.6}, move, according to the generalized Hall law, only in a combined electric and effective magnetic field.

\subsection{Carrollian anyons}\label{CarrAn}

\emph{Carroll particles with  anyonic spin}, $\chi\neq0$ in \eqref{consexoCarr}, can also be studied along the same lines. The Souriau 2-form $\sigma_{\ex}$  \eqref{exoCemS} is generalized consistently with the coupling rule \eqref{Smincoupling} but by replacing the e.m. fields by their effective values,
\beq
\sigma_{any}=
dp_i\wedge dx_i
 + \frac{\theta}{2}\,\epsilon_{ij}dp^i\wedge dp^j
 +\half eB^*\epsilon_{ij}dx^i\wedge dx^j
 +
E_i^*dx^i\wedge ds\,,
\label{any*sigma}
\eeq
where the electric field and the Stern-Gerlach type gradient of the magnetic field are again combined into an effective electric term  \footnote{cf.  \eqref{Elshift} in the Galilean theory.
A constant shift $B \to B + \const$ produces no effect because of the gradient.
},
\besub
\begin{align}
e\bE \to \bE^*=& \; e\bE+ \mu \chi \bnabla B
\label{Estar}
\\
eB \to B^* =& \;eB+ \kappa_{\ma},\,
\label{Bstar}
\end{align}
\label{EBstar}
\esub
The equations of motion take the
 form \eqref{exoCHameq}  with $e\bE$ replaced by $\bE^*$.


The extension to zero mass is obtained along the same lines as above, yielding,

\begin{thm}
\textit{\small Massless Carrollian anyons move by following the generalization of the Hall law~\eqref{Halllaw} with the electric and magnetic fields replaced by the effective values $\bE^*$ and $B^*$ in  \eqref{Estar},}
\beq
\bigbox{
({x}^i)^{\prime}
= \epsilon^{ij}
\left(\frac{eE_{j}
+
\mu\chi \partial_{j}
B}{eB+\kappa_{\ma}}\right)
=
\epsilon^{ij}
\biggl(\frac{E_j^*}{B^*}\biggr)
  \,
\;\aand\;
{p}_i^{\prime}=0\,.\\
}
\label{genHall**}
\eeq
\label{ThmIV.7}
\end{thm}\vskip-5mm
Note here the double role  played by the magnetic field $B$.

The massless limit can also be discussed in the Souriau framework. We start with \eqref{any*sigma} with $m\neq0$ and eliminate $\bp$ in favor of $\bv= \bp/m$. Then letting $m\to 0$ we get a regular Souriau form which, compared to \eqref{Csigmav},  has an additional ``effective electric'' term,
\beq
\sigma_{any}= \half{\kappa_{\ex}}\,\epsilon_{ij}dv^i\wedge dv^j
 +
\half B^*
 \epsilon_{ij}dx^i\wedge dx^j
 +  E^*_idx^i \wedge ds\,.
\label{anyCSm0}
\eeq
The kernel of $\sigma_{any}$ yields the equation of motion \eqref{Emotionm0}
[with $p_i$ replaced by $v_i$].

\begin{table}[thp]
\begin{tabular}{|l|c|c|c|}
\hline
& free extended Carroll
& in  EM  field
& in curved metric
\\
\hline
$m \neq 0$
& no motion
& generalized Hall
& no motion
\\
\hline
$m = 0, \, e \neq 0$
& no motion
& generalized Hall
& no motion
\\
\hline
$m = 0, \, e = 0, \, \mu \chi \neq 0$
& no motion
& generalized Hall
& no motion
\\
\hline
$m^* = 0$
& $\bQ$ fixed, $\bx$ motion undetermined
&&\\
\hline
\end{tabular}\\[4pt]
\caption{\textit{Motion of doubly-extended Carroll particles in $d=2$ spatial dimensions}.}
\label{CarrollTable2}
\end{table}

The generalized Hall law~\eqref{genHall**} also contains an ``anomalous'' anyonic contribution activated by the inhomogeneities of the magnetic field~\eqref{Estar}. For massive particles, the anyonic term can be interpreted as a conventional Zeeman force
\begin{equation}
{\boldsymbol F} = g \mu_B s {\boldsymbol \nabla} B\,,
\label{eq_Zeeman}
\end{equation}
exerted by the spatial gradient of the transverse magnetic field $B \equiv B_z$ on the particle propagating in the plane. Here $g$ is the Land\'e factor, and $\mu_B$ is the Bohr magneton. A comparison of Eqs.~\eqref{Estar} and \eqref{eq_Zeeman} provides us with the identification $e \mu\chi \equiv g \mu_B s$\,.

In spintronic devices, the electrons with different orientations of their spins $s \equiv s_z$ are split by the Zeeman force~\eqref{eq_Zeeman}. This phenomenon lies at the heart of a mesoscopic Stern-Gerlach effect which provides an efficient spin filter for a current traversing a specially fabricated device that incorporates an inhomogeneous magnetic field $B = B_z$~\cite{ref_spin_filters}.

Thus, two-dimensional Carrollian anyons can be interpreted as spin-polarized charged particles with a nonzero Land\'e factor.

\subsection{A toy model}\label{ToyMods}

An ``exotic photon'' is defined by being a massless, uncharged particle, carrying magnetic moment, anyonic spin \footnote{In addition to photons and neutrinos in  particle physics, uncharged and massless quasiparticles also appear in the solid state context as phonon excitations in crystals and fluids.}, and having non-vanishing ``exotic charges''. It has a pure Stern-Gerlach-type Carroll Hamiltonian of a spinning anyon with no kinetic term,
\beq
\mathscr{H}_{SG}=-\mu\chi B\,.
\label{CarrAHam}
\eeq
cf. \eqref{anymagmom}. Although this particle carries a ``bosonic'' name, it can be of a fermionic or even anyonic  nature as well since $\chi$ can be any real number.

A distant physical analogue of the exotic photon in relativistic fundamental field theory can be played by  Dirac or Majorana neutrinos which, being neutral and almost massless, could possess a sizable magnetic moment in certain exotic scenarios~\cite{Bell:2006wi,Bell:2005kz}.  A non-vanishing neutrino magnetic moment would have important astrophysical and cosmological consequences~\cite{Heger:2008er}.

In the condensed matter setup, neutral fermionic excitations appear in strongly correlated electronic systems in Kondo lattice materials such as, for example, the newly found compound YbIr${}_3$Si${}_7$. Neutral fermions cannot, evidently, carry electric charge while they can support thermal energy transfer endowing the charge insulator with metallic thermal conducting properties. This type of exotic materials is sensitive to the background magnetic field implying that neutral fermions couple to the magnetic degrees of freedom~\cite{Kondo}. The corresponding electrically-neutral massless excitations are remote condensed-matter analogues of the exotic photons proposed in Ref.~\cite{MZHLett}.

Now we illustrate this theory by choosing a massless uncharged particle with magnetic moment and anyonic spin with exotic charges,
\beq
m=0, \quad
e=0, \quad
\mu\chi=1\,,
\quad
\kappa_{\ma}=1\,,
\quad
\kappa_{\ex}\neq0\,,
\label{exophoton}
\eeq
 put into the  electromagnetic field
\beq
\bE(\bx)=f(|\bx|)\,\hat{\bx} \aand
B(\bx)=\frac{|\bx|}{(1+|\bx|^2)^3}\,,
\label{ToyEB}
\eeq
where $f$ is some function of the radius. The electromagnetic fields are switched off by $e=0$ but there remains an effective Stern-Gerlach-type electric field $\bE^*=\mu\chi\bnabla B$ coming from \eqref{CarrAHam} [cf. \eqref{anymagmom}],
with $\mu\chi$ behaving as an effective charge and $B$  as an effective potential.
 $\kappa_{\ex}$ (assumed not to vanish) drops out and the general formula  \eqref{genHall**} reduces to,
\beq
\medbox{
({x}^i)^{\prime} = (\mu\chi)\,\epsilon^{ij}\frac{\p_jB}{\kappa_{\ma}}
\aand
{p}_i^{\prime}=0\,,
}
\label{e0m0k}
\eeq
which is indeed the formula we used on the horizon of a black hole  \cite{MZHLett}.

\begin{figure}[!ht]
\includegraphics[scale=0.5]{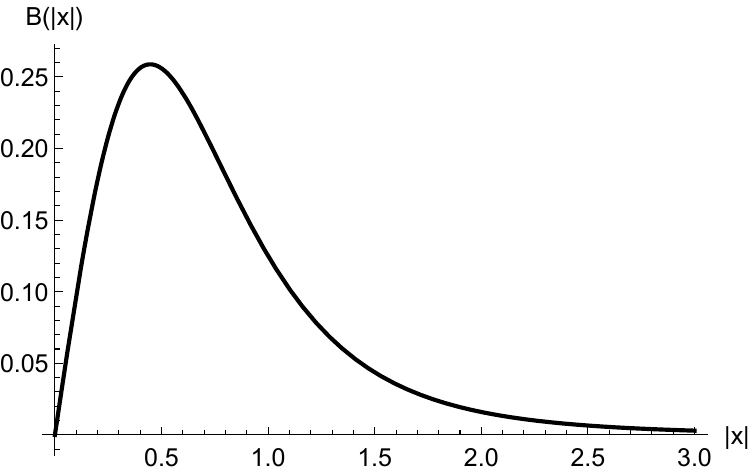}
\qquad\;
\includegraphics[scale=0.52]{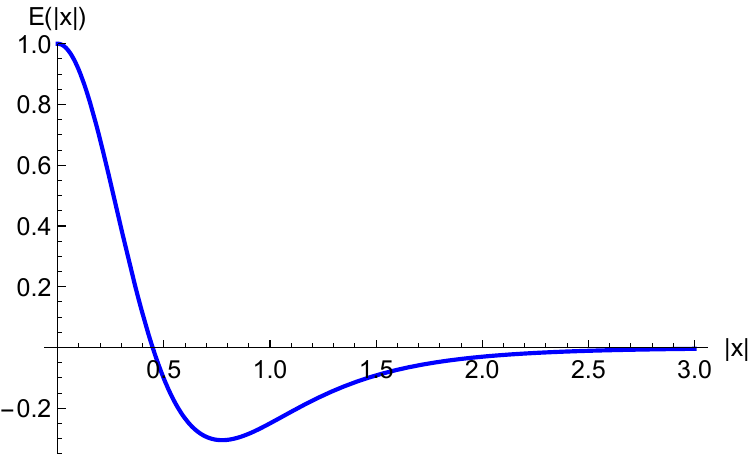}
\\
\hskip-4mm
(a)\hskip 74mm (b)\\
\vskip-3mm
\caption{\small (a) \textit{The planar magnetic field $B(|\bx|)$ in \eqref{ToyEB}} (b) \textit{induces a Stern-Gerlach type radial effective electric field \blue{$\bE^*=\bnabla B$}.}
\label{Toyfig1}
}
\end{figure}

\begin{figure}[!ht]\hskip-4mm
\includegraphics[scale=0.58]{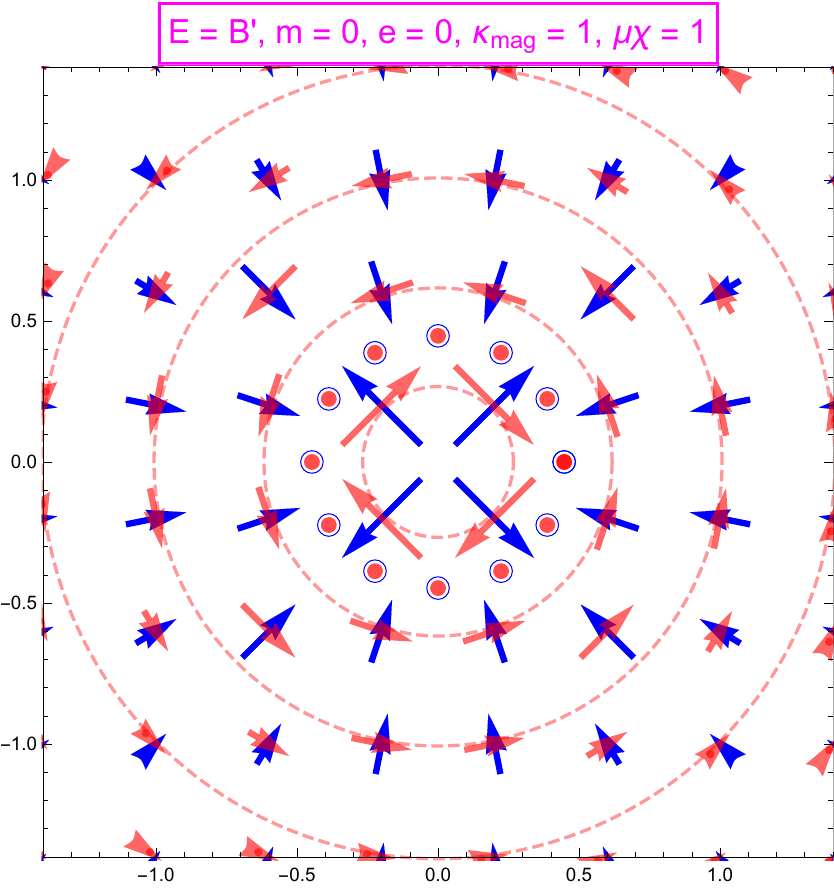}\\
\vskip-3mm
\caption{\small \textit{
For uncharged particles the radial planar magnetic field  $B$ in  \eqref{ToyEB} induces a radial
\blue{\bf effective electric field $\bE^*$}.
 An ``exotic photon'' i.e. a mass and chargeless planar Carroll anyon with nonzero spin $\chi$, magnetic moment $\mu$ and magnetic charge $\kappa_{\ma}$, \eqref{exophoton}, rotates around the origin along concentric circles, consistently with the \emph{generalized Hall Law}, \eqref{genHall**}. The rotation stops on a ring (marked by full \red{\bf dots} in \red{red}) where  the effective electric field vanishes, \blue{$\bE^*=0$}. The direction  of the rotation is reversed after \blue{$\bE^*$} changes sign.}
\label{Toyfige0k1}
}
\end{figure}

The various fields and  the Hall motion  \eqref{e0m0k} are shown in  \ref{Toyfig1}a-b. Fig.\ref{Toyfige0k1}  prefigures Fig.\ref{f:bh_drift}.

\begin{figure}[!ht]
\includegraphics[scale=0.5]{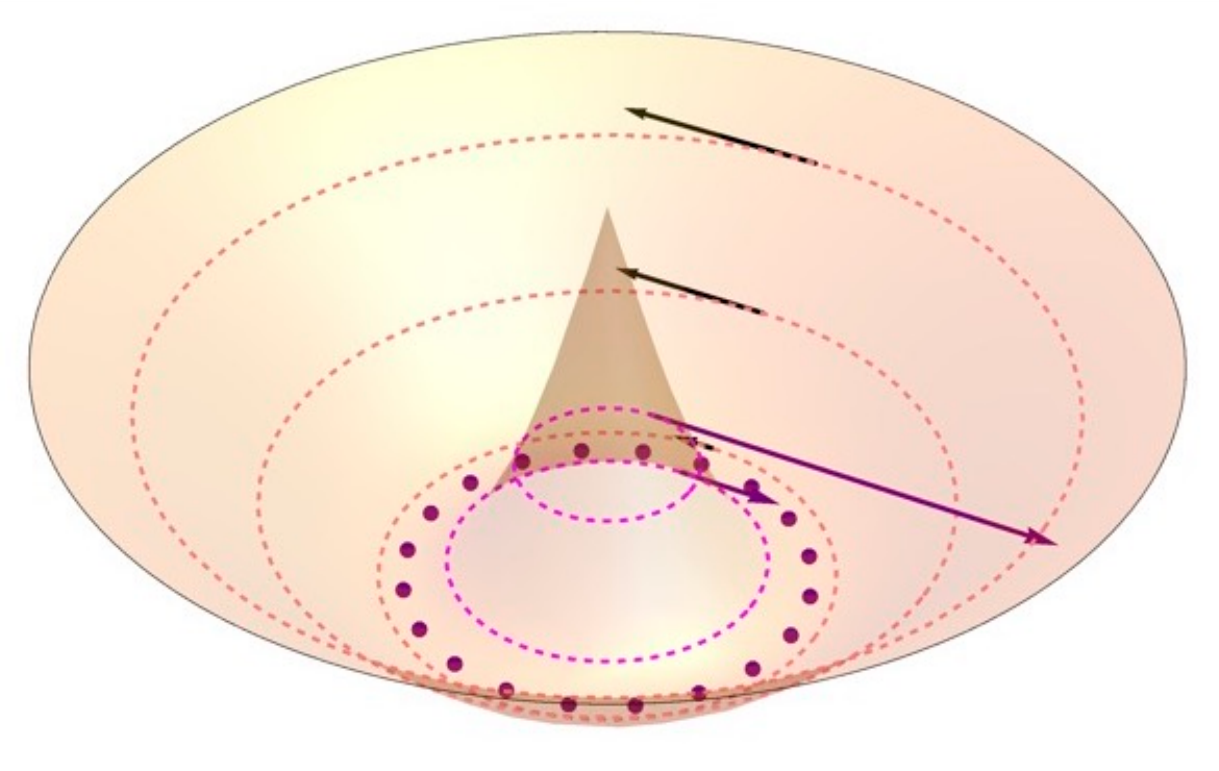}\vskip-5mm
\caption{\textit{\small The motion of an anyon in the effective electric field $\bE^*$ can be lifted to ``Mexican Hat" potential surface $-B(|\bx|)$ above the plane. The trajectories are horizontal ``cords'' (\magenta{\bf circles}) starting near the ``top'' with  high velocities and, then  decreasing  progressively until we arrive at $|\bx|=|\bx_0|$ where the force vanishes, $\bE^*(|\bx_0|)=0$, and the rotation stops. The ``cord" degenerates to \magenta{\bf fixed points} around the ``bottom rim''. Then, as we climb up  the outer ``flange'' by letting $|\bx|$ increase, the anyon restarts to move with ever increasing velocity --- however in the \magenta{\bf opposite direction}. The 3D picture projects to Fig.~\ref{Toyfige0k1}.}
\label{MexicanHat}
}
\end{figure}

An intuitive picture can be obtained by observing that
the motion of a Carroll anyon in the effective electric field $\bE^*$ can  can be pictured also as motion unfolded to
three dimensions and lifted to ``Mexican hat"  potential surface $-B(|\bx|)$ above the plane, as shown in Fig.~\ref{MexicanHat}.

For the sake of comparison, we consider also a \emph{charged} massless particle with \emph{no spin},
\beq
m=0, \quad
e=1, \quad
\chi=0\,,
\quad
\kappa_{\ma}=0\,,
\quad
\kappa_{\ex}\neq0\,,
\label{chargedphoton}
\eeq
where we put $\kappa_{\ma}$ to zero with no loss of generality
 \eqref{Bstar}.
Such quasi-particles might be relevant in Condensed Matter physics  \cite{Zoo}.

Putting both particles \eqref{exophoton} and \eqref{chargedphoton} into  the effective field considered in the uncharged case, \eqref{exophoton},
\beq
\bE(|\bx|) = \bnabla B = \frac{1-5|\bx|^2}{(1+|\bx|^2)^4}\,,\hat{\bx}\,
\label{choosenE}
\eeq
the general formula \eqref{genHall**}  reduces to\footnote{For our choice \eqref{choosenE} the r.h.s. of \eqref{spinlesstoy} is $\epsilon^{ij}\,\p_j\ln{B}$.},
\beq
({x}^i)^{\prime}
= \epsilon^{ij}
\,\frac{E_{j}}{B}\,,
\label{spinlesstoy}
\eeq
which means usual Hall motion \eqref{Halllaw}. The only difference with \eqref{e0m0k} is that $B$ in the denominator is now position-dependent instead of being a constant.

We conclude that an uncharged spinning anyon with magnetic moment moves essentially as a charged scalar particle. What \emph{is} surprising is that while the first  case \eqref{exophoton} involves a Stern-Gerlach dynamics, the second \eqref{chargedphoton} is of the spin-Hall type.
\beq
e \sim \mu \chi
\aand
\bE \sim \bnabla B\,.
\label{chargeSG}
\eeq
Plotting \eqref{spinlesstoy} yields a figure
 similar to fig.\ref{Toyfige0k1} and is not reproduced here therefore.

\begin{figure}[!ht]\hskip-4mm
\includegraphics[scale=0.6]{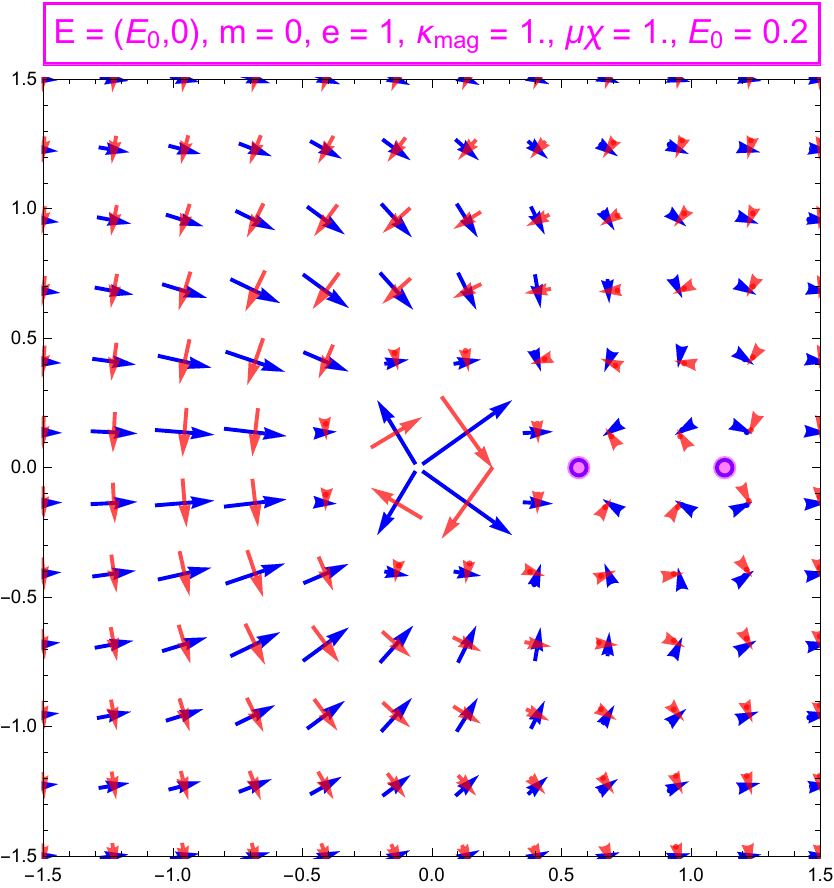} \\
\vskip-3mm
\caption{\small \textit{Combining a constant external electric field $\bE=\bE_0$ with  $\bnabla B$ induced by the spin-field term \eqref{anymagmom} yields \blue{\bf effective electric field} \blue{$\bE^*$} \eqref{Estar}. \red{\bf Velocity} and \blue{\bf force} are perpendicular,  as required by the Hall law. The number of equilibrium points has been considerably reduced.}
\label{Toye1Econst}
}
\end{figure}

The  magnetic and electric fields could plainly be chosen also independently, see \eqref{combCmag}. Pairing, for example, a constant electric field $\bE=\bE_0=\const$, e.g., $\bE_0=(E_0,0)$,
\beq
\bE^*= \bE_0 + \mu\chi \frac{1-5|\bx|^2}{(1+|\bx|^2)^4}\,\hat{\bx}
\label{EBmix}
\eeq
which with $B=B(|\bx|)$ in \eqref{ToyEB} yields  Fig.~\ref{Toye1Econst}.

\section{Contraction from relativistic anyons}\label{CarrPoinSec}

An insight into the physical origin of  exotic Galilean symmetry has been provided by its derivation from relativistic anyons \cite{anyons} by a tricky  contraction \cite{JNExospin,DHExoSpin}. Now we study the Carrollian counterpart \cite{Leblond}: one defines the timelike coordinate $x^0 = s/c$ and then sends $c \rightarrow \infty$. We start  with
 a massive and spinless particle,  described by the Cartan 1-form   \cite{SSD,DHExoSpin},
\begin{equation}
\varpi = p_\mu dx^\mu = p_i dx^i - m \sqrt{1 + \frac{p^2}{m^2 c^2}}\, ds\,.
\label{RelCartform}
\end{equation}
Taking the limit $c \rightarrow \infty$, we readily recover the Cartan 1-form \eqref{genCarrollCS} for Carroll,
\begin{equation}
\label{cartanCarr2}
\varpi_0 = p_i dx^i - m ds\,.
\end{equation}
Notice that the Carroll limit is in fact more straightforward than the Galilean one, for which the limit $c \rightarrow \infty$ of $m c^2 \sqrt{1 + \frac{p^2}{m^2 c^2}}\, dt$ requires  regularization, achieved in \cite{DHExoSpin} by adding a central extension i.e., working with the Bargmann model.

Now consider a massive relativistic model with spin. Generalizing the Jackiw-Nair ansatz $s=\kappa c^2$ \cite{JNExospin}, we set
$ s = \chi + \kappa c^2$\,,
 whose Cartan form is \eqref{cartanCarr2} plus a term describing the spin state of the particle,
 $\varpi = \varpi_0 + \varpi_s$.
The rather complicated  term $\varpi_s$ can be read off from
 \# (4.1) in \cite{DHExoSpin}. The
 result is seen not depend on the time coordinate. \emph{The Carrollian limit of $\varpi_s$ is therefore the same as in the Galilean case}, and we end up with the  Cartan form,
\begin{equation}
\varpi_{any}\equiv
\varpi = p_i dx^i - m ds + \kappa c^2 d\phi + \chi d\phi + \frac{\theta}{2} \epsilon^{ij} p_i dp_j\,.
\end{equation}

The term $\kappa c^2 d\phi$ here diverges when $c\to\infty$, however its exterior derivative drops out,
yielding the exotic part of the symplectic form \eqref{CESympHam} with  $\kappa_{\ex}=\kappa$,
\begin{equation}
\Omega_{\ex} =  dp_i  \wedge dx^i + \frac{\kappa}{2m^2} \epsilon^{ij} dp_i \wedge dp_j \,.
\end{equation}
 Thus the exotic Carroll charge $\kappa_{\ex}$ is deduced from the relativistic anyon model, just as in the Galilean case.
However  we \emph{do not recover (in this model) the second charge}, $\kappa_{\ma}$ by  contraction.

 Some more insight can be gained by regularizing $\varpi$.
We follow  \cite{DHExoSpin}, whose idea is to augment the Poincar\'e group by adding a trivial parameter, which naturally splits into two contributions when computing the limit $c \rightarrow \infty$: one proportional to $c^2$ and thus diverging, which will exactly cancel out the diverging term of $\varpi$, and the next order term, in $c^0$, which will not diverge, and becomes a new physical parameter in the resulting model. Consider a \emph{trivial} $\IR$-central extension to the Poincar\'e group. Its Cartan 1-form will be $\widetilde{\varpi} = \varpi + {\alpha} d\theta$. Now, we choose ${\alpha} = - \kappa c^2$ \footnote{${\alpha} d\theta$ is exact when ${\alpha}$ is either a constant or depends only on  $\theta$.} and $\theta = \phi - w /c^2$ for some $w \in \IR$. This implies that ${\alpha} d\theta = - \kappa c^2 d\phi + \kappa dw$, and thus,
\begin{equation}
\widetilde{\varpi} =  p_i dx^i - m ds + \chi d\phi + \frac{\kappa}{2m^2} \epsilon^{ij} p_i dp_j + \kappa dw\,.
\end{equation}

This expression is to be compared to $\varpi_{\ex}$ in \eqref{exoCarrollCS}.
The presymplectic structure has gained one dimension, spanned by the parameter $w$.
Such a situation arises in the ``non-exotic'' Galilean case, where the infinite energy is regularized similarly  \cite{BGGK,DHExoSpin}.

In conclusion, the exotic term is recovered (up to a total derivative), but the  magnetic term is missing, as it does also from the presymplectic 2-form $d\varpi$, cf. \eqref{exoCarrollCS}.
\goodbreak

\section{Motion in a plane gravitational wave
}\label{GWmotion}

Our first  application to general relativity is to motion in an exact plane gravitational wave \cite{Sou73,GenMemory,Carroll4GW}. In particular, we want
 to clarify the superficial contradiction between the claim that \emph{``particles with Carroll symmetry do not move''} and those  complicated \emph{motions} found in a plane gravitational wave \cite{GenMemory,Carroll4GW,DGHMemory,Soft}, --- which happens to carry a ``sort of'' Carroll symmetry.
The clue is precisely: ``sort of'', as we now explain. We start with an exact gravitational wave written in  Baldwin-Jeffery-Rosen (BJR) \cite{BaJeRo,Sou73} coordinates
$(\bx,t,s)$, in terms of which the  metric is
\begin{equation}
ds^{2}=a_{ij}(t)dx^{i}dx^{j}+2dtds\,,
\label{bjrmetric}
\end{equation}%
where $a(t)=\big(a_{ij}(t)\big)$ is a symmetric positive $2\times2$ matrix, which depends only on the light-cone coordinate $t$ \cite{Sou73,Carroll4GW,DGHMemory,Soft}.
Particular interest will be devoted to the $2+1$ dimensional null submanifolds obtained by fixing the light-cone coordinate $t$,
\beq
\cC_{t_0}=\Big\{\big(\bx,s\big),\; t=t_0 =\const\Big\}\,.
\label{Cell}
\eeq
Restriction  to $\cC_{t_0}$ yields the metric
$ds^2=a_{ij}(t_0)dx^{i}dx^{j}
$
such that
$
\xi = \p_s\,
$
is in its kernel.
Then the triple $(\cC_{t_0}, a_{ij}(t_0), \xi)$  provides us, for each fixed value of $t_0$, with a $2+1$ dimensional Carroll manifold we will call a \emph{Carroll slice} embedded into 4D  \GW spacetime \cite{Carroll4GW,Morand}.

The isometry group of a \GW has long been known to have (generically) 5-parameters  \cite{BoPiRo}, implemented by shifting $\bx$ and $s$  but leaving $t$ (and thus $\cC_{t_0}$)  invariant \cite{Sou73,Carroll4GW},
\begin{subequations}
\begin{align}
\label{xCarrol}
\bx&\to \bx+H(t)\,\bb_1+\bc,
\\
s&\to s-\bb_1\cdot\bx - \2\bb_1\cdot{}H(t)\,\bb_1+b_0,
\label{vCarrol}
\end{align}
\label{genCarr}
\end{subequations}
\!where the  planar vectors  $\bb_1$ and $\bc$, and the real number $b_0$ are constants \cite{Sou73,Carroll4GW}.
The key ingredient here is the $2\times2$ \emph{Souriau matrix}  $H(t)$, obtained by integrating the inverse of the BJR-profile-matrix $a=\big(a_{ij}\big)$ \footnote{ \eqref{Hmatrix} obviously assumes that $a$ is a regular matrix. However this remains true only in a finite time interval \cite{Sou73}: $\det(a)$ necessarily vanishes after a finite $t$-lapse \cite{Sou73,Carroll4GW}. Thus the BJR description is only local; our investigations are limited to a finite time interval.
},
\beq
H(t)
 =\int^t_{0}\!\!a(u)^{-1} du\,.
\label{Hmatrix}
\eeq
 For $t_0=0$ the Souriau matrix is zero, and (for $\bc=b_0=0$) the action \eqref{genCarr}
  reduces, for {arbitrary} BJR profile $(a_{ij})$, to
the Carroll boost \eqref{Cboost}: the transverse coordinate $\bx$ is left invariant and it is only Carrollian time, $s$, which is boosted. For $t_0\neq0$ the implementation is determined by the Souriau matrix $H(t_0)$  \eqref{Hmatrix}, cf. \eqref{genCarr}.

\begin{figure} [ht]
\hskip5mm
\includegraphics[scale=.4]{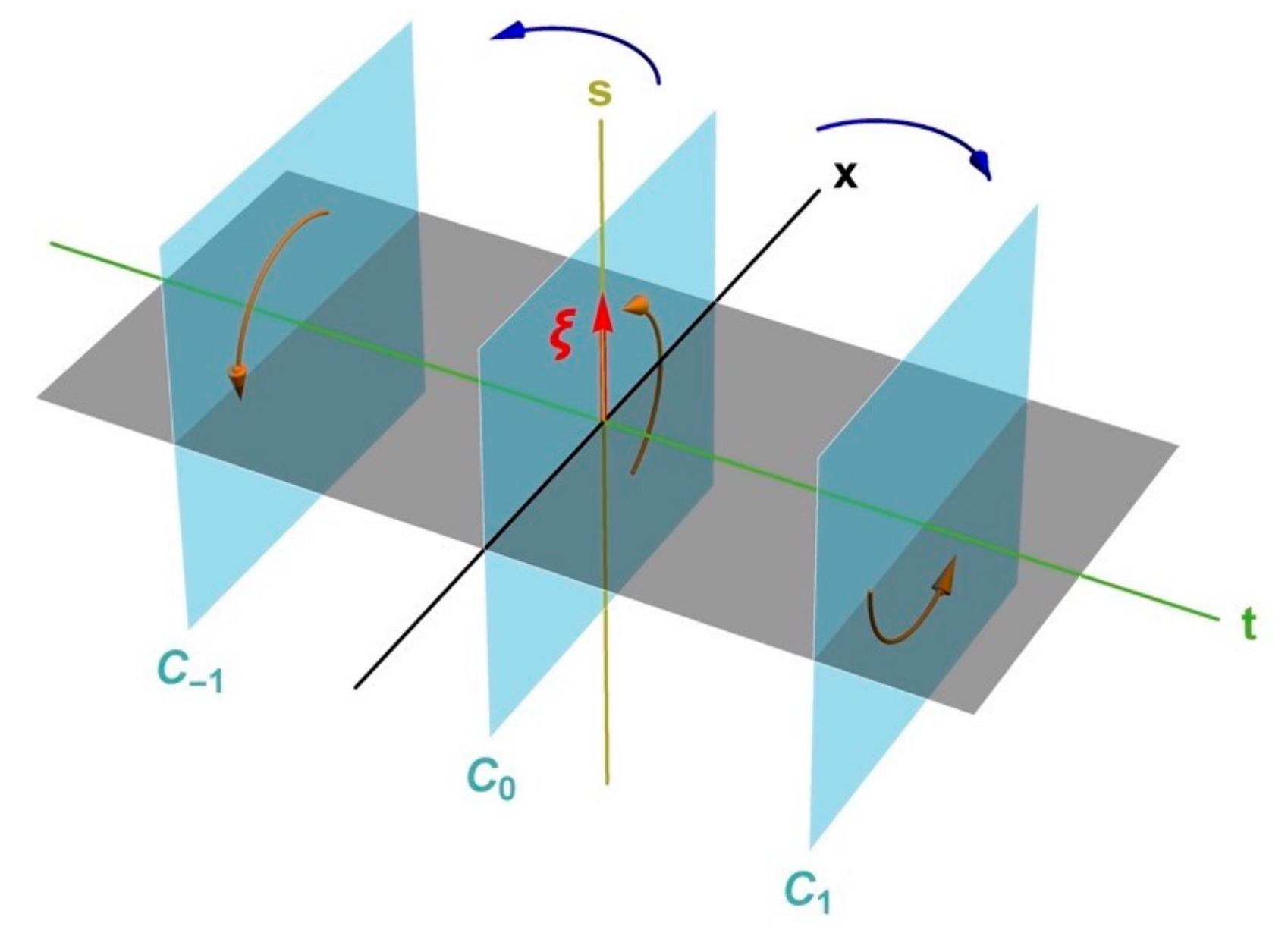}\vskip-6mm
\caption{\small \textit{The (broken) Carroll group \eqref{genCarr} leaves each ``Carroll slice'' $\cC_{t}$ invariant. The implementation is obtained from the broken Carroll one  on $\cC_{0}$ through the Souriau matrix $H(t)$  \eqref{Hmatrix} as indicated in \blue{\bf blue}. The 5 parameter isometry group of the \GW acts (indicated in \orange{\bf orange}) transitively on each slice $\cC_{t}$, but not on the full spacetime.}
\label{CarPlanefigs}
}
\end{figure}

The restriction of the non-transitive 5 parameter isometry group of the \GW to each fixed Carroll manifold \ICl  acts upon  the latter transitively  in a $t_0$-dependent way, as shown in fig.\ref{CarPlanefigs} \footnote{Our illustration uses the
the BJR profile
$ a = \cos^2t\,\diag \big(e^{2t},e^{-2t}\big)$
studied in \cite{DGHMemory}.}.
Conversely,  the plane \GWs  carries a Carroll-type symmetry obtained by ``gluing together'' all those $t=\const$ implementations (but dropping rotations).

\begin{figure} 
\includegraphics[scale=.24]{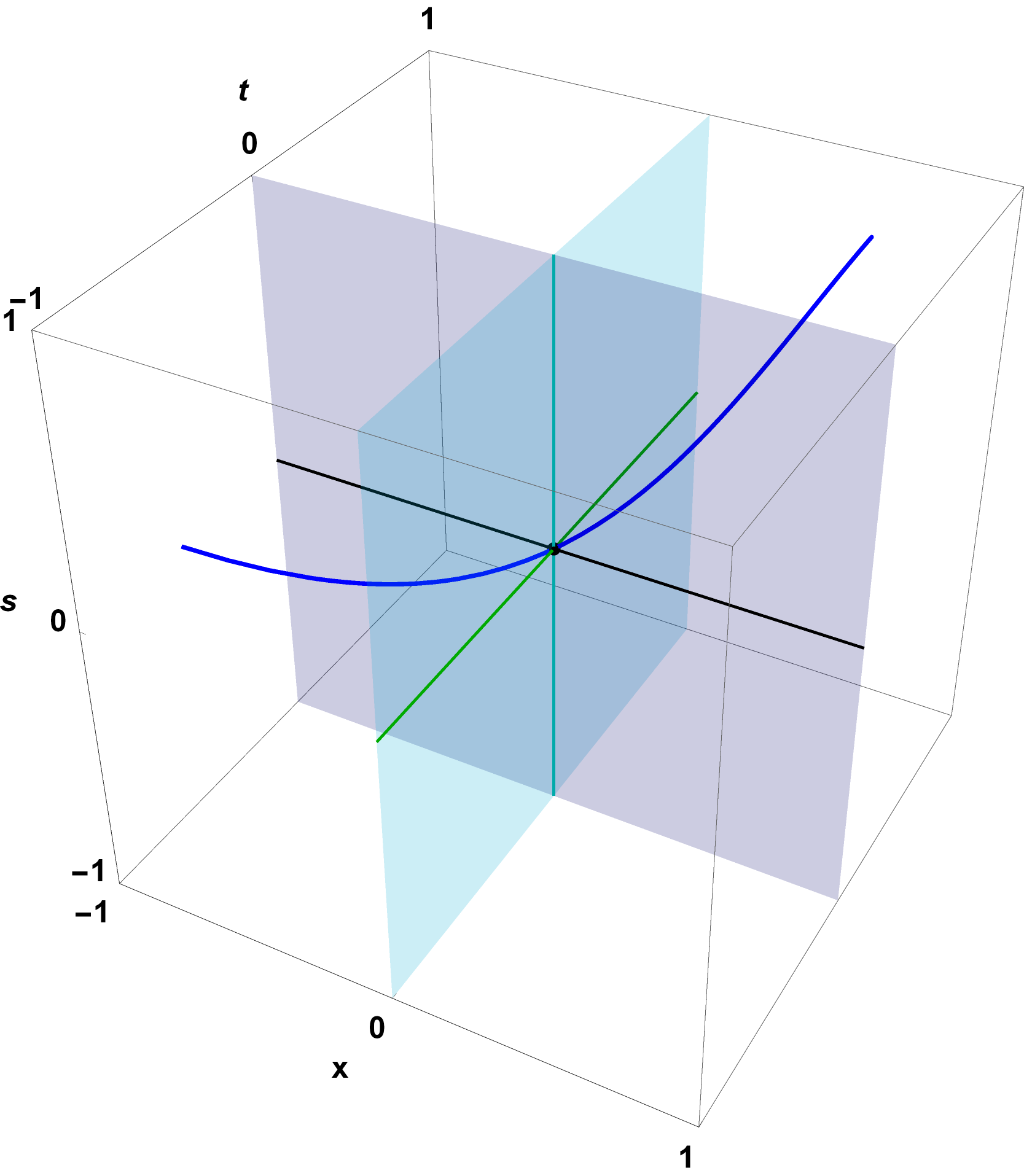}
\qquad
\includegraphics[scale=.24]{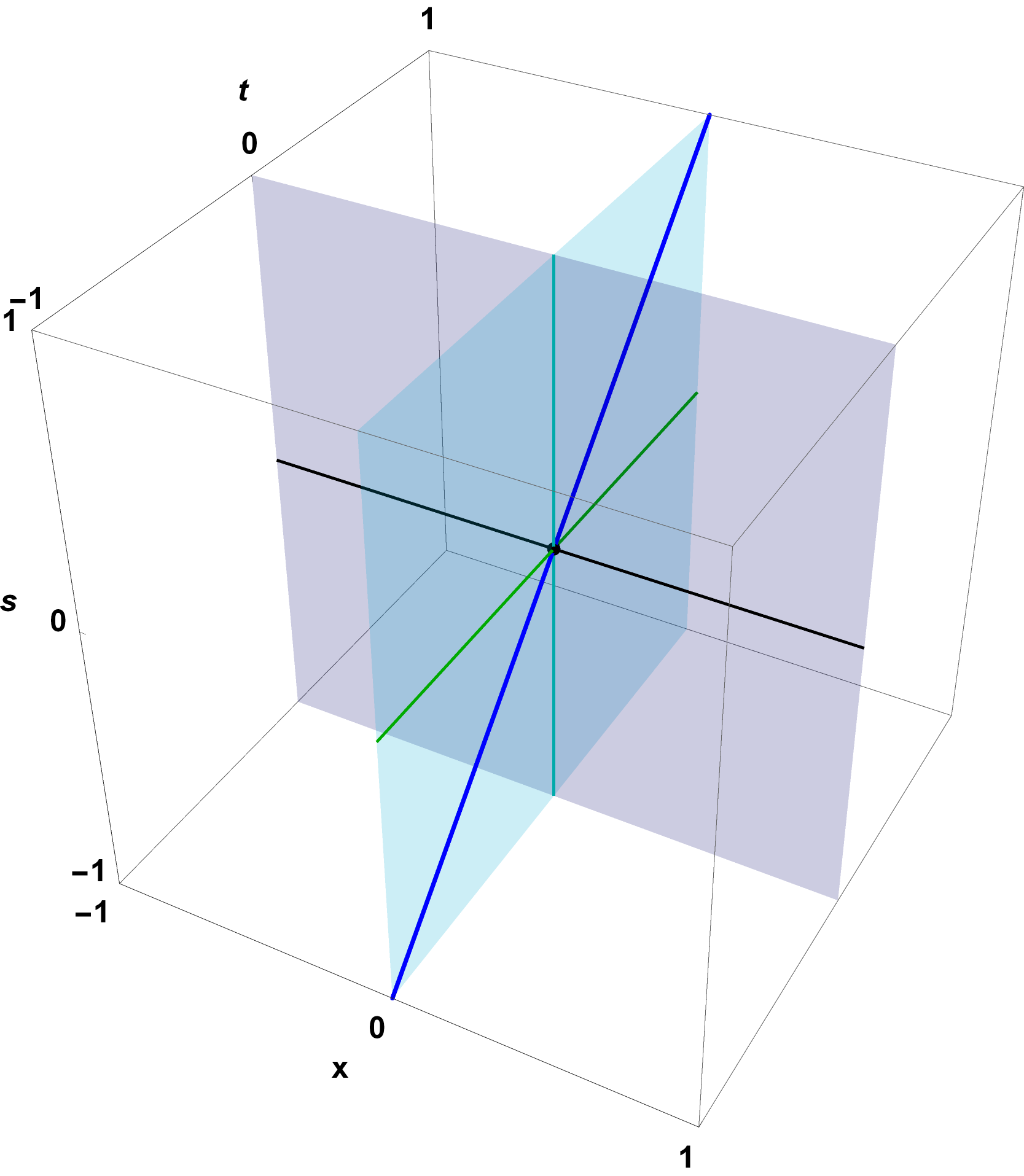}
\\
\hskip-11mm
(a)\hskip 72mm (b)\\
\vskip-3mm
\caption{ (a) {\small \textit{For non-vanishing initial transverse momentum, $\bp\neq0$, a massive particle moves along a timelike geodesic.}}
(b) {\small \textit{For $\bp=0$ instead, the geodesics move, for any profile, only in time, $t$, and have fixed position,
$\bx = \bx_0$.}}
\\
\label{GeoGWfigs}
}
\end{figure}

Now we turn to the geodesics.
If $\lambda$ is any affine parameter then,
\beq
\epsilon=\half\rg_{\mu\nu}\,\dot{x}^\mu\dot{x}^\nu\,,
\label{edef}
\eeq
where $\dot{\{\,\cdot\,\}}= d/d\lambda$,
 is a constant of the motion along any geodesic, is interpreted as the \emph{relativistic mass square}. It is negative/zero/positive for timelike/null/spacelike geodesics.
 The rather complicated-looking general geodesic equations are,
\besub
\begin{align}
&\frac{dx^{2}}{d\lambda^{2}}+\left(\frac{a_{11}^{\prime}a_{22}-a_{12}a_{12}^{\prime}}{\det(a)}\right)\frac{dx}{%
d\lambda}\frac{dt}{d\lambda}+\left(\frac{a_{12}^{\prime}a_{22}-a_{12}a_{22}^{\prime}}{\det(a)}\right)\frac{dy}{%
d\lambda}\frac{dt}{d\lambda}=0\,,
\label{xgeoeq}
\\[8pt]
&\frac{dy^{2}}{d\lambda^{2}}-\left(\frac{a_{11}^{\prime}a_{12}-a_{11}a_{12}^{\prime}}{\det(a)}\right) \frac{dx}{%
d\lambda}\frac{dt}{d\lambda}-\left(\frac{a_{12}^{\prime}a_{12}-a_{11}a_{22}^{\prime}}{\det(a)}\right)\frac{dy}{%
d\lambda}\frac{dt}{d\lambda}=0\,,
\label{ygeoeq}
\\[6pt]
&\frac{d^{2}t}{d\lambda^{2}}=0,
\label{ugeoeq}
\\[6pt]
&\frac{d^{2}s}{d\lambda^{2}}-\frac{1}{2}a_{11}^{\prime}\left(\frac{dx}{%
d\lambda}\right)^{2}-\frac{1}{2}a_{22}^{\prime}\left(\frac{dy}{d\lambda}%
\right)^{2}-a_{12}^{\prime}\frac{dx}{d\lambda}\frac{dy}{d\lambda}=0\,.
\label{vgeoeq}
\end{align}%
\label{lambdaBJR}
\esub
\!\!where $\det(a)= a_{11}a_{22}-a_{12}^{2}$ and the prime means  $d/dt$.
 For a general profile $a_{ij}(t)$ these equations can only be solved numerically. The solution (for the same BJR profile $
a = \cos^2t\,\diag \big(e^{2t},e^{-2t}\big)
$ as in fig.\ref{CarPlanefigs}) is depicted in  fig. \ref{GeoGWfigs}a.
%
Analytic solutions arise in particular cases.
When $t\neq\const$  then it can be chosen itself as affine parameter, $\lambda=t$.
For {arbitrary profile} we get,  trivially \cite{DGHMemory}:

\begin{thm}
 \textit{\small For vanishing initial transverse velocity,
 \beq
 \bx^{\prime}(t_0) =0\Rarrow
 \bx(t) \equiv \bx_0 = \const
\label{lnomotion}
\eeq
 the transverse position $\bx$ does not move (while $s$ is a  linear function of $t$), shown in fig.\ref{GeoGWfigs}b.}
\label{nomotinGW}
\end{thm}
\benu
\item
Timelike [$t\neq\const$]  solutions could be found also by using the \emph{symmetries} \cite{DGHMemory}:
the Noether theorem provides us with 5 conserved quantities,
\beq
\bpi = a\,{\bx}^{\prime},
\qquad
\bd=\bx-H(t)\,\bpi,
\qquad
m=1\,,
\label{CarCons}
\eeq
interpreted as \emph{linear momentum, boost-momentum;
$m=1$ in our para\-metri\-zation}\footnote{
The 4d \GW as the Bargmann manifold  of a $2+1$ dimensional non-relativistic system with
$m$  the ``non-relativistic mass'', which is {\rm a priori} independent of the relativistic mass  $\epsilon$ in \eqref{edef}. }.
 Conversely, our timelike geodesic is determined by these conserved quantities,
\beq
\bx(t)=H(t)\,\bpi+\bd,
\qquad
s(t)=-\half \bpi\cdot H(t)\,\bpi + e\,t+ d,
\label{CarGeo}
\eeq
where  $d$ is an integration constant, as is confirmed
by a straightforward calculation.
This reduces the problem to calculating the Souriau matrix $H(t)$ in \eqref{Hmatrix}, which can be achieved by numerical integration, instead of solving the coupled equations \eqref{lambdaBJR}.

The trivial ``no-motion'' solution  \eqref{lnomotion} is recovered from \eqref{CarGeo} since the chosen initial condition  $\bx^{\prime}=0$ in \eqref{lnomotion} implies (Theorem \eqref{nomotinGW}), by \eqref{CarCons}, that the conserved transverse momentum vanishes \footnote{All those complicated-looking trajectories plotted in \cite{DGHMemory,Soft} come  from the simple ``no-motion'' solution \eqref{dotx0} above by switching from BJR to Brinkmann coordinates \cite{Brinkmann,Gibbons75}. },
\beq
\bpi = 0\,.
\label{p0cond}
\eeq
Thus the boost momentum \eqref{CarCons} becomes, for any profile $a_{ij}$, simply
\beq
\bd = \bx
\label{Cboost4GW}
\eeq
cf. \eqref{mdipolemom} with $m=1$, whose conservation implies, once again, \emph{immobility}, as explained in Corollary \ref{nomotCoroll}.

\item
Let us now turn to a $t=t_{0}=\const$ Carroll slice. \ICl is endowed with the singular metric
$
\tilde{g}= a_{ij}({t_{0}})dx^idx^j
$
whose kernel is generated by $\p_s$, conforming that  it is indeed a Carroll manifold. Carroll boosts, \eqref{Cboost}, act on it as (trivial) isometries.
The choice ${t_{0}}=0$ can and will henceforth be  assumed with no loss of generality, and the shorthand   $\cC\equiv $ \ICl  will be used.

``Vertical'' geodesics  \ie s.t. are constrained to the  Carroll slice $\cC$  can also be considered.  Our first and important observation is that (discarding tachyons for which $\epsilon >0$) they are necessarily \emph{lightlike}. The clue is that  fixing $t$ the relativistic mass \eqref{edef} reduces to
\beq
\epsilon=a_{ij}(0)\dot{x}^i\dot{x}^j \geq 0
\label{negmass}
\eeq
which,  $(a_{ij}(0))$ being a positive matrix,  \emph{rules out a mass} for which $\epsilon < 0$. Thus we have
\beq
\epsilon=\half\rg_{\mu\nu}\,\dot{x}^\mu\dot{x}^\nu=a_{ij}\dot{x}^i\dot{x}^j=0\,
\label{nullgeo}
\eeq
for any geodesic on $\cC$. Moreover,
the transverse velocity vanishes,
\beq
\dot{x}^i=0 \Rarrow \bx =\bx_0=\const
\label{dotx0}
\eeq
cf. \eqref{lnomotion}. Then
 putting $dt/d\lambda=0$ into \eqref{lambdaBJR} reduces the latter to
\besub
\begin{align}
&\frac{d^{2}\bx}{d\lambda^{2}}
=0\,,
\label{xfix}
\\[4pt]
&\frac{d^{2}s}{d\lambda ^{2}}-\frac{1}{2}a_{11}^{\prime}\left(\frac{dx}{%
d\lambda}\right)^{2}-\frac{1}{2}a_{22}^{\prime}\left(\frac{dy}{d\lambda}%
\right)^{2}-a_{12}^{\prime}\frac{dx}{d\lambda}\frac{dy}{d\lambda}=0\,.
\label{vCarroll}
\end{align}%
\label{xvCarrtraj}
\esub
Eqn \eqref{xfix} is automatically satified by \eqref{dotx0}  leaving us with,
\beq
s=s_{0}+ c\,\lambda + d\,,
\label{b10}
\eeq
where $s_0, \, c$ and $d$ are real constants.

One can wonder what happens to a massive particle put on $\cC$? The answer comes from \eqref{negmass}: having non-zero mass is \emph{inconsistent} with staying on $\cC$: the particle will be expelled, as shown in fig.\ref{GeoGWfigs}a. In conclusion,

\begin{thm}
\textit{\small
A particle can stay on a Carroll slice $\cC$ only if it is massless. Then its transverse position $\bx$  is fixed; its trajectory is a ``vertical'' straight line directed along the ``Carroll time axis'', $s$.}
\label{ThmVIII.1}
\end{thm}

\medskip
In the massive case, immobility could also be derived from the conservation of the Carroll moment ${\bd}=m\bx$
\eqref{mCboost}, eqn. \eqref{Cboost4GW} for \GWs. However, massless Carroll particles have \emph{vanishing} C-boost momentum, $\bd=0$ in \eqref{d0form0}, from which \emph{no conclusion can be drawn}. As we have just seen,   the ``no motion'' paradigma still holds on Carroll slices for a massless particle.
Another example is provided by motion on the horizon of a black hole, to be presented in sec. \ref{BHh}.
\eenu

\section{Hall on the Hole: motion on the Horizon}\label{BHh}

Another  physical application of (possibly doubly-extended) Carroll dynamics is to motion on the \emph{horizon of a black hole} \ \cite{Duval2015,MBHhorizon,DonnayM,hydro,Freidel22}.
We first recall the following general properties:

\begin{itemize}

\item
A black hole is characterized by its mass $M$, angular momentum $J$ and charge $Q$:

\item
The horizon $\cH$ is a Carroll manifold \cite{Duval2015,MBHhorizon,DonnayM};

\item
The horizon is a null hypersurface, therefore the geodesics constrained to it are necessarily null \cite{Penrose67},
\beq
\epsilon=\half\rg_{\mu\nu}\,\dot{x}^\mu\dot{x}^\nu=0\,,
\label{masslesonH}
\eeq
see chapters 34.3 and 34.4 in \cite{MTW}.

\item
The geodesic dynamics can  be extended
by adding both exotic \& spin-field (commonly called spin-orbit) terms discussed in sec.\ref{ExoCarrSec}.

\end{itemize}

This section is devoted to the study of an exotic photon  \eqref{exophoton} whose key dynamical ingredients are, by \eqref{anyCSm0},
\besub
\begin{align}
\varpi_{spec} = &\half
{\kappa_{\ex}}\,\epsilon_{ij}v^i dv^j
 +
\half\kappa_{\ma}\,\epsilon_{ij}
x^i dx^j
 +  \mu\chi\,Bdx^i ds\,,
\label{specCartan}
\\
\sigma_{spec} = &\half
{\kappa_{\ex}}\,\epsilon_{ij}dv^i\wedge dv^j
 +
\half\kappa_{\ma}\,\epsilon_{ij}dx^i\wedge dx^j
 +  \mu\chi\p_iBdx^i \wedge ds\,,
\label{specSform}
\\
\cL_{spec} = &\half
{\kappa_{\ex}}\,\epsilon_{ij}v^i \frac{dv^j}{ds}
 +
\half\kappa_{\ma}\,\epsilon_{ij}
x^i \frac{dx^j}{ds}
 +  \mu\chi\,B
\,.
\label{specLag}
\end{align}
\label{specforms}
\esub\vspace{-6mm}
\subsection{Schwarzschild Horizons}\label{Schwarzschild}

We first study the simplest black hole, namely the static and spherically symmetric Schwarzschild metric, described by  using (ingoing) Eddington-Finkelstein coordinates $\big\{v,r,\vartheta,\varphi\big\}$  and natural units with the line element
\begin{equation}
\label{metric_sch}
g \equiv
g_{\mu\nu}dx^\mu dx^\nu = -\left(1 - \frac{2M}{r}\right) dv^2 - 2 dvdr + r^2 (d\vartheta^2 + \sin^2\vartheta d\varphi^2) \, ,
\end{equation}
where $M$ is the mass of the Schwarzschild black hole. The Schwarzschild horizon, $\cH$, is the hypersurface $r = 2M$, \ie,
\beq
\cH = \Big\{r = 2M,\,\vartheta,\varphi,v\Big\} \cong \IS^2 \times \IR\,.
\label{SchHor}
\eeq
The $dv^2$ component of the metric \eqref{metric_sch} induced on $\cH$  vanishes,  whereas the radial component
 disappears on the fixed-$r$ hypersurface $\cH$,
  leaving us with the clearly degenerate ``metric'',
\begin{equation}
\label{metric_sch_deg}
\widetilde{g} = g|_{r = 2M} = 0 \cdot dv^2 + 4M^2(d\vartheta^2 + \sin^2 \vartheta d\varphi^2)\,.	
\end{equation}

On the other hand,
\begin{equation}
\widetilde{\xi} = \partial_{v} \equiv \partial_s
\label{xivector}
\end{equation}
is a nowhere vanishing vector which belongs to the kernel of $\widetilde{g}$ where, to anticipate the role of Carrollian time that it will play, we renamed $v$ to $s$.
 We conclude that the Schwarzschild horizon $(\IS^2 \times \IR, \widetilde{g}, \widetilde{\xi})$  has a Carroll  structure \cite{MBHhorizon}\footnote{Any null hypersurface in a Lorentzian spacetime  is Carroll  \cite{BMSCarr,ConfCarr, Morand,CiambelliNull}.}, as defined in sec. \ref{genCarrollSec}.

The vector $\widetilde{\xi}$ (called the generator of the horizon) is also a Killing vector field for the Carroll metric in that
 $L_{\widetilde{\xi}}\, \widetilde{g} = 0$. It is in fact a Killing vector field of the original  Schwarzschild spacetime, $L_{\widetilde{\xi}}\, g = 0$, which becomes a null vector on the horizon,
\beq
g(\widetilde{\xi}, \widetilde{\xi})|_{r = 2M} = 0\,.
\label{xinull}
\eeq

The Schwarzschild horizon provides us with an insight into the peculiarities of Carrollian dynamics.
One of them is that the horizon of (stationary) black holes can ``trap'' photons \ie, a photon may remain on the horizon forever. As a stationary  horizon is always a null hypersurface, and since photons travel in null directions, a photon could move along the null direction defined by the ($\IR$ component of the) horizon, thus staying on it forever. In the Schwarzschild  case for instance,  this amounts to emit a photon right on the horizon and radially outward, see Fig. \ref{f:eddington}. For details, see \textit{e.g.} \cite[\S 33.6]{MTW}.

\tikzmath{\lc = 0.45;} 
\tikzset{middlearrow/.style={
    decoration={markings,
      mark= at position #1 with {\arrow{{>}[scale=1.2]}} ,
    },
    postaction={decorate}
  }
}
{\Large
\begin{figure}[ht]
\begin{tikzpicture}[line width=1pt,scale=.8
 , every node/.style={transform shape}]
  \draw [line width=1.5pt] (0,0) coordinate (bottomleft) -- node[pos=0.45,rotate=90,yshift=0.4cm]{$r=0$} (0,9) coordinate (topleft);
  \draw [line width=0.4pt] (topleft) -- (9,9) coordinate (topright) -- (9,0) coordinate (bottomright) -- (bottomleft);
  \draw [line width=1.5pt] (2.5,0) coordinate (bottomh) -- node[pos=0.45,rotate=90,yshift=0.4cm]{$r=2M$} (2.5,9) coordinate (toph);

  \draw [black!50!blue](7,6) arc (170:10:{1*\lc} and {0.2*\lc}) coordinate[pos=0] (a);
  \draw [black!50!blue](a) arc (-170:-10:{1*\lc} and {0.2*\lc}) coordinate (b);
  \draw [black!50!blue](a) -- ([yshift={2cm*\lc}]$(b)$) coordinate (c);
  \draw [black!50!blue](b) -- ([yshift={2cm*\lc}]$(a)$) coordinate (d);
  \draw [black!50!blue](d) arc (170:10:{1*\lc} and {0.2*\lc});
  \draw [black!50!blue](d) arc (-170:-10:{1*\lc} and {0.2*\lc});

  \draw [black!50!blue] ([yshift=6cm]bottomh) arc (170:10:{0.5*\lc} and {0.2*\lc}) coordinate[pos=0] (a2);
  \draw [black!50!blue](a2) arc (-170:-10:{0.5*\lc} and {0.2*\lc}) coordinate (b2);
  \draw [black!50!blue](a2) -- ([yshift={2cm*\lc}]$(a2)$) coordinate (c2);
  \draw [black!50!blue](b2) -- ([xshift={-1cm*\lc}]$(c2)$) coordinate (d2);
  \draw [black!50!blue](d2) arc (170:10:{0.5*\lc} and {0.2*\lc});
  \draw [black!50!blue](d2) arc (-170:-10:{0.5*\lc} and {0.2*\lc});

  \draw [black!50!green,middlearrow={0.5}] (6,0) to[out=45,in=-135] (9,3);
  \draw [black!50!green,middlearrow={0.5}] ([xshift=0.5cm]bottomh) to[out=75,in=-135] (a) -- ([xshift=2cm,yshift=2cm]a);
  \draw [black!50!green,middlearrow={0.63}] ([xshift=0.1cm]bottomh) to[out=90,in=-115] (4,9);
  \draw [black!50!green,middlearrow={0.63}] ([xshift=-0.1cm]bottomh) to[out=90,in=-65] (1,9);
  \draw [black!50!green,middlearrow={0.5}] ([xshift=-0.5cm]bottomh) to[out=108,in=-45] (0,3);
\end{tikzpicture}
\caption{\textit{\small Eddington-Finkelstein diagram \cite{MTW} of a Schwarzschild spacetime, with outgoing light rays \dgreen{$u = \const$} in \dgreen{\bf green},
 highlighting key features such as the singularity, the horizon, and  different trajectories of outgoing photons.
Infinitely far away from the black hole, space is Minkowskian. However, the closer we get, the more the gravitational field bends the light cones inward, resulting in outgoing photons taking an ever longer time to escape from the black hole region. Past the horizon, the light cone is completely bent inward such that the only future-pointing directions are toward the singularity.
An outgoing photon emitted right on the horizon would have a vertical trajectory, thus staying on the horizon.}}
\label{f:eddington}
\end{figure}
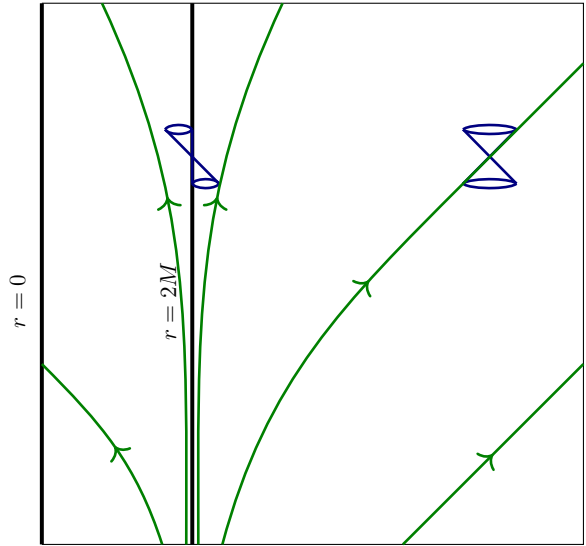
}

``Photon trapping'' is a nice thought experiment to understand Carroll geometries.
Consider a photon  on a Schwarzschild horizon. We can look at its motion (or lack of motion) intrinsically, on the horizon itself. On this surface, the photon is created at some point $(\bx_0,s_0)$ in (Carroll) time (represented by the $\IR$-axis generated by the null vector $\xi = \partial_s$, and $\bx_0 \in \IS^2$). In order to stay on the horizon, the photon must have its momentum directed along the generator $\xi$ of the horizon. Its momentum is then zero in the angular directions, and therefore the photon ``stays in place'' on $\IS^2$. Its ``trajectory'' is precisely the Carrollian ``no-motion", $\bx(s) = \bx_0$.

The situation is illustrated by the Eddington-Finkelstein diagram on Fig.~\ref{f:eddington}, cf. \cite{MTW}.

While this intuitive picture on the Schwarzschild horizon seems to imply no motion on this instance of a Carroll structure, the horizon is $2+1$ dimensional and thus \emph{might} accomodate potentially non-trivial dynamics associated with the double central extension, discussed in sec.\ref{ExoCarrSec}.
 However, it has been shown in \cite{Marsot21} and recalled in section \ref{exoCcoup} that the two charges $\kappa_{\ex}$ and $\kappa_{\ma}$ only couple to the electromagnetic field but \emph{not} the gravitational field. But the Schwarzschild spacetime carries \emph{no} electromagnetic field. Hence,
\bigskip
\goodbreak

\begin{thm}
\textit{An uncharged massless (exotic or not) particle with purely outgoing momentum on the horizon of a Schwarzschild black hole stays fixed, but does not move.}
\label{IX.1}
\end{thm}

This conclusion is valid also for spin-field interaction \eqref{Emotionm0} since $B=0$.

\subsection{Motion on the Kerr-Newman horizon}\label{KNhorizon}

Eqn. \eqref{genHall**} suggests that  non-trivial motion {may} be possible on the horizon of a black hole for a massless particle with magnetic moment $\mu\neq0$,
 non-zero anyonic spin $\chi\neq0$ and
non-zero magnetic charge $\kappa_{\ma}\neq0$,
when the black hole has a non-uniform magnetic field $B\neq0$ on the horizon.
\goodbreak

Below we show that  these conditions are met for a \emph{Kerr-Newman black hole}, described by its mass $M$, angular momentum $J$, and  charge $Q$.
In terms of Eddington-like coordinates $(v,r,\vartheta,\varphi)$ \cite{Newman65}, a Kerr-Newman  metric  can be written in the form
\footnote{The metric and its inverse are regular at $\Delta = 0$, as the $dr^2$ terms cancel each other out.},
\besub
\begin{align}
\label{kn_g}
g = & - \frac{\Delta}{\Sigma}\left(-dv + \frac{\Sigma}{\Delta} dr + a\sin^2\vartheta d\varphi\right)^2 + \frac{\sin^2\vartheta}{\Sigma} \left(adv-(r^2+a^2) d\varphi\right)^2 + \Sigma d\vartheta^2
+ \frac{\Sigma}{\Delta} dr^2
\\
\Sigma =&\; r^2 + a^2 \cos^2\vartheta,
\quad
\Delta = r^2 + a^2 + Q^2 - 2Mr \,,
\label{KNcoeff}
\end{align}
\esub
where $a = J/M$.
The (outer) horizon of a Kerr-Newman black hole is the a $r = r_+=\const$ hypersurface $\cH$ defined by requiring\footnote{In fact, there are two hypersurfaces which satisfying $\Delta = 0$. We are only interested in the behavior at the outer radius $r_+$ .
}
\beq
\Delta = 0\,
\quad\text{i.e.}\quad
r=M+\sqrt{M^2-(a^2+Q^2)}=r_+\,.
\label{Delta0}
\eeq
Then we consider again the $2+1$ dimensional
 structure \cite{DonnayM,MBHhorizon}
whose key ingredients are the induced metric and a vector,
 \besub
\begin{align}
\label{kn_g_h}
\widetilde{g} = &\;g|_{\Delta = 0} = \frac{\sin^2 \vartheta}{\Sigma}\left(a\,dv - (r_+^2+a^2) d\varphi\right)^2 + \Sigma d\vartheta^2\,,
\\[4pt]
\label{kn_xi}
\xi = &\;\partial_v + \Omega_H \partial_\varphi
\where \Omega_H = \frac{a}{r_+^2+a^2}\,.
\end{align}
\esub
Here $\Omega_H$ is the angular velocity of the horizon.

The metric \eqref{kn_g_h} is again singular, as seen from shifting the coordinates as
 $(\vartheta, \varphi, v) \mapsto (\vartheta, \widetilde{\varphi} = \varphi - \Omega_H v, v)$,
\begin{equation}
\widetilde{g} = \frac{(r_+^2+a^2)^2 \sin^2 \vartheta}{\Sigma}\, d\widetilde{\varphi}^2 + \Sigma d\vartheta^2 \,
\qquad \& \qquad
\xi = \partial_v\,.
\label{KNhmetric}
\end{equation}
The kernel is generated by the vector  $\xi$ which takes now the familiar form,  $\widetilde{g}(\xi) = 0$.

Thus we have a degenerate metric and a vector field in its kernel allowing us to conclude that the horizon of a Kerr-Newman black hole carries indeed a Carroll structure $(\IS^2 \times \IR, \widetilde{g}, \widetilde{\xi})$. Viewed in the  original spacetime, $\xi$ is again null on the horizon and generates the ($\IR$ component of the) horizon.

Now if we were to study the trapping of an ``ordinary'' photon (meaning one without extended charge) on the horizon $\cH$ of a Kerr-Newman black hole as we did for Schwarzschild, we would arrive again at the same negative conclusion: we would find no motion.
However the Carroll manifold $\cH$ has $2+1$ dimensions and therefore can accommodate the 2-parameter extension yielding an ``exotic photon" \eqref{exophoton}, which may again be coupled to the electromagnetic field and which \emph{can} move, as we will see it below.

 In Eddington-like coordinates $(v,r,\vartheta,\varphi)$  \cite{Newman65}\footnote{$0\leq \vartheta\leq \pi$ is oriented  with $\vartheta=0$ at the north pole.} the electromagnetic tensor in the Kerr-Newman spacetime is given by,
\beqa
F &= &\;\frac{2aQr \sin \vartheta \cos \vartheta}{(r^2+a^2 \cos^2\vartheta)^2}
\Big(adv-(r^2+a^2) d\varphi\Big) \wedge d\vartheta \nn
\\[8pt]
&& +\, \frac{Q (r^2-a^2\cos^2\vartheta)}{(r_+^2+a^2 \cos^2\vartheta)^2} \Big(a \sin^2\vartheta d\varphi-dv\Big)\wedge dr\,.
\eeqa

The magnetic field is conveniently derived by using the Hodge operator which interchanges the electric and magnetic fields,  $\star F = B_i dx^i \wedge dv + \ldots$. We then identify the magnetic field induced on the horizon ($r=r_+$) as $B = B^r |_{\Delta = 0} = (\star F)^{rv}$, which yields, in our case,
\begin{equation}
\label{kn_b_h}
B = \left.\frac{F_{\theta\varphi}}{\sqrt{|\det g|}}\right|_{\Delta = 0} =
\left(2 a Q r_+(r_+^2+a^2)\right) \,\frac{\cos\vartheta}{\left(r_+^2 + a^2 \cos^2\vartheta\right)^3}\,.
\end{equation}
We note for later reference that the magnetic field is a function of $\vartheta$ only, $B=B(\vartheta)$,
as it follows from the axial symmetry of the Kerr-Newman spacetime. Non-vanishing $B$ is obtained for non-zero  angular momentum $J=aM$ and charge $Q$.
The electromagnetic field induced on the horizon,
\begin{equation}
\label{kn_f_h}
\widetilde{F} =  F\vert_{\Delta = 0} = \frac{a Q r_+ \sin2\vartheta}{(r_+^2+a^2 \cos^2\vartheta)^2}
\Big(adv \wedge d\vartheta + \left(r_+^2+a^2\right) d\vartheta \wedge d\varphi\Big)\,,
\end{equation}
 seems to have a non-vanishing  electric field  component,
\begin{equation}
E_{\vartheta} = \;F_{u \vartheta} = (a^2Q r_+) \,
\dfrac {\sin 2 \vartheta} {(r_+^2 + a^2 \cos^2 \vartheta)^2}\,.
\label{Eonhor}
\end{equation}
However switching to co-moving coordinates, \eqref{KNhmetric}, the coordinate transformation $\varphi \to \tilde{\varphi} = \varphi - \Omega_H v$ with $\Omega_H$ given in \eqref{kn_xi},
\beq
\widetilde{F} \to \hat{F}  = \left(aQr_+(r_+^2+a^2)\right)
\dfrac{\sin 2 \vartheta}{(r_+^2 + a^2\cos^2 \vartheta)^2}\,d\vartheta \wedge d\tilde{\varphi}\,,
\label{ttF}
\eeq
eliminates the electric field \eqref{Eonhor}:
 ${\bE}=0$ on the horizon and we are left with the pure magnetic field \eqref{kn_b_h}, which is indeed left invariant under this coordinate transformation.
\goodbreak

\kikezd{Dynamics on the Kerr-Newman Black Hole horizon}

\begin{figure}[ht]
\includegraphics[scale=.52]{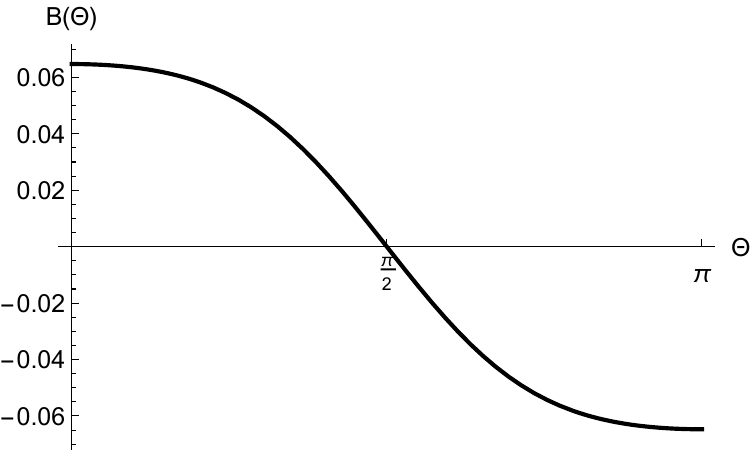}
\qquad\quad
\includegraphics[scale=.52]{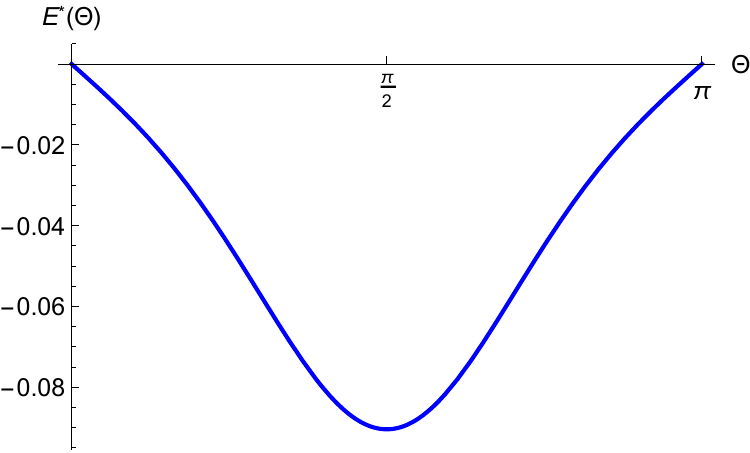}
\\\vspace{-8mm} \null
\hskip2mm(a)\hskip75mm (b)\\\vskip-4mm
\caption{\textit{\small (a) The magnetic and  (b) the effective electric field, $B=B(\vartheta)$ and
\blue{$\bE^* = \bnabla B$}, respectively, on the  horizon of a Kerr-Newman black hole.
}
\label{KNBE*}
}
\end{figure}

Turning to the dynamics of an ``exotic photon'' \eqref{exophoton} we recall that it has no electric charge, and therefore the terms $e\bE$ and  $eB$ are  switched off, leaving us with the effective fields,
\besub
\begin{align}
B^* = &\;\kappa_{\ma}\,,
\label{Bkmag}
\\
E_{\vartheta}^* = &\; \mu\chi\,\p_{\vartheta} B =
- \mu\chi\,\left(2aQr_+ (r_+^2+a^2)\right)\,
 \frac{(r_+^2 - 5 a^2 \cos^2 \vartheta)}{(r_+^2 + a^2\cos^2 \vartheta)^4} \sin \vartheta\,,
\label{E*hori}
\end{align}
\label{B*E*}
\esub
respectively. The fields $B$ and $\bE^* = \bnabla B$ are shown on Fig.~\ref{KNBE*}.\\

Thus the general ({Hall}) equations of motion \eqref{genHall**}
reduce to\footnote{Here the coefficient $\mu\chi$ is included into $\bE^*$ for convenience, cf. \eqref{Estar}.}
\begin{equation}
\label{kn_eom_h}
\hskip-5mm\bigbox{
({x}^\vartheta)^{\prime} = 0,
\;\;\;\;
({x}^\varphi)^{\prime} =
\underbrace{\big(\mu\chi (2aQr_+\big)(r_+^2+a^2)\big)
\frac{(r_+^2 - 5a^2 \cos^2\vartheta)}{(r_+^2 + a^2\cos^2\vartheta)^4}\sin\vartheta
}_{E_{\vartheta}^*}\;\cdot\;\underbrace{\;\big(\frac{1}{\kappa_{\ma}}\big)\;}_{(B^*)^{-1}}
\\
}
\end{equation}
\begin{figure}[!ht]
\includegraphics[scale=0.54]{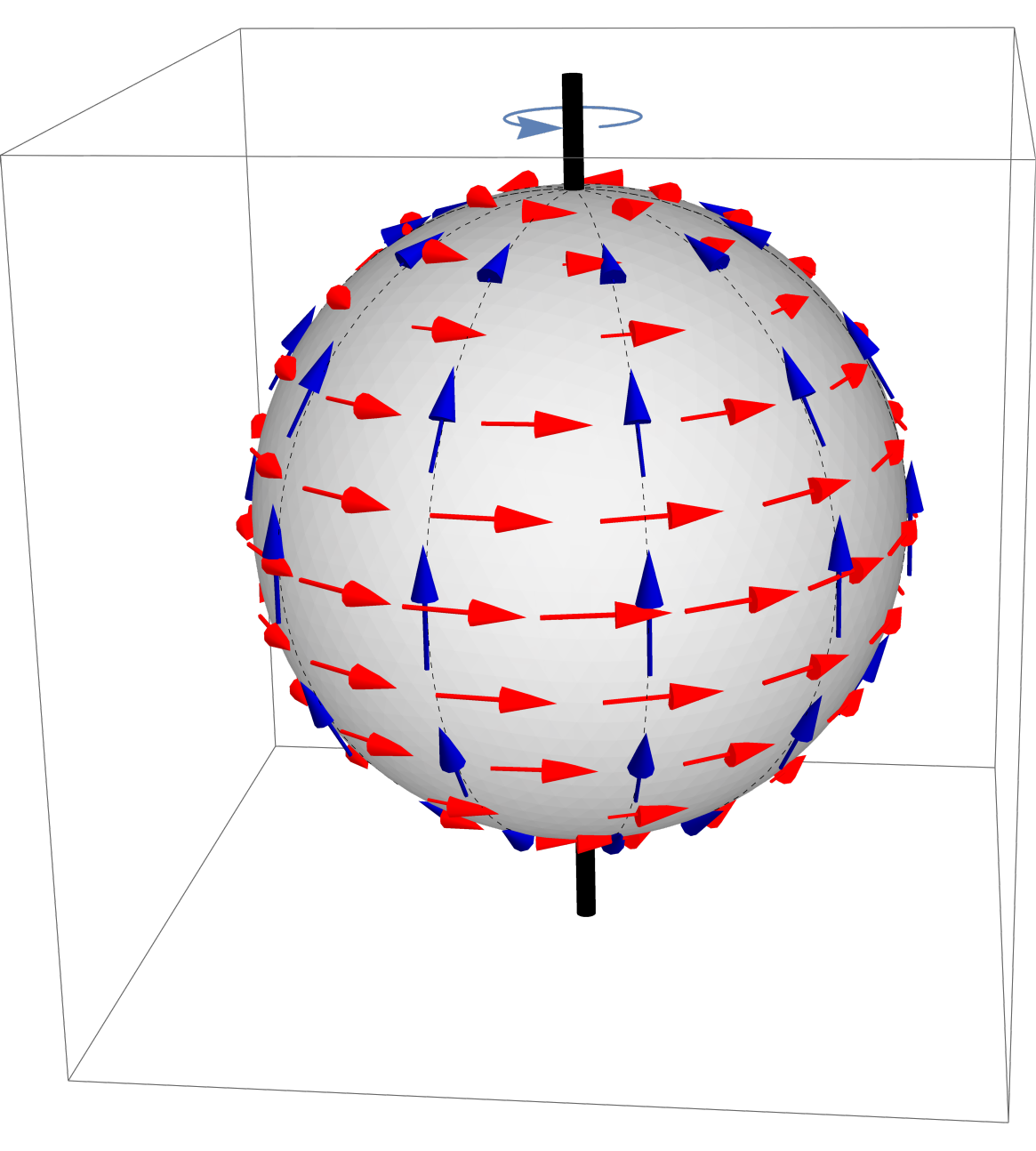}\vskip-3mm
\caption{\textit{\small
The horizon of a Kerr-Newman black hole (topologically a two-sphere) rotates around its vertical axis. For an uncharged particle the usual  EM  fields are switched off but the spin-field term induces an
 \blue{\bf effective electric field} \blue{$\bE^*$} $\propto \bnabla B$ indicated by vertical \blue{\bf blue arrows}.
The horizontal \red{\bf red arrows} depict the  \red{\bf velocity}, \eqref{kn_eom_h}, of a massless, uncharged  doubly-extended anyon  on the horizon. The \blue{\bf force} and the \red{\bf velocity} follow the Hall law~\eqref{Halllaw}
which, in the case of the Kerr-Newman black hole, has the form of \eqref{kn_eom_h}.
 }
\label{f:bh_drift}
}
\end{figure}
\!which is the general Hall law with the cast \eqref{B*E*}. The forces and velocities are depicted in Fig.~\ref{f:bh_drift}.
The trajectories are horizontal ($\vartheta=\const$) circles; the velocity is determined by $E_{\vartheta}^*$, shown in Fig.~\ref{KNBE*}b.
 The angular velocity is maximal on the equator and goes smoothly to zero as we approach the poles.
The angular velocity depends on the radius of the horizon roughly as $r_+^{-3}$, which implies  higher speed for a smaller black hole.
\goodbreak

In conclusion, we get:

\goodbreak
\begin{thm}
\textit{A massless uncharged anyon with non-vanishing exotic charges $\kappa_{\ex}$ and $\kappa_{\ma}$ moves on the horizon of a Kerr-Newman black hole
by following the Hall law \eqref{kn_eom_h}.
}
\label{IX.2}
\end{thm}\vspace{-5mm}

The rotation we  found is reminiscent of the known frame-dragging by a rotating black hole. It is however \emph{unrelated} to it: frame-dragging is hidden in the coordinates we used in \eqref{kn_eom_h}. The ``Carroll time'' coordinate $s$ is defined indeed through $\partial_s = \partial_v + \Omega_H \partial_\varphi$,  see \eqref{kn_xi}, with $\Omega_H$ the angular velocity of the horizon: the coordinates are comoving with the horizon.
Equations \eqref{kn_b_h}-\eqref{kn_eom_h} and Figs.~\ref{KNBE*}-\ref{f:bh_drift} should be compared with those in flat space, discussed in sec.\ref{ExoCarrSec}.

One could, theoretically, consider massless particles with electric charge, extending our flat-case study in sect.\ref{ExoCarrSec} to the curved spaces.
Combining the effective fields would yield a distorted version.
Massless charged (quasi)particles are considered in condensed matter physics \cite{Zoo}.

The general Carrollian equations of motion used to derive \eqref{kn_eom_h} depend on the spin-orbit Hamiltonian \eqref{anymagmom},
inspired from its flat spacetime form. However we cannot be sure of its precise expression in curved spacetime until 2+1 dimensional dynamics are derived from an established theory in the ambient spacetime --- which does not exists yet, to our knowledge.

One could consider a general Hamiltonian of the form $\cH = - \mu \chi \mB(\vartheta)$, for some yet unknown $\mB(\vartheta)$. It is straightforward to see that the conclusion of this section would not change: there would still be motion on the horizon of the black hole, of the same qualitative form: particles would still display an azimuthal motion.

\subsection{Partially broken Carroll and BMS symmetry on the horizon}\label{Carronhorsec}

Carroll manifolds were discussed in connection with
BMS symmetry \cite{BMS} -- for the good reason that the BMS group is the conformal extension of the Carroll group \cite{Bagchi,BMSCarr,ConfCarr}. Let us briefly summarize how the relation goes. For more details, the reader is advised to consult \cite{BMSCarr,ConfCarr}.
The conformal symmetries of a (general) Carroll structure $(\cC = \Sigma \times \IR, g, \xi)$ are
vectorfields on $\cC$ s.t.,
\beq
L_X g = \lambda g,\qquad  L_X \xi = \mu \xi
\with \lambda + 2 \mu = 0\,.
\label{confCarrTr}
\eeq
Such an $X$ has the form,
\beq
X = Y + \left(\frac{\lambda}{2} s + \mT(x)\right) \partial_s
\label{confCarrX}
\eeq
where $(x, s)$ are coordinates on $\cC$ and $Y$ is a conformal vector field of $(\Sigma, g_{\Sigma})$. The $\mT(\bx)$ (called ``the super translations'') are arbitrary functions of the coordinate $\bx$.
So the symmetry group of $(\cC = \Sigma \times \IR, g, \xi)$ is  the semi-direct product $Conf(\Sigma, g_{\Sigma}) \ltimes \cT$ with $\cT$ the group of supertranslations $\mT$.
When $\lambda$ and thus also $\mu$ vanish,  such a transformation is a Carrollian isometry, whose form is
\beq
X = Y + \mT(x)\partial_s
\label{isoCarrX}
\eeq
where $Y$ is a Killing vector of the singular Carroll ``metric'' $g$ on $\Sigma$.

For the round sphere $\Sigma = \IS^2$ with its usual  metric, for example, one has that $\Conf(\IS^2, d\Omega^2) = \SL(2, \IC)$, and thus we get the BMS group: $ \SL(2, \IC) \ltimes \cT$.

For the Kerr-Newman horizon $(\IS^2, \widetilde{g})$ with $\widetilde{g}$ as in \eqref{KNhmetric}, the conformal isometries are found by solving the system of PDE,
\beqa
2 a^2 Y^\vartheta \sin \vartheta \cos \vartheta + (r^2 + a^2 \cos^2 \vartheta)(\lambda - 2 \partial_\vartheta Y^\vartheta) = 0\,,
\\
2(r^2+a^2) Y^\vartheta \cos \vartheta + (r^2+a^2 \cos^2 \vartheta) \sin \vartheta (2 \partial_{\widetilde{\varphi}} Y^{\widetilde{\varphi}} - \lambda) = 0 \,,
\\
(r^2+a^2)^2 \sin^2 \vartheta \partial_\vartheta Y^{\widetilde{\varphi}}  + (r^2+a^2 \cos^2 \vartheta)^2 \partial_{\widetilde{\varphi}} Y^\vartheta = 0
\label{KHconf}
\eeqa
with $Y^\vartheta, Y^{\widetilde{\varphi}}$ and $\lambda$ functions of $(\vartheta, \widetilde{\varphi})$.
For isometries  $\lambda$ (and thus $\mu$) vanish. The Killing vector of the (spatial part of the) horizon is a rotation around the  axis of the black hole \cite{MTW},
\begin{equation}
Y = \partial_{\widetilde{\varphi}}\,.
\end{equation}
The spatial isometries of the Kerr-Newman horizon  $\cH$ are thus the same as that of the full Kerr-Newmann spacetime,  $\SO(2) \ltimes \cT$, generated by the
\begin{equation}
X = \partial_{\widetilde{\varphi}} +
\mT(\vartheta, \widetilde{\varphi}) \partial_s\,.
\end{equation}

The symmetries of the equations of motion \eqref{genHall**},
$
({x}^i)^{\prime} =  \epsilon^{ij}
\big(\frac{E_j^*}{B^*}\big)\,,
$
and ${p}_i^{\prime}=0$,
spelled out on the horizon are conveniently  encoded in the
Lagrangian~\eqref{specLag} [or equivalently, in the Cartan form \eqref{specCartan}]:
\beq
\cL_{spec} = \half
{\kappa_{\ex}}\,\epsilon_{ij}v^i \frac{dv^j}{ds}
 +
\half\kappa_{\ma}\,\epsilon_{ij}
x^i \frac{dx^j}{ds} + \mu\chi B
\,dx^i\,,
\label{specLagbis}
\eeq
where $\bx = (\vartheta, \varphi)$. The magnetic field is a function of $\vartheta$ only, $B=B(\vartheta)$ as seen from \eqref{kn_b_h}.

Recall now the mechanical version of the Noether theorem : a vector field $X$ is a symmetry  if it changes the Cartan form / Lagrangian  by a surface term,
$\delta\cL = df_X$.
Then
\beq
Q_X=
\frac{\p\cL}{\p(x^i)^{\prime}}X^i-f_X
\label{LNoether}
\eeq
is a conserved quantity (see the Appendix).

\begin{itemize}
\item
The only isometries of the black hole spacetime  are ``horizontal" rotations generated by $X = \p_\varphi$ for which we find
$\delta\cL= -d\big(\kappa_{\ma}\vartheta/{2}\big)$, which combines   to yield
\beq
\pi_\varphi = \kappa_{\ma} \vartheta\,.
\label{phishift}
\eeq
Its  conservation is manifest: trajectories of \eqref{kn_eom_h} are ``horizontal'' ($\vartheta = \const$).
This unusual form agrees with the (3.18c) in \cite{Marsot21}.
\item
The zeroth order term of super translations, \ie a  (Carrollian) time-translation $X = \partial_s$ leaves the Lagrangian / Cartan form invariant, providing us with the conserved quantity
\begin{equation}
Q_0=
\mu \chi B = -\mH\,,
\label{CarrAHambis}
\end{equation}
which is indeed the Hamiltonian \eqref{CarrAHam} of a Carroll anyon.

\item
For a general super-translation $X = \mT(\vartheta, \varphi) \partial_s$ we have, instead,
\begin{align}
L_X \varpi =& \mu \chi B \,\partial_i \mT dx^i,
\label{Tthetaphi}
\end{align}
which is \emph{not} closed in general,
$
d\big(B \,d\mT) = dB\wedge d\mT \neq0\,.
$
Thus the magnetic field breaks the supertranslation symmetry in general. However remembering that $B=B(\vartheta)$ on the horizon, cf. \eqref{kn_b_h}, the obstruction vanishes when the supertranslation is ``vertical'',
\beq
\mT=\mT(\vartheta)\,.
\eeq
For a C-boost for example, the
 ${\varphi}$-component must vanish,
$
\mT({\vartheta})=-\beta_1^{\vartheta}{\vartheta}\,.
$
The associated Noether quantity
\beq
Q_{\vartheta} = \mu\chi\left(-B{\vartheta}+ \int\!Bd\vartheta\right)
\label{thetaboostcharge}
\eeq
cf. \# (V.8) in \cite{MZHLett},
 is a combination of the Carroll Hamiltonian $\cH$ \eqref{CarrAHambis}  and the  integral of $B$.
Complicated as it is, $Q_{\vartheta}$ depends on ${\vartheta}$ only, therefore its conservation implies, once again, that the trajectories \emph{must have} ${\vartheta}=\const$ (as we had found it).
On the contrary, ${\varphi}$-boost are broken and thus do not obstruct motion in the ${\varphi}$ direction.

The ``allowed" and ``forbidden'' directions are correlated with the C-boost symmetry. In the example above; the allowed lines are plainly the ``horizontal'' motions
depicted in fig.\ref{f:bh_drift}. The  allowed directions follow the \emph{broken}  C-boost directions and are perpendicular to the residual conserved component of $\bd$. The ``partial immobility", is analogous to the one found  for fractons -- e.g. for ``lineons''  \cite{JainJensen,Bidussi}, or for
``X-cube'' models \cite{Vijay}.

\item
The last class of notable super translation symmetry is when the supertranslation is proportional to  a power of  the magnetic field itself,  e.g.
$
\mT \equiv \mT_n \propto B^n
$
for some  $n=0,\dots$, then,
we get a conserved quantity for each $n$,
$Q_n \propto \big(\mu \chi B\big)^{n+1}=\mH^{n+1}.$

While these symmetries are interesting to notice, they bring no new physical information.

Related investigations of magnetized black hole horizons are reported in~\cite{Kubiznak}.

\end{itemize}

\bigskip
\section{Conclusion}\label{Concl}

Our investigations are devoted to  particles with Carroll symmetry and to the r\^ole of the latter  in Hall-type effects  in various physical instances.

A fundamental question raised from  the very beginning \cite{Leblond,SenGupta} concerns the \emph{mobility} of such particles \cite{Leblond,Carrollvs,MZHLett,Bidussi,JainJensen,Grosvenor:2021hkn}. In the massive case, the  ``no-motion'' statement follows directly from the eqns of motion, \eqref{primeds} and \eqref{emHam} and is due to the absence of a kinetic term in the Hamiltonian. Alternatively, it follows from the conservation of the quantity  \eqref{mCboost} associated with Carroll boost symmetry.

C-boost symmetry can also be partial: it may be broken in one, but not in another direction. In the first case, the particle is ``liberated'' and can move, while in the other direction motion remains forbidden. ``Pinning'' to a fixed point is then replaced by being constrained to one-dimensional submanifolds (``lineons'' \cite{JainJensen})  --- leading to a Hall-type behavior \eqref{Halllaw}.
Typical examples arise in an external electromagnetic field, or on the horizon of a Kerr-Newman black hole.

After having long limited interest in the subject,
the ``no-motion property" of Carroll particles \cite{Leblond,SenGupta,BacryLL,SouriauSDSIV} is now, conversely, \emph{attracting new attention}: for quasiparticles called \emph{fractons} (sect. \ref{fractonSec}) studied in condensed matter physics, immobility is attributed to \emph{conserved dipole momentum} \cite{JainJensen,Bidussi}. The
 relation of the latter to Carroll symmetry has been noticed  --- up to terminology: in ref. \cite{Bidussi} only the time-based one is called ``Carroll'', but it is considered different from a rather mysterious dipole symmetry. The  relation is  clarified in  sec.\ref{fractonSec}:
lifting the problem to one higher dimension we get indeed, that

\benu
\item
In the free case we have  \emph{two, differently implemented} Carroll symmetries: one time ($t$)-based, the other ``internal'' i.e. $s$-based. The
 associated conserved quantities curiously coincide.
 They both extend, in the free case, to an infinite BMS-type symmetry.

\item
Interactions break the $t$-based ``external'', but \emph{not} the $s$-based ``internal'' Carroll implementation. The higher order symmetries are generically broken, though.

\eenu

Which implementation is the ``true'' one? It depends on the viewpoint.
In $d+1$ dimensions the second one is indeed ``internal''
while in the Kaluza-Klein-type $d+2$ - dimensional framework they are both on the same footing up to being built upon different ``time'' coordinates : $t$ or $s$, respectively. But in a gravitational wave, all 4 dimensions are \dots just space-time dimensions of the Universe where natural phenomena happn, and it is the projection to $2+1$ dimensional Newtonian spacetime which becomes a mere technical trick.

The ``no-motion'' property changes for \emph{massless particles} (sec.\ref{s:massless_carr}). The C-boost momentum  vanishes, $\bd=0$, and immobility can not be deduced from its conservation. For a spinless (``Fermat'')  photon (sec.\ref{s:massless_carr}) there is (instantaneous) motion; for a gravitational wave, or on the black hole horizon (sec.\ref{BHh}), it does prevent motion in the  unbroken C-boost direction, as it does for massive particles.

The Carroll group has, in $d=2$ dimensions, a double central extension \cite{Azcarraga,Ancille,Marsot21} with additional parameters we denote by $\kappa_{\ex}$ and $\kappa_{\ma}$, respectively. They are  related to \emph{non-commuting positions} (for $\kappa_{\ex}$),  and to \emph{non-commuting momenta} (for $\kappa_{\ma}$), respectively. The first one was familiar before from the Galilean theory recalled in sec.\ref{GalSec}. The second one is in turn explained by the particular role played by the euclidean subgroup.
Taking them into account, and also that in $d=2$ we can have anyons with any real spin $\chi$ and magnetic moment $\mu$, yields an extended  dynamics, which is however still consistent with the general formula~\eqref{genHall**} in Theorem \ref{ThmIV.7}. The motion of an extended Carroll particle follows a generalized (anomalous) Hall law which includes a Zeeman force acting on the particle in inhomogeneous fields.

After presenting intuitively appealing toy models in sec.\ref{ToyMods} and explaining the relation to the Poincar\'e invariant relativistic theory (sec.\ref{CarrPoinSec}), we illustrated this theory in the gravitational context. Firstly, for the ``memory effect'' of gravitational waves \cite{GenMemory,Carroll4GW,DGHMemory} (sec.\ref{GWmotion}), and then for motion of an hypothetical massless particle with anyonic magnetic moment on the black hole horizon (sec. \ref{BHh}). See also recent investigations of magnetized black hole horizons~\cite{Kubiznak}.

Concerning the physical relevance of these investigations, recent progress with metamaterials makes it conceptually possible to create \emph{condensed matter analogs} ---  including Kerr-Newman black holes \cite{MetaMat}. This hints at the  possibility to \emph{test experimentally} what happens there. Other  laboratory tests might exist in condensed matter physics.

\begin{acknowledgments}\vskip-4mm
The authors are grateful to P.~Chru\'sciel, G.~Gibbons, J.~Hartong, F.~Pe\~{n}a-Ben\'\i tez, M.~Stone, P.~Surowka and F.~Ziegler for discussions and correspondence. LM is grateful to Prof.~T.~Harko for hospitality and discussions at Babe\c{s}-Bolyai University. PMZ was partially supported by the National Natural Science Foundation of China (Grant No. 11975320).
\end{acknowledgments}
\goodbreak


\newpage

\subsection{Appendix}
\appendix{\bf Conservation laws in the presymplectic framework}

\renewcommand{\theequation}{A\thesection.\arabic{equation}}
\renewcommand
\appendix{\appendix}{\setcounter{equation}{0}}

Cartan's formula
$
L_X = d i_X +i_X d
$
implies that the Lie derivative of the Souriau 2-form $\sigma$ vanishes when $i_X\sigma$ is exact,
\beq
i_X\sigma = -df
\label{sigmaconserved}
\eeq
for some function $f$, which is a constant of the motion
\cite{SSD}.

As an example, we consider a NR (Galilean) particle in $(1+1)$ spacetime dimension, with
\beq
\sigma^G =  dp \wedge dx - (p/m) dp \wedge dt\,,
\label{Gal1+1}
\eeq
defined on 1+2 dimensional Galilean evolution space
$\cE=\big\{t,x,p\}$.
A Galilei boost by $b$ is generated by the vector field on $\cE$:
\beq
X^G = (b t)\p_x + mb \p_p\,.
\label{Gboost1+1}
\eeq
Thus \eqref{sigmaconserved}
yields the conserved G-boost momentum:
\beq
\medbox{
f^G =- mx + {p}\,t\,.}
\label{Gbmom}
\eeq
Assume that $\sigma$ is exact, $\sigma = d\omega$. Then Cartan's formula implies that $\omega$ is invariant up to a surface term (physically a gauge transformation),
\beq
L_X\omega = 
dK
\with
K = \omega(X)-f\,,
\label{Cartanvar}
\eeq
providing us with  the conserved quantity, calculated also as
\beq\medbox{
f = \omega(X) - K .}
\label{CarCquant}
\eeq

Eqn \eqref{Cartanvar} can be checked also independently, by calculating
 Lie derivative of the 1-form $\omega = \omega_k dx^k$ using
\beq
L_X\omega = \Big(X^j \frac{\p \omega_i}{\p x^j} +\omega_j \frac{\p X^j}{dx^i}\Big)dx^i\,.
\label{LXomega}
\eeq
In the Galilean example above:
\beq
\omega^G = p dx - (p^2/2m) dt
\Rarrow
L_X\omega^G = d(mb x) \equiv dK
\eeq
yielding again \eqref{Gbmom}.

Coming to Carroll in $(1+1)$ d, the Cartan form and C boost are,
\beq
\omega^C = pdx - mds\,,
\quad
X^C = mb\p_p - b x \p_s\,,
\aand
L_X\omega^C =   2mb dx = dK\,,
\label{CCartnboost}
\eeq
where we used \eqref{LXomega}.
Thus we get the conserved quantity
\beq
\medbox{
f^C = \omega (X) - K = m bx\,.}
\label{Cboostmom}
\eeq
The same result can be recovered also from \eqref{sigmaconserved} with $\sigma^C = dp \wedge dx$,
\beq
i_X\sigma^C = -d(mx) b \,.
\label{Cboostsymm}
\eeq

\end{document}